\documentclass[aps,rmp,reprint,amsmath,amssymb,graphicx,longbibliography]{revtex4-1}

\usepackage{graphicx,amssymb,amsmath,physics,braket,caption,subcaption,bm}
\usepackage{xspace}

\usepackage[final]{showkeys}

\newcommand{\ee}{\mathrm{e}}
\newcommand{\ii}{\mathrm{i}}
\newcommand{\Hham}{\hat{H}_\textrm{H}}
\newcommand{\Heff}{\hat{H}_\textrm{eff}}
\newcommand{\Hsys}{\hat{H}_\textrm{sys}}
\newcommand{\Ps}{P_\textrm{surv}}
\newcommand{\comp}{\hat{I}_\infty}
\newcommand{\iden}{\hat{I}}
\newcommand{\zero}{\hat{O}}
\newcommand{\Sch}{Schr\"{o}dinger\xspace}
\newcommand{\hatA}{\hat{A}}
\newcommand{\hatB}{\hat{B}}
\newcommand{\hatC}{\hat{C}}
\newcommand{\hatD}{\hat{D}}
\newcommand{\hatH}{\hat{H}}
\newcommand{\hatL}{\hat{L}}
\newcommand{\hatP}{\hat{P}}
\newcommand{\hatQ}{\hat{Q}}
\newcommand{\hatT}{\hat{T}}
\newcommand{\hatU}{\hat{U}}
\newcommand{\hattilU}{\hat{\tilde{U}}}
\newcommand{\hatX}{\hat{X}}
\newcommand{\hatY}{\hat{Y}}
\newcommand{\hatZ}{\hat{Z}}
\newcommand{\PHP}{\hatP\hatH\hatP}
\newcommand{\PHQ}{\hatP\hatH\hatQ}
\newcommand{\QHP}{\hatQ\hatH\hatP}
\newcommand{\QHQ}{\hatQ\hatH\hatQ}

\begin{document}

\title{Non-Hermitian Quantum Mechanics of Open Quantum Systems: Revisiting The One-Body Problem}

\author{Naomichi Hatano}
\affiliation{Institute of Industrial Science, The University of Tokyo, Kashiwa, Chiba 277-8574, Japan}
\author{Gonzalo Ordonez} 
\affiliation{Department of Physics and Astronomy, Butler University, Gallahue Hall, 4600 Sunset
Avenue, Indianapolis, Indiana 46208, USA}

\date{\today{}}

\begin{abstract}
We review analyses of open quantum systems within the one-body problem.
Open quantum systems are systems, possibly complicated but with a finite number of degrees of freedom, to which systems, possibly with an infinite number of degrees of freedom but simple, are coupled.
We show how non-Hermiticity arises in an open quantum system with an infinite environment, focusing on the one-body problem.
One of the reasons for taking the present approach is that we can solve the problem completely, making it easier to see the structures of problems involving open quantum systems.
We show that this results in the discovery of a new complete set, which is one of the main topics of the present article.

Another reason for focusing on the one-body problem is that the theory permits the strong coupling between the system and the environment.
For systems with interactions, particularly within the environment, the Born-Markov approximation is quite often used, although the dynamics of open quantum systems is generally non-Markovian.
Since the Born approximation is valid only in a weak-coupling regime, phenomena that would emerge uniquely in strong-coupling regimes are yet to be pursued. 
In the current research landscape, it is valuable to revisit the one-body problem for open quantum systems, which can be solved accurately for arbitrary strengths of the system-environment couplings.
A rigorous understanding of the problem structures in the present approach will be helpful when we tackle problems with many-body interactions.

First, we consider potential scattering and directly define the resonant state as an eigenstate of the Schr\"{o}dinger equation under the Siegert outgoing boundary condition.
We show that the resonant eigenstate can have a complex energy eigenvalue, even though the Hamiltonian is seemingly Hermitian.
We resolve common puzzlementss about the resonant states, including the appearance of complex eigenvalues and the divergence of the eigenfunctions.
In this direct formalism, the non-Hermiticity of open quantum systems is hidden in the boundary condition.

Second, we introduce the Feshbach formalism, which eliminates the infinite degrees of freedom of the environment and represents its effect as a complex potential.
The non-Hermiticity hidden in the direct formulation manifests as the complexity in the Feshbach formalism.
The resulting effective Hamiltonian is explicitly non-Hermitian.
By unifying these two ways of defining resonant states, we obtain a new complete set of bases for the scattering problem that contains all discrete eigenstates, including resonant states.

We finally mention the non-Markovian dynamics of open quantum systems.
In particular, we show that the system's openness emerges as non-Markovianity in both the short- and long-time regimes.
We emphasize the time-reversal symmetry of the dynamics that continuously connects the past and the future.
We can capture it using the new complete set that we develop here.
\end{abstract}


\maketitle

\tableofcontents{}

\section{Introduction: Target of the present article}
\label{sec1}

\begin{figure}
\includegraphics[width=0.8\columnwidth]{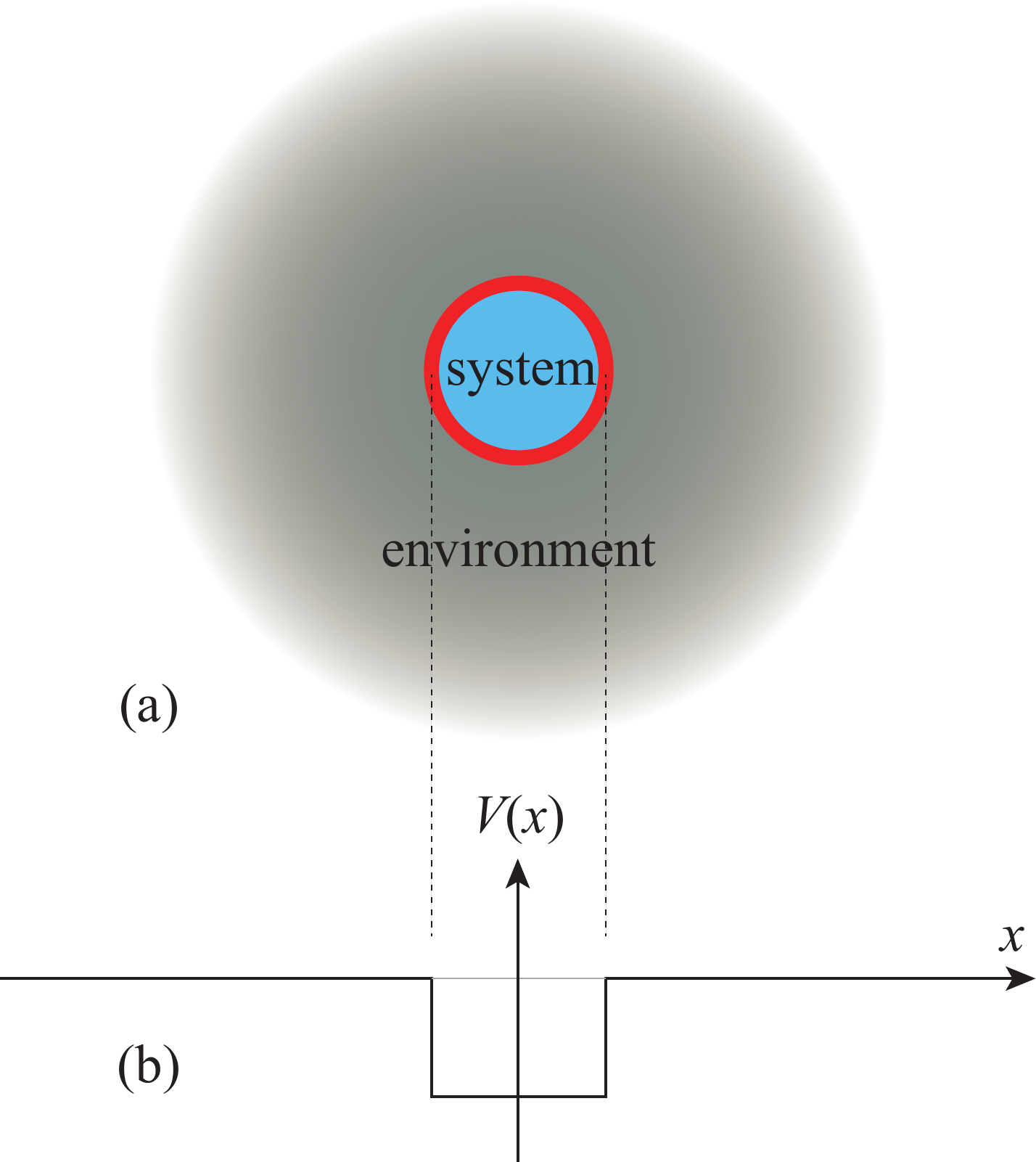}
\caption{(a) A schematic view of the system and the environment of an open quantum system. The system indicated by a small circle is embedded in the environment indicated by a big circle, with a coupling between them indicated by a solid red curve.
(b) In the problem of potential scattering, the potential area is identified as the system, while the rest, flat space is identified as the environment.}
\label{fig1}
\end{figure}

In this article, we review analyses of open quantum systems within the one-body problem.
Open quantum systems (see \textit{e.g.} Refs.~\cite{Breuer-Petruccione,rivas11, Vacchini} for general textbooks) are systems, possibly complicated but with a finite number of degrees of freedom, to which systems, possibly with an infinite number of degrees of freedom but simple, are coupled.
We hereafter refer to the former simply as the system, while to the latter as the environment; see Fig.~\ref{fig1}(a).
In this article, we repeatedly stress that non-Hermiticity in an open quantum system arises only when the environment has an infinite number of degrees of freedom.

We can classify problems in open quantum systems into three categories, depending on whether interactions occur within the system or in the environment.
If we assume that there are no interactions within the system or the environment, the problem reduces to a one-body problem and is mostly solvable.
When we assume interactions in the system but not in the environment, the problem may still be solvable if we can successfully diagonalize the system Hamiltonian.
Finally, if we want a heat bath for the environment, we would need interactions within it as well.
We would most likely need approximations for the third category.

We here focus on the problem in the first category.
One of the reasons for taking the present approach is that we can solve the problem completely,
which makes it easy to see the structures of problems involving open quantum systems.
Indeed, this results in the discovery of a new complete set~\cite{Hatano14}, which we will review below.

Another reason is that the theory permits the strong coupling between the system and the environment.
For the problems in the third category in particular, because of various difficulties in solving them,
the Gorini-Kossakowski-Sudarshan-Lindblad (GKSL) equation~\cite{Gorini76, Lindblad76} after Born-Markov approximations~\cite{Breuer-Petruccione} is quite often used, although the dynamics of open quantum systems is generally non-Markovian.
Since this approximation is valid only in a weak-coupling regime, phenomena that would emerge uniquely in strong-coupling regimes are yet to be pursued. 

In the current research landscape, it is valuable to revisit the one-body problem for open quantum systems, which can be solved accurately for arbitrary strengths of the system-environment couplings.
A rigorous understanding of the problem structures in the present approach will be helpful when we tackle problems in higher categories.

The problem of open quantum systems dates back to Gamow in 1928~\cite{Gamow28}, when he introduced a non-Hermitian component to the Hamiltonian to explain alpha decay.
The theory was then developed primarily in nuclear physics~\cite{Humblet61, Rosenfeld61, Humblet62, Humblet64-1, Jeukenne64, Humblet64-2, Mahaux65, Rosenfeld65} and related mathematical physics~\cite{Peierls59, leCouteur60, Zeldovich60, Hokkyo65, Romo68}, although it was not referred to as an open quantum system at the time.

There are two storylines.
On one hand, the definition of the resonant state with a complex energy eigenvalue was developed, particularly by Siegert in 1939~\cite{Siegert39}.
The resonant state was then defined as an eigenstate of the \Sch equation under a specific boundary condition.
On the other hand, Feshbach in 1958~\cite{Feshbach58, Feshbach62} justified the introduction of a non-Hermitian component to the Hamiltonian as he eliminated the components of the environmental infinite degrees of freedom.

In the present article, we provide an overview of these results. 
We unify the two storylines and, as a consequence, find a new complete set of bases for the scattering problem based on resonant states.

Accordingly, we define resonant states in two ways.
First, in Sec.~\ref{sec2}, we consider potential scattering and directly define the resonant state as an eigenstate of the \Sch equation under the Siegert boundary condition.
We show that the resonant eigenstate can have a complex energy eigenvalue, even though the Hamiltonian is seemingly Hermitian.
In this direct formalism, the non-Hermiticity of the open quantum system is hidden in the boundary condition.

Second, in Sec.~\ref{sec3}, we introduce the Feshbach formalism, with which we eliminate the infinite degrees of freedom of the environment, and represent its effect as a complex potential.
The non-Hermiticity hidden in the direct formulation manifests as the complexity in the Feshbach formalism.
The resulting effective Hamiltonian is explicitly non-Hermitian.

We then mention the non-Markovian dynamics of open quantum systems in Sec.~\ref {sec4} before summarizing the article in Sec.~\ref{sec5}.
In particular, we show that the system's openness emerges as non-Markovianity in both the short- and long-time regimes.
We emphasize the time-reversal symmetry of the dynamics around $t=0$, which we can capture using the new complete set that we develop in Sec.~\ref{sec3}.

\section{Resonant states in open quantum systems}
\label{sec2}

As schematically shown in Fig.~\ref{fig1}(b), potential scattering is one of the simplest nontrivial versions of the open quantum system.
We identify the potential area of a finite range as the system and the flat outer space that extends infinitely as the environment.
In this section, we show the appearance of resonant eigenstates with complex energy eigenvalues, indicating the problem's non-Hermiticity.
We primarily work in continuous space, but we also translate it into a lattice problem to provide a more straightforward explanation in the next Sec.~\ref{sec3}.

\subsection{Potential scattering as an open quantum system}
\label{subsec2-1}

Let us consider the problem of potential scattering with the standard time-independent \Sch equation in one dimension:
\begin{align}\label{eq10}
\qty(-\frac{\hbar^2}{2m}\dv[2]{x}+V(x))\psi(x)=E\psi(x).
\end{align}
Here and hereafter, we assume that the potential function $V(x)$ is real and its support is compact;
it takes nonzero real values only within a finite range:
\begin{align}
V(x)\equiv 0 \quad\mbox{for}\quad \abs{x}>\ell ,
\end{align}
where $\ell$ is a finite positive number.
In other words, we exclude potentials of infinite range.
At this moment, we do not know how we can extend the analyses below to the case of infinite-range potentials, such as the Coulomb potential.

In this and the following subsections, we will define the resonant state, which indeed has a complex energy eigenvalue, as a solution of the \Sch equation~\eqref{eq10} under the boundary condition of out-going waves only:
\begin{align}\label{eq30}
\psi(x)\propto \ee^{\ii K |x|} \quad\mbox{for}\quad \abs{x}>\ell,
\end{align}
where $E=(\hbar K)^2/(2m)$. 
This was first proposed by Siegert in 1939~\cite{Siegert39}; hence, we refer to it as the Siegert boundary condition.
We explain the rationale for this definition of the resonant state below.

Note that we will show below that $K$ and $E$ can be complex under the Siegert boundary condition.
Here and hereafter, we use the capital $K$ to indicate that the value can be complex, whereas we use the lowercase $k$ to indicate that it is either real or the real part of $K$.
For $\Re K>0$, the boundary condition~\eqref{eq30} means that the wave is right-going for $x>\ell$ and left-going for $x<-\ell$.
We will refer to such an eigenstate as the resonant state.
For $\Re K<0$, the boundary condition~\eqref{eq30} means that the wave is left-going for $x>\ell$ and right-going for $x<-\ell$.
We will refer to such an eigenstate as the anti-resonant state.

We begin our explanation with the usual scattering problem.
We assume an incident wave on the left of the potential, a reflection wave on the left, and a transmission wave on the right:
\begin{align}\label{eq40}
\psi(x)=\begin{cases}
A\ee^{+\ii k x}+B\ee^{-\ii k x}&\mbox{for $x<-\ell$},\\
C\ee^{+\ii k x}&\mbox{for $x>+\ell$}.
\end{cases}
\end{align}
We then derive the relations among the amplitudes $A$, $B$, and $C$ from the connection conditions.

\begin{figure}
\includegraphics[width=0.8\columnwidth]{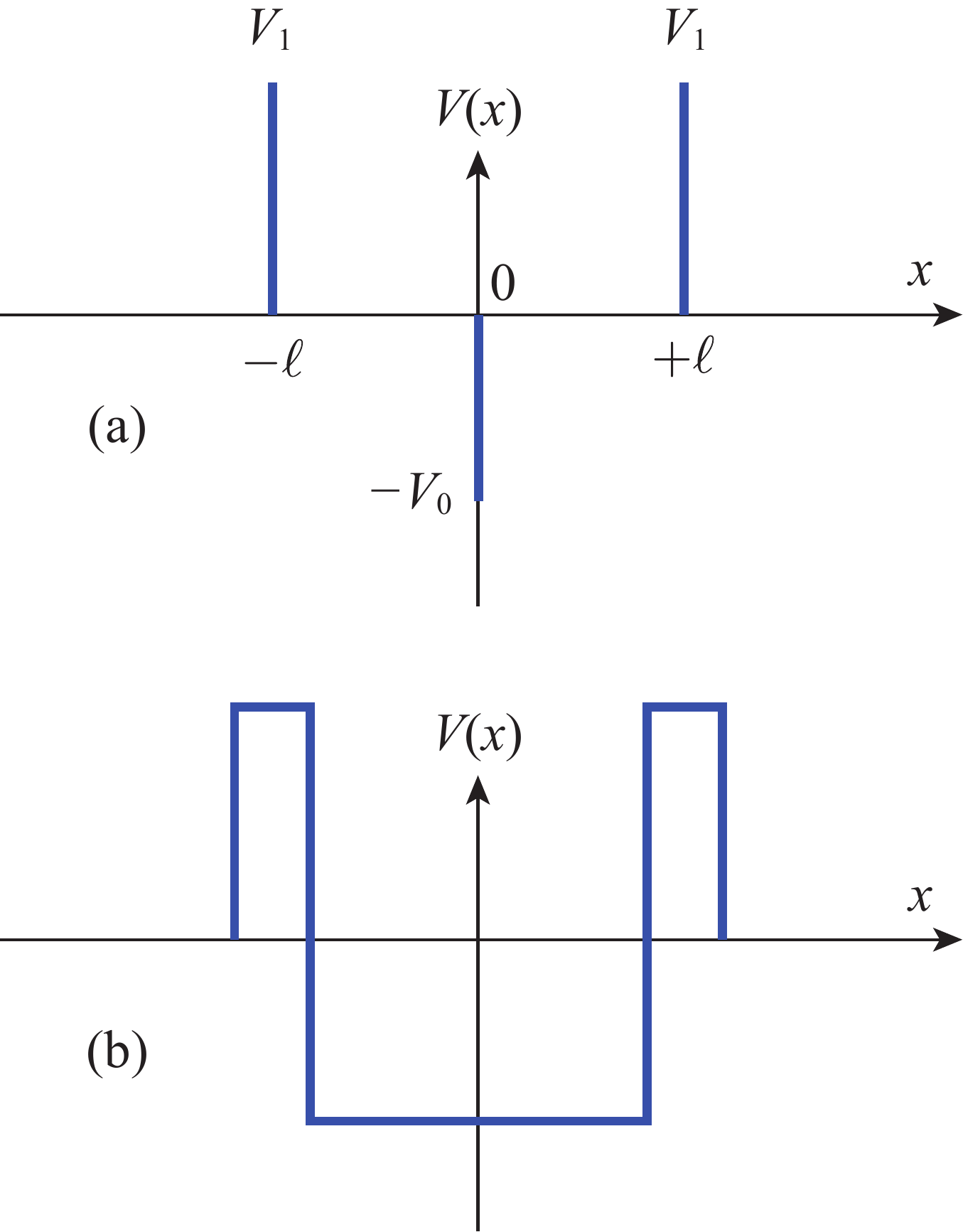}
\caption{(a) The potential of three delta functions in Eq.~\eqref{eq50}. (b) A schematic potential we have in mind.}
\label{fig2}
\end{figure}

For example, let us analyze the case of a set of delta potentials 
\begin{align}\label{eq50}
V(x)=-V_0\delta(x)+V_1\qty(\delta(x+\ell)+\delta(x-\ell)),
\end{align}
where $V_0>0$ and $V_1>0$, as in Fig.~\ref{fig2}(a), which would mimic the situation in Fig.~\ref{fig2}(b), which would further mimic a shell-model potential in nuclear and atomic physics.
We present details of the calculation in App.~\ref{appA-1}.
We  obtain the transmission amplitude in the form
\begin{align}\label{eq60}
t_\textrm{amp}&:=\frac{C}{A}=\frac{1}{T_{11}},
\end{align}
where the $(1,1)$ component of the transfer matrix, $T_{11}$, is given by Eq.~\eqref{eqA80},
and the transmission coefficient in the form
\begin{align}\label{eq70}
\mathcal{T}&:=\abs{\frac{C}{A}}^2=\frac{1}{\abs{T_{11}}^2}.
\end{align}
We will plot this along with the resonance poles below in Figs.~\ref{fig4} and~\ref{fig6}.

In the standard textbook of quantum mechanics, a resonant state is often defined as a pole of the transmission coefficient.
The expression~\eqref{eq60} then implies that the pole of the transmission amplitude $t_\textrm{amp}$ is given by the zero of the amplitude $A$ and that the wave function that corresponds to the pole is found by putting $A$ to zero from the very beginning in Eq.~\eqref{eq40}.
This is the rationale of setting the boundary condition in the form~\eqref{eq30} to find the resonance poles.

\subsection{Finding the resonant states using the Siegert boundary condition}
\label{subsec2-2}

Setting the Siegert boundary condition
\begin{align}\label{eq80}
\psi(x)=\begin{cases}
B\ee^{-\ii K x}&\mbox{for $x<-\ell$},\\
C\ee^{+\ii K x}&\mbox{for $x>+\ell$}
\end{cases}
\end{align}
makes the calculation of the resonance poles quite easier than the one presented in App.~\ref{appA-1}.
One point to note here is the reduction of the number of unknown variables.

In solving the scattering problem, as in the previous Subsec.~\ref{subsec2-1}, we have the unknown variables $k$, $A$, $B$, and $C$ in Eq.~\eqref{eq40} as well as $J$, $M$, $F$, and $G$ in Eq.~\eqref{eqA10}.
Since we need only the ratios of the amplitudes, the number of unknown variables is seven.
To determine them, we have six conditions given in Eq.~\eqref{eqA20}, and hence there is one variable that we cannot determine.
In fact, we make the wave number $k$ of the incident wave a control parameter in solving the scattering problem.

On the other hand, we set $A=0$ in the Siegert boundary condition~\eqref{eq80}, thereby reducing the number of unknown variables by one. 
We now have six unknown variables to be determined by six conditions~\eqref{eqA20}.
This allows us to find discrete solutions to $K$, which we will show in the present subsection.

Since the potential function~\eqref{eq50} is an even function, we can now classify the resonant states of the form~\eqref{eq80} into those with even and odd parities.
This simplifies the computation considerably, as we show below.

Let us first consider the simple case of $V_0=0$, which leaves us only the positive delta peaks at $x=\pm \ell$~\cite{Hatano08}.
In this case, we can generally assume the following with the Siegert boundary condition~\eqref{eq80}:
\begin{align}\label{eq91}
\psi(x)=\begin{cases}
B \ee^{-\ii K x}&\mbox{for $x<-\ell$},\\
F\ee^{+\ii K x} \pm G \ee^{-\ii K x}&\mbox{for $-\ell\leq x\leq \ell$},\\
C\ee^{+\ii K x}&\mbox{for $x>+\ell$},
\end{cases}
\end{align}
but for a fixed parity, we can further assume
\begin{align}\label{eq93}
B=\pm C,\qquad G=\pm F,
\end{align}
where the upper sign corresponds to even-parity solutions and the lower sign to odd-parity ones throughout the present subsection.
Since we need only the ratios of the amplitudes, the unknown variables are reduced to the following two: $K$ and $F/C$.

We determine them with the following two connection conditions:
\begin{subequations}\label{eq92}
\begin{align}
\label{eq100b}
\psi(+\ell-\epsilon)&=\psi(+\ell+\epsilon),\\
\label{eq100c}
\psi'(+\ell-\epsilon)&=\psi'(+\ell+\epsilon)-2v_1\psi(+\ell),
\end{align}
\end{subequations}
\newcounter{remcount}
\setcounter{remcount}{\value{equation}}
where
\begin{align}
v_1:=\frac{mV_1}{\hbar^2}.
\end{align}
The second condition~\eqref{eq100c} is found by integrating the \Sch equation over an infinitesimally narrow region $[+\ell-\epsilon, +\ell+\epsilon]$. 

Using Eq.~\eqref{eq91} in Eq.~\eqref{eq92}, we obtain
\begin{subequations}
\begin{align}
F \qty(\ee^{+\ii K\ell}\pm\ee^{-\ii K\ell})&=C\ee^{+\ii K\ell}\\
\ii K F \qty(\ee^{+\ii K\ell}\mp\ee^{-\ii K\ell})&=\ii C K\ee^{+\ii K\ell}-2v_1 C\ee^{+\ii K\ell}
\end{align}
\end{subequations}
Multiplying both sides by $\ee^{-\ii K\ell}$ and inserting $C$ on the right-hand side of the first equation into that of the second one, we end up with
\begin{align}\label{eqA120}
\ee^{2\ii K\ell}=\mp\qty(1-\ii\frac{K}{v_1}),
\end{align}
In fact, these equations are found from setting $T_{11}$ in Eq.~\eqref{eqA80} to zero with $v_0:=mV_0/\hbar^2=0$, as in
\begin{align}
T_{11}=\frac{1}{k^2}\qty[-\qty(1-\ii\frac{k}{v_1})^2+\ee^{4\ii k \ell}]=0,
\end{align}
although we do not immediately know the parity of the solutions from it.

\begin{figure}
\vspace{\baselineskip}
\includegraphics[width=0.7\columnwidth]{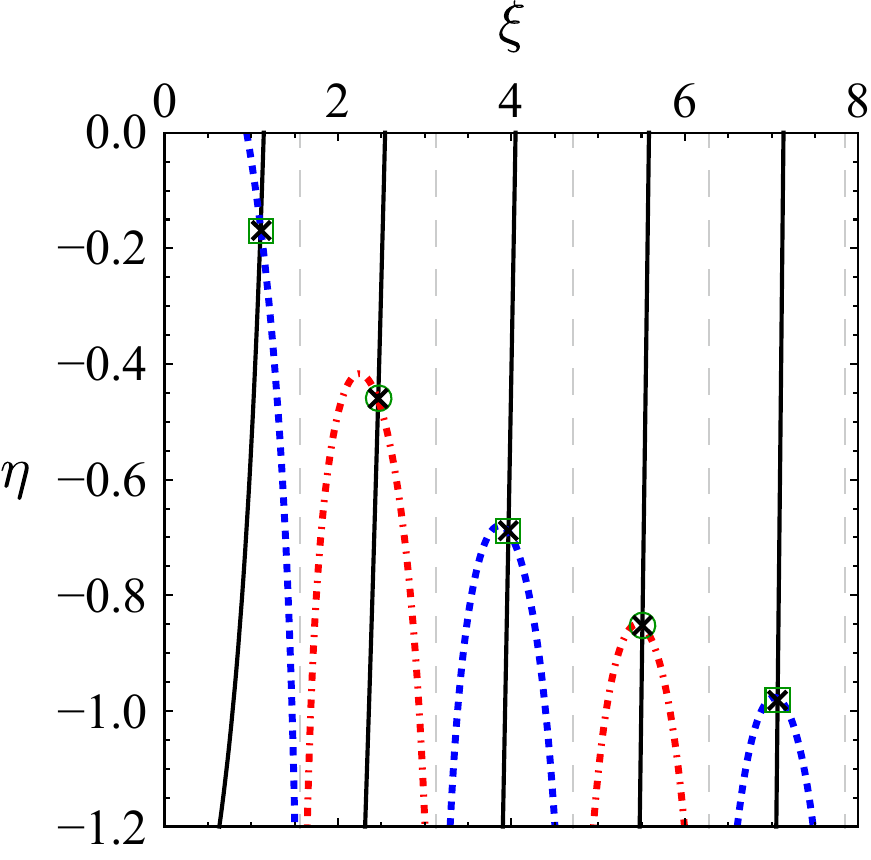}
\caption{Plots of the two curves Eq.~\eqref{eq110a} (dashed blue curves for even solutions and dot-dashed red curves for odd solutions) and Eq.~\eqref{eq110b} (solid black curves).
The vertical broken gray lines indicate multiples of $\pi/2$ on the $\xi$ axis.
The crossing points indicated by crosses with green squares indicate resonant states of even parity and those indicated by crosses with green circles indicate resonant states of odd parity.
The example is given for $\alpha_1=1$ with $V_0=0$.}
\label{fig3}
\end{figure}

Writing $K=k+\ii\kappa$ and introducing the dimensionless variables $\xi:=k\ell$, $\eta:=\kappa \ell$ and $\alpha_1:=\ell v_1$, we can tranform Eq.~\eqref{eqA120} into
\begin{align}\label{eq105}
\ee^{2\ii\xi-2\eta}=\mp\qty(1-\frac{\ii \xi-\eta}{\alpha_1}).
\end{align}
Taking the real and imaginary parts of both sides of Eq.~\eqref{eq105}, we have 
\begin{subequations}\label{eqA130}
\begin{align}
\ee^{-2\eta}\cos(2\xi)&=\mp\qty(1+\frac{\eta}{\alpha_1}),\\
\ee^{-2\eta}\sin(2\xi)&=\pm\frac{\xi}{\alpha_1}.
\end{align}
\end{subequations}
which we can transform into
\begin{subequations}\label{eq110}
\begin{align}\label{eq110a}
\eta&=-\frac{1}{2}\ln\qty[\pm\frac{\xi}{\alpha_1}\csc(2\xi)],\\
\label{eq110b}
\eta&=-\xi\cot(2\xi)-\alpha_1.
\end{align}
\end{subequations}
We can find the solutions $\{\xi_n,\eta_n\}$ as the crossing points of the two curves, as shown in Fig.~\ref{fig3};
note that the crossing points also exist on the negative side of $\xi$ symmetrically because the right-hand sides of both of Eq.~\eqref{eq110} are even functions of $\xi$.
We can numerically find accurate solutions using the Newton-Raphson method, with an initial guess provided by a rough estimate that we read off from Fig.~\ref{fig3}~\cite{Hatano08}.

\begin{figure}
\includegraphics[width=0.8\columnwidth]{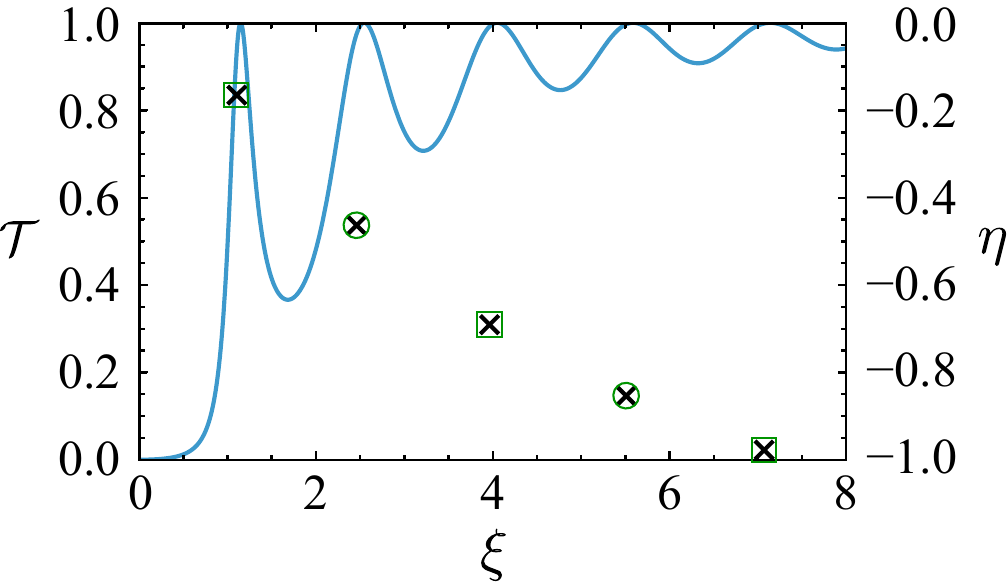}
\caption{The transmission coefficient $\mathcal{T}$ (blue curve) for $v_0=0$ and the locations $\{\xi_n,\eta_n\}$ of the resonance poles, which are the same as the ones in Fig.~\ref{fig3}.
For the values of the former and the latter, see the left and right vertical axes, respectively.
The example is given for $\alpha_1=1$ with $V_0=0$.}
\label{fig4}
\end{figure}

Figure~\ref{fig4} shows good agreement between the transmission coefficient~\eqref{eq70} with $v_0=0$ and the locations of the resonance poles.
For each resonance eigen-wave-number, the real part gives the peak location, and the imaginary part roughly gives the peak width.

In fact, this agreement is not always true.
We can confirm that the real wave number
\begin{align}\label{eq115}
k\ell \cot(2k\ell)+\ell v_1=0,
\end{align}
which we obtain by setting $\eta$ to zero in Eq.~\eqref{eq110b}, gives the point of perfect transmission $\mathcal{T}=1/\abs{T_{11}}^2=1$.
This means that, in this particular case, the agreement is good for a small value of $\eta$.
We will see in the next example that this agreement breaks.

To see it, we introduce a positive value of $V_0$.
This actually does not affect the solutions of odd parity, because they do not couple to the potential at the origin.
For the solutions of even parity, we replace Eq.~\eqref{eq110} with
\begin{align}\label{eq90}
\psi(x)=\begin{cases}
C \ee^{-\ii K x}&\mbox{for $x<-\ell$},\\
F\ee^{-\ii K x} + G \ee^{+\ii K x}&\mbox{for $-\ell\leq x\leq 0$},\\
F\ee^{+\ii K x} + G \ee^{-\ii K x}&\mbox{for $0\leq x\leq +\ell$},\\
C\ee^{+\ii K x}&\mbox{for $x>+\ell$}.
\end{cases}
\end{align}
The two connection conditions~\eqref{eq92} are also supplemented by
\begin{align}
\label{eq100a}
\psi'(-\epsilon)&=\psi'(+\epsilon)+2v_0\psi(0), 
\tag{\arabic{remcount}c}
\end{align}
where
\begin{align}
v_0:=\frac{mV_0}{\hbar^2}.
\end{align}
With these three connection conditions~\eqref{eq100b}, \eqref{eq100c}, and~\eqref{eq100a}, we fix the ratios $F/C$ and $G/C$ along with the point spectra of $K$.

Using the form~\eqref{eq90} in the three connection conditions, we have
\begin{subequations}
\begin{align}
\qty(F\ee^{+\ii K\ell}+G\ee^{-\ii K\ell})&=C\ee^{+\ii K\ell}\\
\ii K  \qty(F\ee^{+\ii K\ell}-G\ee^{-\ii K\ell})&=\ii C K\ee^{+\ii K\ell}-2v_1 C\ee^{+\ii K\ell},\\
\label{eq22c}
-\ii K (F-G)&=\ii K(F-G)+2v_0(F+G),
\end{align}
\end{subequations}
Eliminating $C$ from the first two equations, we have
\begin{align}
\ii K  \qty(F\ee^{2\ii K\ell}-G)&=(\ii K-2v_1)\qty(F\ee^{2\ii K\ell}+G),
\end{align}
or
\begin{align}\label{eq24}
v_1\ee^{2\ii K\ell}F=\qty(\ii K-v_1)G.
\end{align}
From the third condition~\eqref{eq22c}, on the other hand, we have
\begin{align}\label{eqA150}
(\ii K+v_0)F=(\ii K-v_0)G.
\end{align}
Combining Eqs.~\eqref{eq24} and~\eqref{eqA150}, we obtain
\begin{align}
(\ii K-v_0)\ \ee^{2\ii K \ell}=-(\ii K+v_0)\qty(1-\ii \frac{K}{v_1}),
\end{align}
Expressing this equation in terms of the dimensionless quantities, we have 
\begin{align}\label{eq120}
(\ii \xi-\eta-\alpha_0)\ \ee^{2\ii \xi-2\eta}=-(\ii \xi-\eta +\alpha_0)\qty(1-\frac{\ii\xi-\eta}{\alpha_1}),
\end{align}
where $\alpha_0=v_0 \ell$.
This equation is more complicated than in the case of $v_0=0$ in Eq.~\eqref{eq105},
and it seems impossible to find the explicit equations for the real and imaginary parts of the form~\eqref{eq110}.

\begin{figure}
\includegraphics[width=0.7\columnwidth]{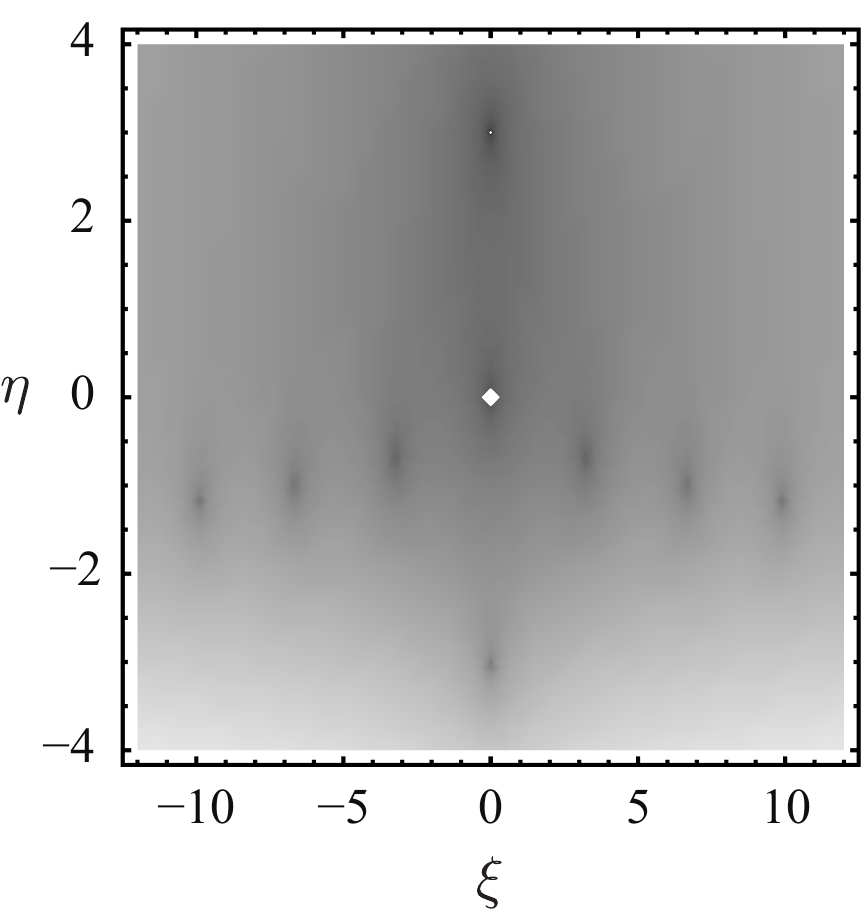}
\caption{The density plot of the function~\eqref{eq150}. The black spots indicate negative infinity. The example is given for $\alpha_0=3$ and $\alpha_1=1$.}
\label{fig5}
\end{figure}

In such a case, we propose the following approach to obtain numerically accurate solutions.
We first make a density plot of the following function in the complex $K$ plane:
\begin{widetext}
\begin{align}\label{eq150}
f(\xi,\eta):=\log\abs{(\ii \xi-\eta-\alpha_0)\ \ee^{2\ii \xi-2\eta}+(\ii \xi-\eta +\alpha_0)\qty(1-\frac{\ii\xi-\eta}{\alpha_1})}.
\end{align}
\end{widetext}
Then the solutions of Eq.~\eqref{eq120} should appear as dark spots that represent negative infinity, as is exemplified in Fig.~\ref{fig5}.
We can then use the Newton-Raphson method in two dimensions to obtain numerically accurate solutions $\{\xi_n,\eta_n\}$, equating the real and imaginary parts of Eq.~\eqref{eq120}, with an initial guess set as a rough estimate that we read off from Fig.~\ref{fig5}.

\begin{figure}
\includegraphics[width=0.8\columnwidth]{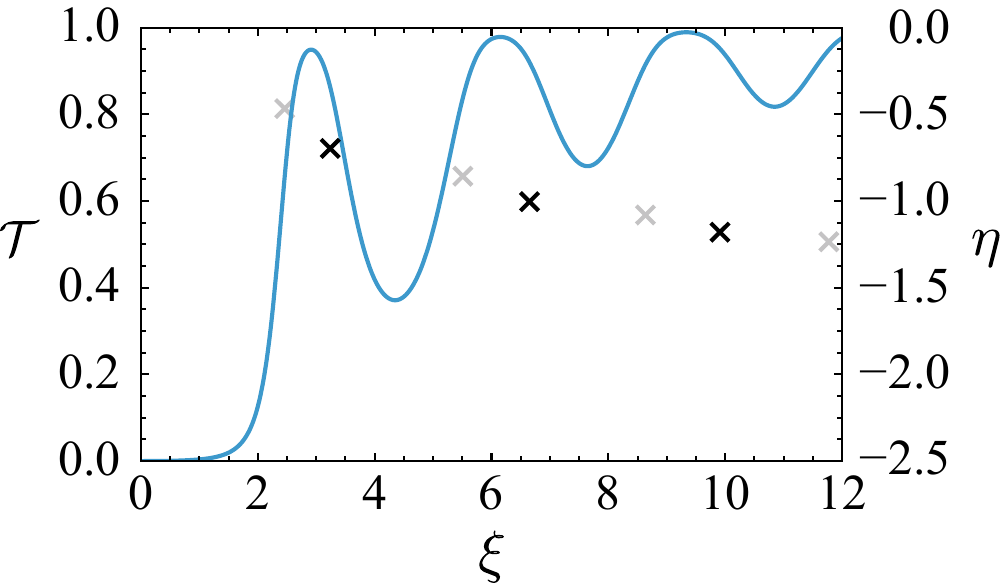}
\caption{The transmission coefficient $\mathcal{T}$ (blue curve) and the locations $\{\xi_n,\eta_n\}$ of the resonance poles of even parity (black crosses) as well as the unchanged poles of odd parity (gray crosses).
For the values of $\mathcal{T}$ and the location of both poles, see the left and right vertical axes, respectively.
The example is given for $\alpha_0=3$ and $\alpha_1=1$.}
\label{fig6}
\end{figure}

We plot the transmission coefficient~\eqref{eq70} and the locations of the resonance poles in Fig.~\ref{fig6} for $V_0>0$.
We note that there is only one broad resonance peak for a pair of odd-parity and even-parity resonance poles in this case; see App.~\ref{appB} for more details regarding the transition from Fig.~\ref{fig4} to Fig.~\ref{fig6}.

The introduction of $V_0$ breaks the good agreement between the pole locations and the resonance peaks.
This implies that only checking the resonance peaks may miss hidden resonance poles.
Note that resonance poles are more physical entities than resonance peaks;
unstable nuclides with finite lifetimes, such as nihonium~\cite{nihonium}, are nothing but resonant states.
This underscores the importance of pursuing the resonant states with complex eigenvalues.

\begin{figure}
\includegraphics[width=0.8\columnwidth]{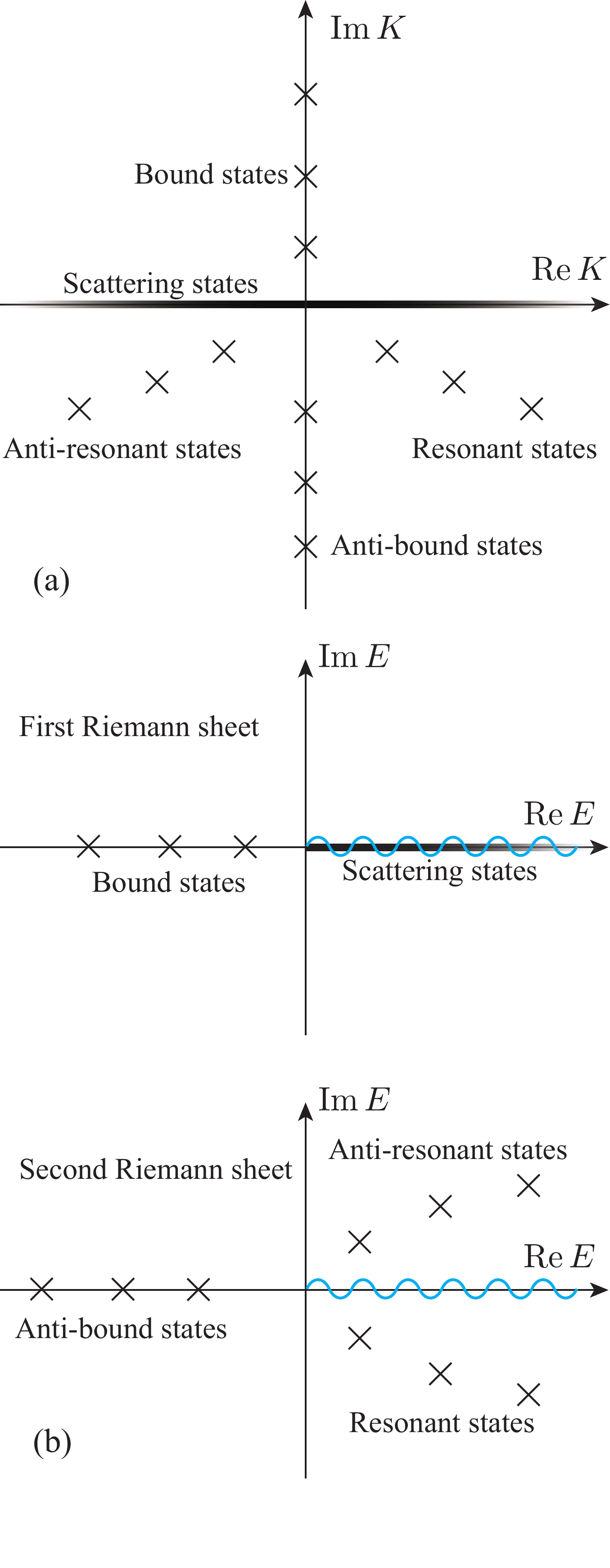}
\caption{Schematic plots of the distributions of the eigenvalues (a) in the complex $K$ plane and (b) in the complex $E$ plane with two Riemann sheets.}
\label{fig7}
\end{figure}

To summarize the calculations for the example above, we generally have the distribution of point spectra in the complex $K$ plane, including resonant states, as shown in Fig.~\ref{fig7}(a).
The states on the positive imaginary axis are in fact bound states, which we can realize by putting $K=\ii\kappa$ with $\kappa>0$ into Eq.~\eqref{eq80}. 
There is a continuum of scattering states over the entire real axis of $k$.
R.~Newton~\cite{Newton60, Newton82} proved that all bound states together with the entire continuum of the scattering states constitute a complete set of the Hilbert space.

The states below the real axis of $k$ are therefore outside the Hilbert space.
Indeed, these states are not normalizable, as can be seen by taking $K=\ii\kappa$ with $\kappa<0$ in Eq.~\eqref{eq80}; the eigenstates diverge exponentially on both sides of the space as $x\to\pm\infty$.
This may be why the resonant states are often called unphysical, but we will show in the next Subsec.~\ref{subsec2-3} that we need this spatial divergence for probability conservation.
(We refer the readers to a textbook~\cite{Moiseyev11} and a recent review~\cite{Myo20} for a method called complex scaling for making the divergent eigenfunctions convergent.)
The states in the fourth quadrant of the complex $K$ plane are called the resonant states, whereas those in the third quadrant are called the anti-resonant states.
We will also show in the next Subsec.~\ref{subsec2-3} that every resonant state and its corresponding anti-resonant state form a pair, with the two states being time-reversed.
The states on the negative real axis are called anti-bound states.

Moving over to the complex $E$ plane, we have two Riemann sheets corresponding to the upper and lower half parts of the complex $K$ plane, because of the dispersion relation $E\propto K^2$.
The two sheets are connected on the branch cut indicated by a wavy line in Fig.~\ref{fig7}(b).

The first Riemann sheet of the complex $E$ plane has all the bound states on the negative real axis and all the scattering states on the positive real axis.
All states that belong to R.~Newton's complete~\cite{Newton60, Newton82} set are therefore located in the first Riemann sheet.

The second Riemann sheet of the complex $E$ plane, on the other hand, has resonant states in the lower half-plane, anti-resonant states in the upper half-plane, and anti-bound states on the negative real axis.
B.~Simon~\cite{Simon00} proved a specific arrangement of the bound and anti-bound states for a semi-infinite space, but no arrangement is known for the infinite space, as far as we are aware.

\subsection{Resolving puzzlements about the resonant state}
\label{subsec2-3}
At this point, readers may have several questions about the physical relevance of states with complex energy eigenvalues and their diverging eigenfunctions.
We first present physical interpretations of the resonant and anti-resonant states, and then address possible questions.

\begin{figure}
\includegraphics[width=0.8\columnwidth]{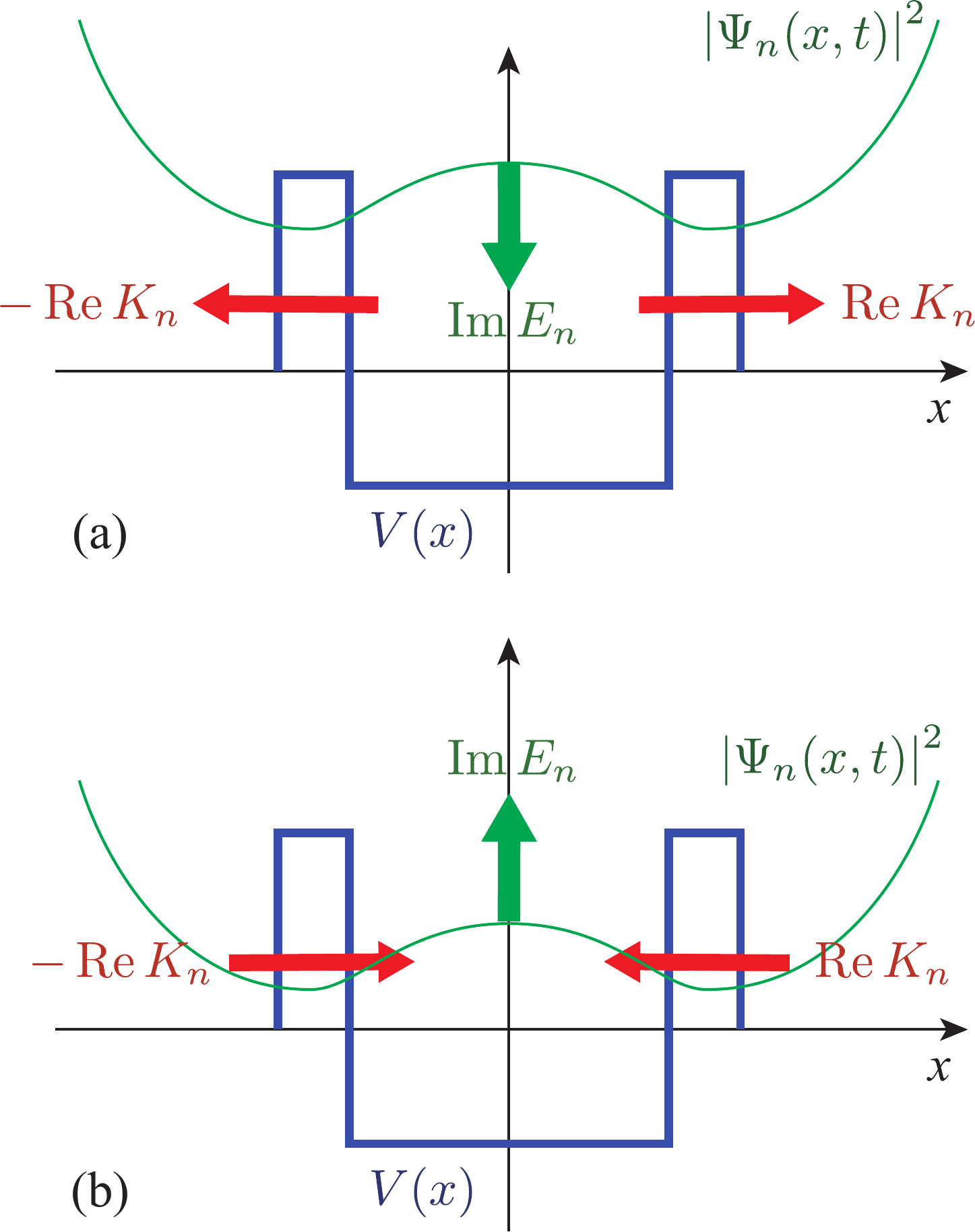}
\caption{Schematic views of (a) the resonant state and (b) the anti-resonant state.}
\label{fig8}
\end{figure}

Consider the Siegert boundary condition~\eqref{eq80} with the time-dependent part:
\begin{align}\label{eq160}
\Psi_n(x,t)=\ee^{-\ii E_nt/\hbar}
\times\begin{cases}
B\ee^{-\ii K_nx} & \mbox{for $x<-\ell$} ,\\
C\ee^{+\ii K_nx} & \mbox{for $x>+\ell$},
\end{cases}
\end{align}
where $n$ denotes each resonant or anti-resonant eigenstate.
In Fig.~\ref{fig7}, we note $\Re K_n>0$ and $\Im E_n<0$ for the resonant states.
This means, as schematically shown in Fig.~\ref{fig8}(a), that the probability flux escapes from the central potential area (red horizontal arrows), and correspondingly the probability in the central area decays in time (green vertical arrow).

On the other hand, we have $\Re K_n<0$ and $\Im E_n>0$ for the anti-resonant states.
This means, as schematically shown in Fig.~\ref{fig8}(b),  that the probability flux is injected into the central area (red horizontal arrows), and consequently the probability there grows in time (green vertical arrow).
(Note, however, that each of the resonant and anti-resonant states never exists as a single state; it only exists as a component of a superimposed state, as we show in the next Sec .~\ref{sec3}. 
They either decay or grow from the infinite past to the infinite future, which should not happen in reality.)

We now provide answers to the following questions~\cite{Hatano21}:
\renewcommand{\theenumi}{(\roman{enumi})}
\renewcommand{\labelenumi}{\theenumi}
\begin{enumerate}
\item Why can the \Sch equation~\eqref{eq10} with the ``Hermitian" Hamiltonian have eigenstates with complex energy eigenvalues?
\label{q-i}
\item How can we understand the spatially diverging wave functions?
\label{q-ii}
\item Why can the time-reversal symmetric eigenvalue problem~\eqref{eq10} have eigenstates that break time-reversal symmetry?
\label{q-iii}
\end{enumerate}

The answer to the question~\ref{q-i} will also lead to the answer to the question~\ref{q-ii}.
Let us remember how we proved the ``Hermiticity" of the Hamiltonian
\begin{align}\label{eq180}
\hatH=-\frac{\hbar^2}{2m}\dv[2]{x}+V(x).
\end{align}
If we can prove that the expectation value $\braket{\psi|\hatH|\psi}$ is real for an ``arbitrary" function $\ket{\psi}$, then the Hamiltonian $\hatH$ is a Hermitian operator.
Since we assume the potential $V(x)$ is real, and its support is compact,
\begin{align}
\braket{\psi|V|\psi} =\int_\infty^\infty \psi(x)^\ast V(x)\psi(x) \dd x
\end{align}
should be real for an arbitrary function $\psi(x)$ which is not singular at least within the support of the potential.

The issue arises from the expectation value of the kinetic term.
We first consider its expectation value under the integration over a finite range $[-L, L]$, where $L>\ell$, and check whether we can take the limit $L\to\infty$.
Following the strategy, we define, for an ``arbitrary" function $\psi(x)$, 
\begin{align}\label{eq190}
\braket{\psi|{\qty(-\frac{h^2}{2m}\dv[2]{x})}|\psi}\hspace{-0.8ex}\raisebox{-0.8ex}{}_L:=-\frac{h^2}{2m}\int_{-L}^L \psi(x)^\ast \dv[2]{x} \psi(x)\dd x.
\end{align}
After partial integration, we can transform the right-hand side of Eq.~\eqref{eq190} as in
\begin{align}\label{eq200}
-\frac{\hbar^2}{2m}\qty[\psi(x)^\ast\dv{x}\psi(x)]_{x=-L}^{+L}
+\frac{\hbar^2}{2m}\int_{-L}^L \abs{\dv{x}\psi(x)}^2\dd x.
\end{align}
The second term of Eq.~\eqref{eq200} has no imaginary part even in the limit of infinite $L$, but the first term is generally complex.
In the elementary level of quantum mechanics, readers might have been told that the first term would vanish if the function $\psi(x)$ vanishes quickly enough in the limit $L\to\infty$, being square-integrable, and hence Eq.~\eqref{eq190} would be real, and the Hamiltonian operator is Hermitian.
Therefore, all of its energy eigenvalues must be real.

In fact, this argument means that the proof of the Hermiticity is valid only when the ``arbitrary" function $\psi(x)$ is chosen from a specific functional space of functions that vanish quickly enough in the limit $L\to\infty$, and the Hamiltonian~\eqref{eq180} is Hermitian only within such a functional space.
The Hamiltonian, or more specifically the kinetic term $-(\hbar^2/2m)\dv*[2]{x}$, can be non-Hermitian outside such a functional space.
In other words, whether the Hamiltonian~\eqref{eq180} is Hermitian or non-Hermitian could not be determined without specifying the functional space that we work in.

Indeed, in explaining the eigenvalue distribution in the complex $K$ plane in Fig.~\ref{fig7}(a), we stressed that the states in the lower half plane of the complex $K$ plane diverge exponentially on both sides of $x\to\pm\infty$.
Hence, the first term of Eq.~\eqref{eq200} does not vanish in the limit $L\to\infty$ for the resonant and anti-resonant states.
The Hamiltonian $\hatH$ can be non-Hermitian in the functional space of these eigenstates, and therefore they can legitimately have complex energy eigenvalues.
This is the answer to the question~\ref{q-i}.

It also underscores that resonant states with complex eigenvalues can appear only when the environmental system extends to infinity.
Indeed, we would not be able to set the Siegert boundary condition~\eqref{eq160} if the environmental system is of a finite size;
the outgoing waves would bounce back to the system.
As we emphasized at the beginning of the article, the non-Hermiticity of the open quantum system arises only when the environmental system is infinite.

As we mentioned earlier, the spatial divergence of its wave function may be why the resonant state is often called unphysical. 
We now show that the spatial divergence is, in fact, necessary for probability conservation.
We analyze the imaginary part of the expectation value $\braket{\Psi|\hatH|\Psi}_L$, where we let $\ket{\Psi}$ contain the time-dependent factor $\ee^{-\ii Et/\hbar}$.
The imaginary part of the element comes only from the first term of Eq.~\eqref{eq200}, and hence
\begin{align}\label{eq210}
\Im\braket{\Psi|\hatH|\Psi}_L=-\frac{h^2}{2m}\Im\qty[\Psi(x,t)^\ast\dv{x}\Psi(x,t)]_{x=-L}^{+L}.
\end{align}
The left-hand side of Eq.~\eqref{eq210} is transformed as follows.
Consider the time-dependent \Sch equation
\begin{align}
\ii\hbar\dv{t}\Psi(x,t)=\hatH\Psi(x,t)
\end{align}
and its complex conjugate
\begin{align}
-\ii\hbar\dv{t}\Psi(x,t)^\ast=\hatH\Psi(x,t)^\ast.
\end{align}
Therefore, we have
\begin{widetext}
\begin{subequations}\label{eq240}
\begin{align}
\ii\hbar \dv{t}\braket{\Psi|\Psi}_L
&=\ii\hbar \int_{-L}^L \dd x \qty[\qty(\dv{t}\Psi(x,t)^\ast)\Psi(x,t)+\Psi(x,t)^\ast\qty(\dv{t}\Psi(x,t))]
\\
&=\int_{-L}^L \dd x \qty[-\qty(\hatH\Psi(x,t)^\ast)\Psi(x,t)+\Psi(x,t)^\ast \qty(\hatH\Psi(x,t))]
\\
&=\qty(\braket{\Psi|\hatH|\Psi}_L-{\braket{\Psi|\hatH|\Psi}_L}^\ast)
=2\ii\Im\braket{\Psi|\hatH|\Psi}_L
\end{align}
\end{subequations}
\end{widetext}

The right-hand side of Eq.~\eqref{eq210}, on the other hand,  is given in terms of the momentum operator $\hat{p}=(\hbar/\ii)\dv*{x}$ as in
\begin{align}\label{eq250}
-\frac{\hbar}{2m}
\Re\qty(\Psi(L,t)^\ast \hat{p} \Psi(L,t)-\Psi(-L,t)^\ast \hat{p} \Psi(-L,t))
\end{align}
This gives, apart from the coefficient, the momentum flux going out from the right boundary $x=L$ and from the left boundary $x=-L$, as is schematically shown in Fig.~\ref{fig8}(a).
Combining Eqs.~\eqref{eq240} and~\eqref{eq250}, we realize that Eq.~\eqref{eq210} yields an equation of continuity of the probability density,
\begin{align}
&\dv{t}\braket{\Psi|\Psi}_L
\nonumber \\
&=-\frac{\hbar}{m}
\Re\qty(\Psi(L,t)^\ast \hat{p} \Psi(L,t)-\Psi(-L,t)^\ast \hat{p} \Psi(-L,t)),
\end{align}
which implies the probability conservation.

Let us explicitly demonstrate probability conservation for each resonant state as follows~\cite{Hatano09}.
We define the probability in the following form:
\begin{subequations}\label{eq260}
\begin{align}
P_n(t)&:=\braket{\Psi_n|\Psi_n}_{L_n(t)}\\
&=\int_{-L_n(t)}^{L_n(t)}\Psi_n(x,t)^\ast \Psi_n(x,t) \dd x.
\end{align}
\end{subequations}
Here $\Psi_n(x,t)$ is one specific resonant eigenstate of the form~\eqref{eq160} and
\begin{align}
L(t):=\Re \frac{\hbar K_n}{m} t+\ell
\end{align}
for $t>0$, where $\hbar K_n/m$ is the phase velocity defined by $\dv*{E_n}{p}=\dv*{E(K_n)}{(\hbar K_n)}$.

\begin{figure}
\includegraphics[width=1.0\columnwidth]{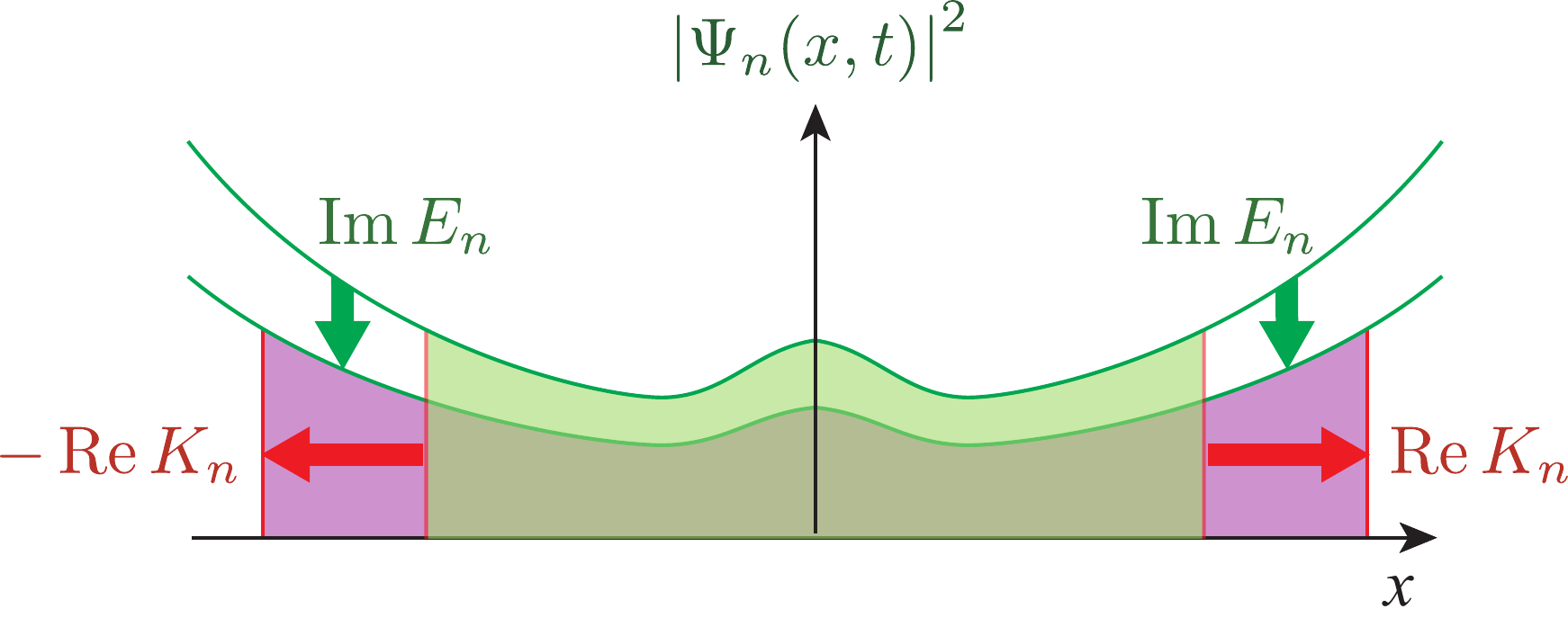}
\caption{The exponential decrease of the wave function over time due to $\Im E_n<0$ is canceled out by the exponential increase of the wave function due to $\Im K_n<0$ when we expand the integral range at the speed of the flux.}
\label{fig9}
\end{figure}

The idea behind this definition is shown in Fig.~\ref{fig9}.
The schematic view of Fig.~\ref{fig8}(a) suggests that the temporal decrease of the probability must be equal to the probability fluxes going out of the potential area.
The probability, integrated over a fixed range, would decrease exponentially over time. 
In Eq.~\eqref{eq260}, we chase the escaping fluxes by expanding the integral range at the speed of the fluxes.
We will indeed find that the probability~\eqref{eq260} is time-independent.
In other words, the exponential decrease in time due to $\Im E<0$ balances with the exponential increase in space due to $\Im K<0$. 
This implies that the exponential divergence of the wave function of the resonant state is \textit{not} unphysical;
On the contrary, it is essential in probability conservation.

We now prove that the time derivative of the probability~\eqref{eq260} vanishes.
The time derivative of $P_n(t)$ consists of two terms: the time derivative of the integrand $\Psi_n(x,t)^\ast \Psi_n(x,t)$ and the time derivative of the integral range $\pm L(t)$.
The calculation of the former is similar to Eq.~\eqref{eq240}, and we obtain
\begin{align}\label{eq280}
&\int_{-L_n(t)}^{L_n(t)}\dv{t}\qty(\Psi_n(x,t)^\ast \Psi_n(x,t)) \dd x
=\frac{2}{\hbar}\Im\braket{\Psi_n|\hatH|\Psi_n}_{L(t)}
\nonumber \\
&=-\frac{h}{m}\Im\qty[\Psi_n(x,t)^\ast\dv{x}\Psi_n(x,t)]_{x=-L_n(t)}^{+L_n(t)},
\end{align}
where we used Eq.~\eqref{eq210} for the second equality.

On the other hand, the calculation that involves the time derivative of $\pm L(t)$ is given by
\begin{align}
&\dv{L_n(t)}{t}\qty(\abs{\Psi_n(L_n(t),t)}^2+\abs{\Psi_n(-L_n(t),t)}^2)
\end{align}
This quantity is related to the right-hand side of Eq.~\eqref{eq210} through Eq.~\eqref{eq160}.
Since $L>\ell$, we have
\begin{align}
\dv{x}\Psi_n(x,t)=\Psi_n(x,t)\times
\begin{cases}
-\ii K_n & \mbox{for $x=-L$},\\
+\ii K_n & \mbox{for $x=+L$},
\end{cases}
\end{align}
and hence
\begin{subequations}
\begin{align}
&\frac{\hbar^2}{2m}\Im \qty[\Psi(x,t)^\ast \dv{x}\Psi(x,t)]_{x=-L_n(t)}^{+L_n(t)}
\nonumber\\
&=\frac{\hbar^2}{2m}\Im \qty(\ii K_n)\qty(\abs{\Psi(+L_n(t),t)}^2+\abs{\Psi(-L_n(t),t)}^2)
\\
&=\Re \frac{\hbar^2 K_n}{2m}\qty(\abs{\Psi(+L_n(t),t)}^2+\abs{\Psi(-L_n(t),t)}^2).
\end{align}
\end{subequations}
Therefore, we find
\begin{align}\label{eq320}
&\dv{L_n(t)}{t}\qty(\abs{\Psi_n(L_n(t),t)}^2+\abs{\Psi_n(-L_n(t),t)}^2)
\nonumber\\
&=\frac{\hbar}{m}\Im \qty[\Psi(x,t)^\ast \dv{x}\Psi(x,t)]_{x=-L_n(t)}^{+L_n(t)}.
\end{align}
This equation exactly cancels out Eq.~\eqref{eq280}, and hence we find $\dv*{P_n(t)}{t}\equiv0$ for the probability~\eqref{eq260}.

The quantity in Eq.~\eqref{eq280} decreases exponentially over time, whereas the quantity in Eq.~\eqref{eq320} increases exponentially over time because of the spatial divergence of the wave function.
This is why the probability~\eqref{eq260} is conserved.
If there were no spatial divergence of the wave function of the resonant state, the probability would not be conserved, which \textit{would be} unphysical.
This is the answer to the question~\ref{q-ii}.

A more physical answer is as follows.
For example, consider a quantum dot connected to electrodes.
A quantum dot can harbor Coulomb interactions, whereas electrodes confined to two-dimensional boundaries between two different semiconductors can be nearly free.
Hence, this is a typical situation of the open quantum system.
The number of electrons in the quantum dot is microscopic, whereas the number of electrons in the electrodes can be macroscopic, on the order of Avogadro's number.
It is then plausible to observe exponential growth as we move from the quantum dot towards the electrodes over a relatively large distance.


The question \ref{q-iii} is formulated as follows.
The time-reversal operation $\hat{T}$ in quantum mechanics not only puts $t\to -t$ but also takes the complex conjugation $\ii\to-\ii$.
Since the Hamiltonian of the \Sch equation~\eqref{eq10} is real and time-independent, it commutes with the time-reversal operation:
\begin{align}\label{eq370}
[\hatH,\hat{T}]=0.
\end{align}
This appears to indicate that the Hamiltonian and the time-reversal operation share the eigenstates, and hence all the eigenstates must be taken to be time-reversal symmetric.
This is, however, not true; we saw in Fig.~\ref{fig8} that the resonant states decay in time while the anti-resonant states grow in time, both of which break the time-reversal symmetry.

The origin of the seeming contradiction is the fact that the time-reversal operation $\hat{T}$ is \textit{not} a linear operator but an anti-linear one.
Suppose that an eigenstate $\ket{\psi_n}$ of the Hamiltonian $\hatH$ has an energy eigenvalue $E_n$:
\begin{align}
\hatH\ket{\psi_n}=E_n\ket{\psi_n}.
\end{align}
Operating $\hat{T}$ on both sides from the left, we have
\begin{align}
\hat{T}\hatH\ket{\psi_n}=\hat{T}E_n\ket{\psi_n}.
\end{align}
On the left-hand side, we can commute $\hat{T}$ and $\hatH$ with each other, but on the right-hand side, $\hat{T}$ and $E_n$ do \textit{not} generally commute with each other.
What results in is
\begin{align}\label{eq360}
\hatH\hat{T}\ket{\psi_n}={E_n}^\ast \ \hat{T}\ket{\psi_n},
\end{align}
because of the complex-conjugate operation of $\hat{T}$.

We therefore have two cases.
If $E_n$ is a real energy eigenvalue, Eq.~\eqref{eq360} reduces to 
\begin{align}
\hatH\qty(\hat{T}\ket{\psi_n})=E_n\qty(\hat{T}\ket{\psi_n}).
\end{align}
This means that $\hat{T}\ket{\psi_n}$ is also an eigenstate of the Hamiltonian $\hatH$ with the same energy eigenvalue $E_n$, and hence, if we assume no degeneracy, $\hat{T}\ket{\psi_n}$ must be proportional to $\ket{\psi_n}$ and the proportionality constant is the eigenvalue of the operator $\hat{T}$ for the eigenstate $\ket{\psi_n}$. It also means that the state $\ket{\psi_n}$ is time-reversal symmetric.
This is the case for the bound and anti-bound states in Fig.~\ref{fig7}.

If $E_n$ is a complex energy eigenvalue, on the other hand, Eq.~\eqref{eq360} indicates that for the eigenstate $\ket{\psi_n}$, there is always a partner $\hat{T}\ket{\psi_n}$ with the energy eigenvalue ${E_n}^\ast$.
This relation is precisely what we saw in Fig.~\ref{fig7}(b) between the resonant and anti-resonant states.
The whole system of the eigenstates conserves the time-reversal symmetry of the Hamiltonian $\hatH$.
Nonetheless, each eigenstate can break the time-reversal symmetry because $\hat{T}$ is not a linear operator.
This is the answer to question \ref{q-iii}.

\subsection{Solutions of the scattering problem for the tight-binding model}
\label{subsec2-4}

We hereafter use the tight-binding model to discuss the resonance poles.
(For readers unfamiliar with the tight-binding model, we provide a brief introduction to it from two perspectives in App.~\ref{appC}.)
The tight-binding model has only a finite number of poles, and hence is much more tractable than the case of continuum space, which has a countably infinite number of poles.
The tractability of the tight-binding model lets us discover a new complete set of bases for scattering problems, which is the main topic of Sec.~\ref{sec3}.
For introductory purposes, we here formulate the scattering problem in the tight-binding model, stressing that the way to solve it appears to be different from the case of the continuum space but is essentially the same.

\begin{figure}
\includegraphics[width=0.8\columnwidth]{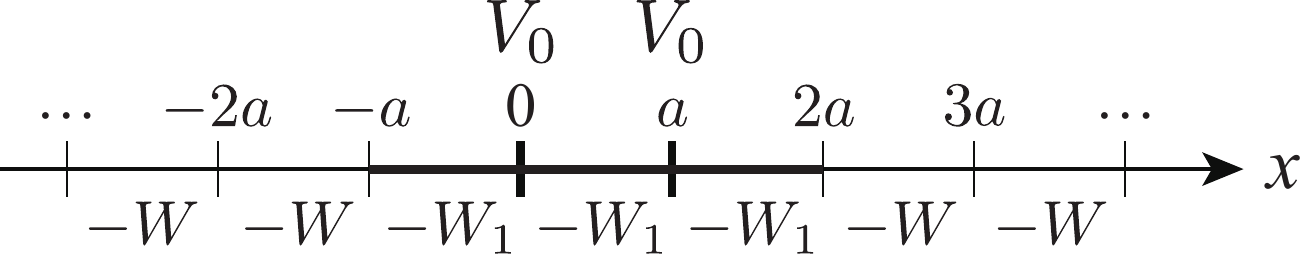}
\caption{A schematic view of the tight-binding Hamiltonian~\eqref{eq420}.}
\label{fig10}
\end{figure}

Let us consider the potential-scattering problem of the following specific Hamiltonian of the tight-binding model, for example:
\begin{align}\label{eq420}
\hatH&=-W\underset{n\neq -1,0,1}{\sum_{n=-\infty}^\infty} \qty(\ketbra{n+1}{n}+\ketbra{n}{n+1})
\nonumber\\
&-W_1\sum_{n=-1}^1\qty(\ketbra{n+1}{n}+\ketbra{n}{n+1})
\nonumber\\
&+V_0\qty(\ketbra{0}+\ketbra{1}),
\end{align}
where $W>0$ and $W_1>0$; 
see Fig,~\ref{fig10}.


Let us first show how to solve the eigenvalue equation for the Hamiltonian~\eqref{eq420}.
Each component of the eigenvalue equation reads
\begin{align}
\braket{n|\hatH|\psi}=E\braket{n|\psi}.
\end{align}
Introducing the notation $\psi_n:=\braket{n|\psi}$, we have
\begin{subequations}\label{eq440}
\begin{align}\label{eq440a}
-W\qty(\psi_{n-1}+\psi_{n+1})&=E\psi_n\\
\intertext{for $n\leq-2$ or $n\geq3$, whereas for $-1\leq n\leq 2$, we have}
\label{eq440b}
-W\psi_{-2}-W_1\psi_{0}&=E\psi_{-1} ,\\
\label{eq440c}
-W_1\psi_{-1}-W_1\psi_{1}+V_0\psi_0&=E\psi_0 ,\\
\label{eq440d}
-W_1\psi_{0}-W_1\psi_{2}+V_0\psi_1&=E\psi_1 ,\\
\label{eq440e}
-W_1\psi_{1}-W\psi_{3}&=E\psi_{2}.
\end{align}
\end{subequations}
These conditions correspond to the connection conditions in solving the potential-scattering problem in the continuum space, such as Eq.~\eqref{eqA30}.

We now assume the scattering wave function in the form
\begin{align}
\psi_n=\begin{cases}\label{eq450}
A \ee^{+\ii k na}+B\ee^{-\ii k na} &\mbox{for $n\leq -1$},\\
\psi_0,\\
\psi_1,\\
C \ee^{+\ii k na} &\mbox{for $n\geq 2$}.
\end{cases}
\end{align}
Using Eq.~\eqref{eq440a} for Eq.~\eqref{eq450}, we find the dispersion relation
\begin{align}
E(k)=-2W\cos(ka).
\end{align}
The unknown variables in the solution~\eqref{eq450} are the following five: the wave number $k$, the amplitude ratios $B/A$ and $C/A$, as well as the wave functions $\psi_0/A$ and $\psi_1/A$.
Since there are only four conditions~\eqref{eq440b}--\eqref{eq440e}, we cannot determine one variable.
As we explained below Eq.~\eqref{eq80}, we use the wave number $k$ of the incident wave as a control parameter.

Straightforward algebras of inserting Eq.~\eqref{eq450} into the conditions~\eqref{eq440b}--\eqref{eq440e} yield
\begin{subequations}
\label{eq470}
\addtocounter{equation}{1}
\begin{align}
-W(A+B)&=-W_1\psi_0,\\
-W_1\qty(A\ee^{-\ii ka}+B\ee^{+\ii ka}+\psi_1)+V_0\psi_0&=E\psi_0,\\
-W_1\qty(\psi_0+C\ee^{+2\ii ka})+V_0\psi_1&=E\psi_1,\\
-WC\ee^{+\ii ka}&=-W_1\psi_1.
\end{align}
\end{subequations}
They are summarized in the matrix forms
\begin{widetext}
\begin{subequations}
\begin{align}
\begin{pmatrix}
W & W \\
W_1\ee^{-\ii ka} & W_1\ee^{+\ii ka}
\end{pmatrix}
\begin{pmatrix}
A \\ B
\end{pmatrix}
&=\begin{pmatrix}
W_1 & 0 \\
V_0-E & -W_1
\end{pmatrix}
\begin{pmatrix}
\psi_0 \\ \psi_1
\end{pmatrix},
\\
\label{eq480b}
\begin{pmatrix}
-W_1 & V_0-E \\
0 &  W_1
\end{pmatrix}
\begin{pmatrix}
\psi_0 \\ \psi_1
\end{pmatrix}
&=
\begin{pmatrix}
W_1 \ee^{+\ii ka} & W_1 \ee^{-\ii ka}\\
W & W
\end{pmatrix}
\begin{pmatrix}
C\ee^{+\ii ka}\\ 0
\end{pmatrix},
\end{align}
\end{subequations}
\end{widetext}
where we have taken the right-hand side of the second equation~\eqref{eq480b} in its present form to obtain the transfer matrix in a full form.
We thereby obtain the transfer matrix $T$ as in
\begin{align}\label{eq490}
\begin{pmatrix}
A \\ B
\end{pmatrix}
&=\begin{pmatrix}
T_{11} & T_{12} \\
T_{21} & T_{22}
\end{pmatrix}
\begin{pmatrix}
C\ee^{+\ii ka}\\0
\end{pmatrix},
\end{align}
where 
\begin{align}
T_{11}=\frac{\qty[\theta\lambda^2+(v_0+w_1)\lambda+1]\qty[\theta\lambda^2+(v_0-w_1)\lambda+1]}{{w_1}^3(\lambda^2-1)}.
\end{align}
Here, we introduced the variable
\begin{align}\label{eq523}
\lambda:=\ee^{\ii Ka},
\end{align}
which is related to the wave number and the energy in the forms
\begin{subequations}\label{eq535}
\begin{align}\label{eq535a}
K&=-\frac{\ii}{a}\log\lambda,\\
\label{eq535b}
E&=-W\qty(\lambda+\lambda^{-1}).
\end{align}
\end{subequations}
We also introduced dimensionless constants
\begin{align}\label{eq537}
\theta:=1-\frac{{W_1}^2}{W^2},\quad w_1:=\frac{W_1}{W},\quad\mbox{and}\quad v_0:=\frac{V_0}{W}.
\end{align}
Equation~\eqref{eq490} produces the transmission amplitude in the form
\begin{align}
t_\textrm{amp}=\frac{C}{A}=\frac{1}{T_{11}}.
\end{align}
The resonance poles, therefore, should be given by the equation $T_{11}=0$, 
which have the four solutions
\begin{subequations}\label{eq530}
\begin{align}
\lambda&=\frac{-(v_0+w_1)\pm\sqrt{(v_0+w_1)^2-4\theta}}{2\theta},
\\
\lambda&=\frac{-(v_0-w_1)\pm\sqrt{(v_0-w_1)^2-4\theta}}{2\theta}.
\end{align}
\end{subequations}
The corresponding eigen-wave-numbers and the energy eigenvalues are given by Eq.~\eqref{eq535}.

\subsection{Solutions of the resonant states for the tight-binding model}
\label{subsec2-6}

We now show that finding the resonance poles is significantly easier by using the Siegert boundary condition.
We find the Siegert boundary condition by omitting $A$ from Eq.~\eqref{eq450}:
\begin{align}\label{eq540}
\psi_n=\begin{cases}
B\ee^{-\ii Kna} &\mbox{for $n\leq -1$},\\
C\ee^{+\ii Kna} &\mbox{for $n\geq2$}.
\end{cases}
\end{align}
As we emphasized below Eq.~\eqref{eq80}, the reduction of the number of unknown variables leads to discrete solutions of $K$.

Instead of Eq.~\eqref{eq470}, we have
\begin{subequations}
\addtocounter{equation}{1}
\label{eq550}
\begin{align}\label{eq550b}
-WB&=-W_1\psi_0,\\
\label{eq55c}
-W_1\qty(B\ee^{+\ii Ka}+\psi_1)+V_0\psi_0 &= E\psi_0,\\
\label{eq550d}
-W_1\qty(\psi_0+C\ee^{+2\ii Ka})+V_0\psi_1 &= E\psi_1,\\
\label{eq550e}
-WC\ee^{+\ii Ka}&=-W_1\psi_1.
\end{align}
\end{subequations}
We eliminate $B$ and $C$, obtaining
\begin{align}\label{eq560}
\begin{pmatrix}
V_0+\Sigma & -W_1 \\
-W_1 & V_0+\Sigma
\end{pmatrix}
\begin{pmatrix}
\psi_0 \\ \psi_1
\end{pmatrix}
=E
\begin{pmatrix}
\psi_0 \\ \psi_1
\end{pmatrix},
\end{align}
where
\begin{align}\label{eq570}
\Sigma:=-\frac{{W_1}^2}{W}\ee^{+\ii Ka}
\end{align}
is sometimes called the self-energy.
In the next Sec.~\ref{sec3}, we will refer to the $2\times2$ matrix on the left-hand side of Eq.~\eqref{eq560} as the effective Hamiltonian:
\begin{align}\label{eq580}
\Heff:=\begin{pmatrix}
V_0+\Sigma & -W_1 \\
-W_1 & V_0+\Sigma
\end{pmatrix}.
\end{align}
Note, however, that this matrix depends on $E$ through the wave number $K$ in Eq.~\eqref{eq570}.
The condition that Eq.~\eqref{eq560} has nontrivial solutions,
\begin{align}
\det\qty(\Heff-E)=0,
\end{align}
reads
\begin{align}\label{eq600}
\lambda^2\qty[\theta\lambda^2+(v_0+w_1)\lambda+1]\qty[\theta\lambda^2+(v_0-w_1)\lambda+1]=0,
\end{align}
which produces the same four solutions as in Eq.~\eqref{eq530} despite that the matrix is $2\times 2$.
This is because the eigenvalue equation~\eqref{eq560} is nonlinear in $E$.

In fact, solving Eq.~\eqref{eq550} is further simplified by assuming the parity symmetry.
For resonant states with even parity, we assume $B=C$ and $\psi_0=\psi_1$ in Eq.~\eqref{eq550d}, which is followed by 
\begin{align}
\qty(-W_1\psi_1-\frac{{W_1}^2}{W}\psi_1\ee^{+\ii Ka})+V_0\psi_1=E\psi_1,
\end{align}
and we immediately obtain the equation
\begin{align}\label{eq620}
\theta\lambda^2+(v_0-w_1)\lambda+1=0.
\end{align}
Assuming the odd parity similarly produces the other equation
\begin{align}\label{eq630}
\theta\lambda^2+(v_0+w_1)\lambda+1=0
\end{align}
because we assume $\psi_0=-\psi_1$ in Eq.~\eqref{eq550d}.
Equations~\eqref{eq620} and~\eqref{eq630} separately give the solutions of Eq.~\eqref{eq600}.

\begin{figure*}
\includegraphics[width=0.75\textwidth]{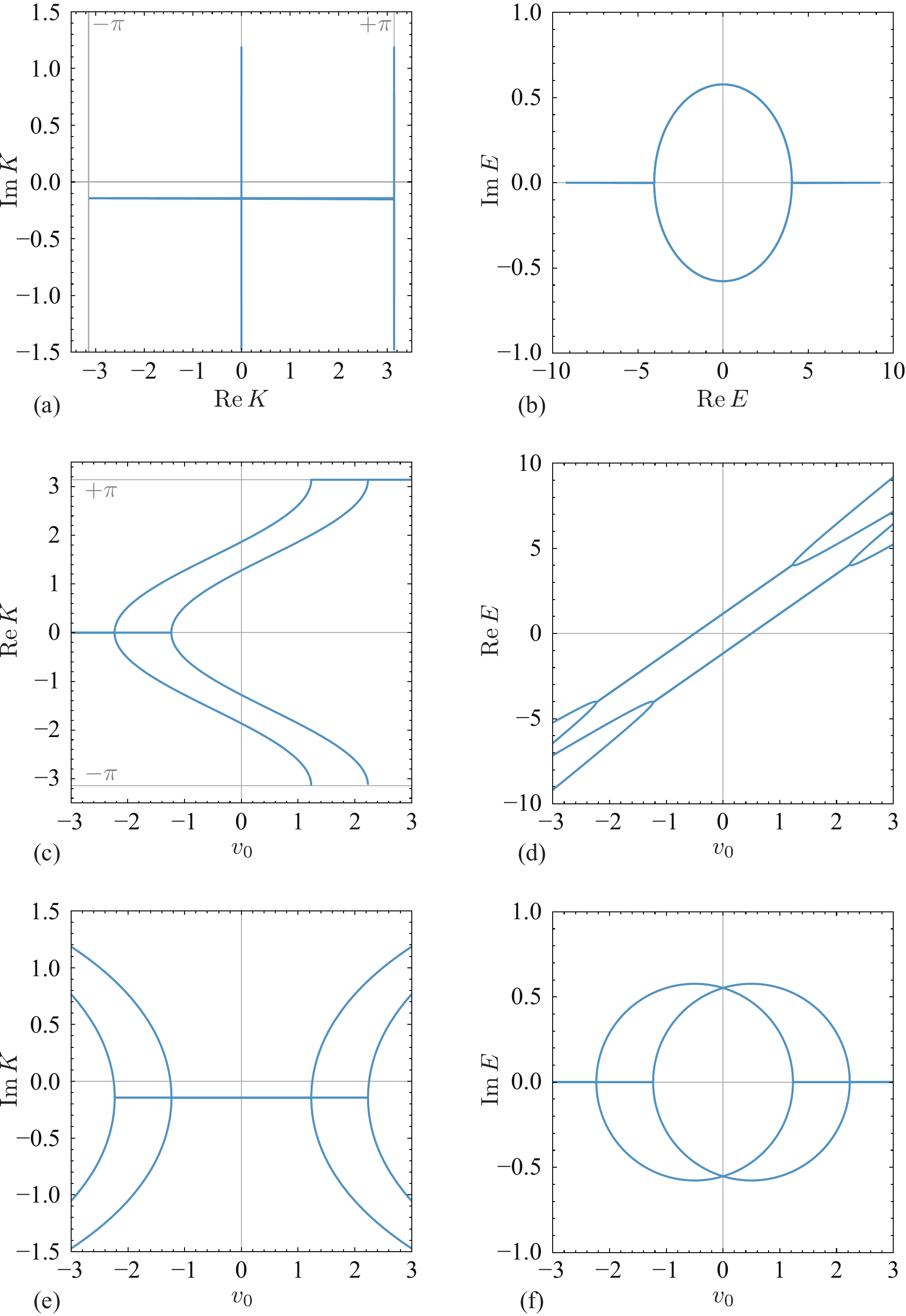}
\caption{The changes of the eigen-wave-numbers and the energy eigenvalues. 
We put $w_1=1/2$ and $\theta=3/4$ with $\hbar=W=a=1$, and vary the potential $v_0$ from $-3$ to $3$.
(a): The trajectories of the four eigen-wave-numbers on the complex wave-number plane, as we change $v_0$.
(c) and (e): The $v_0$-dependence of the real and imaginary parts of the eigen-wave-numbers, respectively.
(b) The trajectories of the four energy eigenvalues on the complex energy plane, as we change $v_0$.
(d) and (f): The $v_0$-dependence of the real and imaginary parts of the energy eigenvalues, respectively.}
\label{fig11}
\end{figure*}

We plot in Fig.~\ref{fig11} how the four eigen-wave-numbers~\eqref{eq535a} and the four energy eigenvalues~\eqref{eq535b} change under the variation of parameters.
For explanatory purposes, in Fig.~\ref{fig11}, we fix $w_1=W_1/W$ to $1/2$, which also fixes $\theta$ to $3/4$, and vary the strength of the potential $v_0=V_0/W$ from negative values to positive values.

When $v_0$ is largely negative enough, the square roots in the solutions~\eqref{eq530} are both real.
All four solutions are therefore positive, and hence the corresponding eigen-wave-numbers due to Eq.~\eqref{eq535a} are all pure imaginary and the corresponding energy eigenvalues due to Eq.~\eqref{eq535b} are all negative.
These are bound and anti-bound states.

When $v_0$ approaches zero from the negative side, or in other words, when the potential becomes shallower, the bound states are lost, collide with the anti-bound states on the negative imaginary axis of the complex wave-number plane when the square root in each solution of Eq.~\eqref{eq530} vanishes, and then turn to resonant and anti-resonant states.
Then the real and imaginary parts of each solution are respectively given by
\begin{subequations}
\begin{align}
\Re\lambda&=-\frac{v_0\pm w_1}{2\theta},\\
\Im\lambda&=-\frac{\sqrt{4\theta-(v_0\pm w_1)^2}}{2\theta},
\end{align}
\end{subequations}
and hence we can find the expression
\begin{align}
\lambda=\theta^{-1/2}\ee^{\ii (\Re K) a}
\end{align}
because
$\qty(\Re\lambda)^2+\qty(\Im\lambda)^2=\theta^{-1}$.
The eigen-wave-numbers~\eqref{eq535a} are  given by
\begin{align}
\Re K&=\frac{1}{a}\arctan{\frac{\sqrt{4\theta-(v_0\pm w_1)^2}}{v_0-w_1}},\\
\Im K&=\frac{1}{2a}\log \theta.
\end{align}
Note that $\Im K<0$ because $\theta=1-(W_1/W)^2<1$.

As we further increase $v_0$ into large enough positive values, the square roots in the solutions~\eqref{eq530} turn back to real values.
All solutions this time take negative values, and hence the corresponding eigen-wave-number has the real part $\Re K=\pi/a$ and the corresponding energy eigenvalues are positive.
Note that for the tight-binding model, $\Re K=\pm\pi/a$ are equivalent to each other and the positive energy eigenvalues can exist; see App.~\ref{appC}.
These observations explain what is happening in Fig.~\ref{fig11}.

\begin{figure}
\includegraphics[width=0.95\columnwidth]{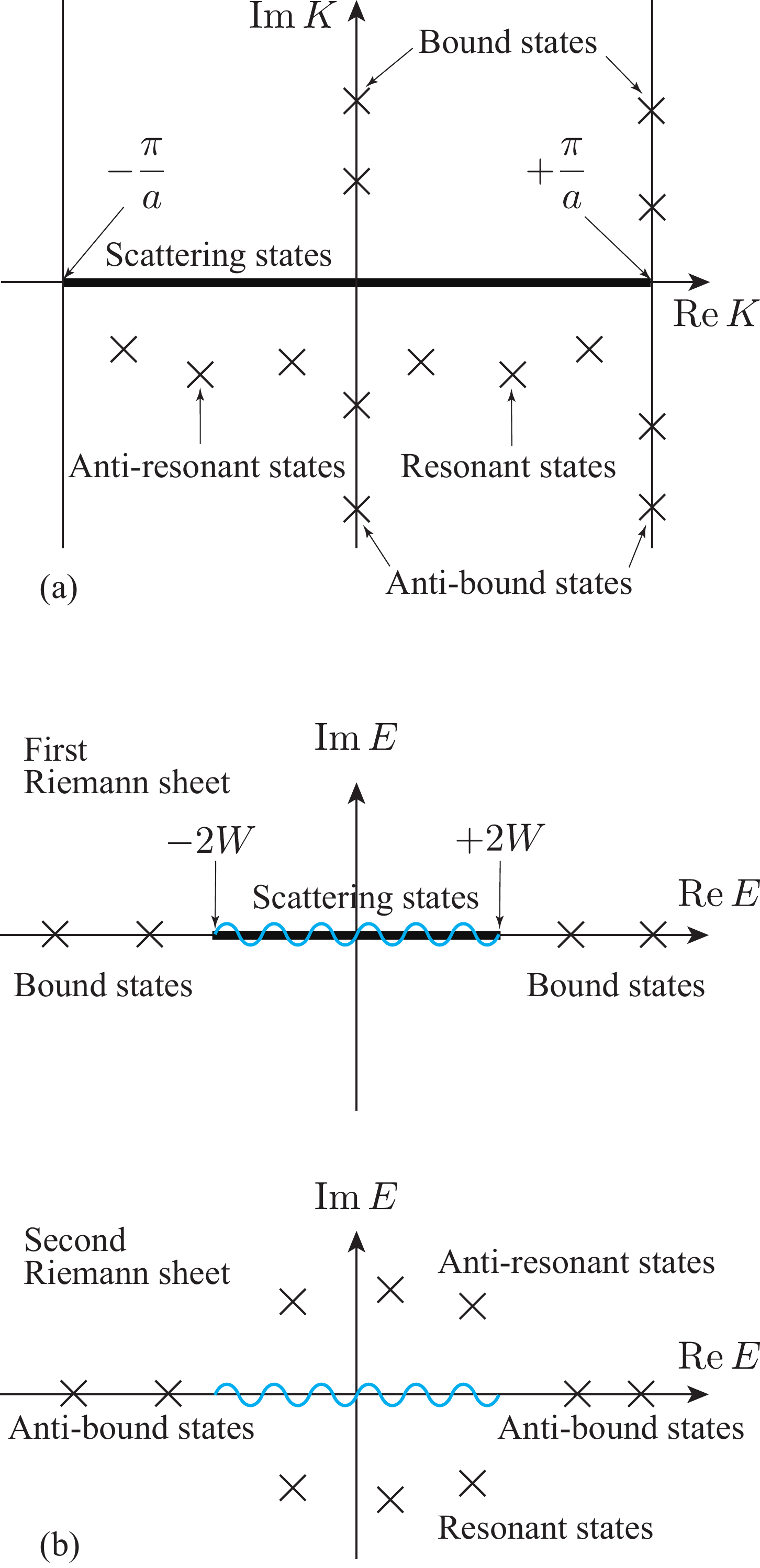}
\caption{Schematic plots of the distributions of the eigenvalues of the tight-binding model (a) in the complex $K$ plane and (b) in the complex $E$ plane with two Riemann sheets.}
\label{fig12}
\end{figure}

To summarize the results for the tight-binding model, they are generally consistent with those shown in Fig.~\ref{fig7} for the continuum-space case, but there are also differences; see Fig.~\ref{fig12}.
First, the tight-binding model has only a finite number of poles in contrast to the continuum-space model, which has a countable but infinite number of poles.
Second, only the first Brillouin zone, $-\pi/a<\Re K\leq\pi/a$, is relevant to the tight-binding model.
Third, bound and anti-bound states can exist not only on the imaginary axis $\Re K=0$ but also on the line $\Re K=\pi/a$.
Finally, the scattering states extend over a finite range $[-2W,2W]$, which overlaps the branch cut (the wavy line in Fig.~\ref{fig12}) that connects the first and second Riemann sheets.

\section{Feshbach formalism: Reduction to an effective Hamiltonian}
\label{sec3}

In the previous Sec.~\ref{sec2}, we analyzed the potential scattering in infinite space.
We had complex eigenvalues because of the Siegert boundary condition~\eqref{eq540}. 
In this sense, the non-Hermiticity of the open quantum system is hidden in infinity.

In this section, we use the Feshbach projection formalism~\cite{Feshbach58,Feshbach62} to eliminate the infinite degrees of freedom of the environment, thereby obtaining an explicitly non-Hermitian effective Hamiltonian for the system.
The Feshbach projection method has often been used to project out some of the eigenstates, but below we will use it to project out the environmental spatial degrees of freedom~\cite{Hatano14}.

In fact, it is much easier to formulate the separation of the space by projecting out the environment and finding an effective Hamiltonian for the central system in tight-binding models than in the continuum space.
Through Subsecs.~\ref{subsec3-1}--\ref{subsec3-3}, we thus give details of the application of the Feshbach formalism to tight-binding models.
Finally, in Subsec.~\ref{subsec3-4}, we will show the correspondence between the Feshbach formalism and a previous analysis in the continuum space~\cite{Tolstikhin98}.

\subsection{Feshbach formulation in a tight-binding model}
\label{subsec3-1}

Let us demonstrate the Feshbach formalism for the specific example of the tight-binding model~\eqref{eq420}; see also Fig.~\ref{fig10}.
We will find the effective Hamiltonian in the form~\eqref{eq580}.
The non-Hermiticity explicitly emerges in the self-energy term~\eqref{eq570}.

We begin by splitting the complete set of the problem,
\begin{align}
\comp=\sum_{n=-\infty}^{+\infty}\dyad{na}
\end{align}
into two projection operators
\begin{subequations}\label{eq710}
\begin{align}
\hatP&:=\dyad{0}+\dyad{a},\\
\hatQ&:=\comp-\hatP=\underset{n\neq0,1}{\sum_{n=-\infty}^{+\infty}}\dyad{na}.
\end{align}
\end{subequations}
Note the following properties of the projection operators:
\begin{subequations}\label{eq720}
\begin{align}
\hatP^2&=\hatP,\\
\hatQ^2&=\hatQ,\\
\hatP+\hatQ&=\comp,\\
\hatP\hatQ&=\hatQ\hatP=0.
\end{align}
\end{subequations}

We apply these two projection operators to the eigenvalue equation
\begin{align}\label{eq730}
\hatH\ket{\psi}=E\ket{\psi},
\end{align}
obtaining
\begin{subequations}
\begin{align}
\hatP\hatH\ket{\psi}&=E\hatP\ket{\psi},\\
\hatQ\hatH\ket{\psi}&=E\hatQ\ket{\psi}.
\end{align}
\end{subequations}
Using the properties~\eqref{eq720}, we find
\begin{subequations}
\begin{align}\label{eq750a}
\PHP\qty(\hatP\ket{\psi})+\PHQ\qty(\hatQ\ket{\psi})&=E\qty(\hatP\ket{\psi}),\\
\label{eq750b}
\QHP\qty(\hatP\ket{\psi})+\QHQ\qty(\hatQ\ket{\psi})&=E\qty(\hatQ\ket{\psi}).
\end{align}
\end{subequations}
We now try to eliminate $\hatQ\ket{\psi}$ from the first equation~\eqref{eq750a} and make it an equation of $\hatP\ket{\psi}$ so that we can write it down in the form
\begin{align}\label{eq760}
\Heff\qty(\hatP\ket{\psi})=E\qty(\hatP\ket{\psi}).
\end{align}
We transform the second equation~\eqref{eq750b} into
\begin{align}
\hatQ\ket{\psi}=\frac{1}{E-\QHQ}\QHP\qty(\hatP\ket{\psi}),
\end{align}
and insert it into the  first equation~\eqref{eq750a}, which results in
\begin{align}
\qty[\PHP+\PHQ\frac{1}{E-\QHQ}\QHP]\qty(\hatP\ket{\psi})=E\qty(\hatP\ket{\psi}).
\end{align}
Comparing this with Eq.~\eqref{eq760}, we identify the effective Hamiltonian as
\begin{align}\label{eq790}
\Heff(E)=\PHP+\PHQ\frac{1}{E-\QHQ}\QHP.
\end{align}

\begin{figure}
\includegraphics[width=0.8\columnwidth]{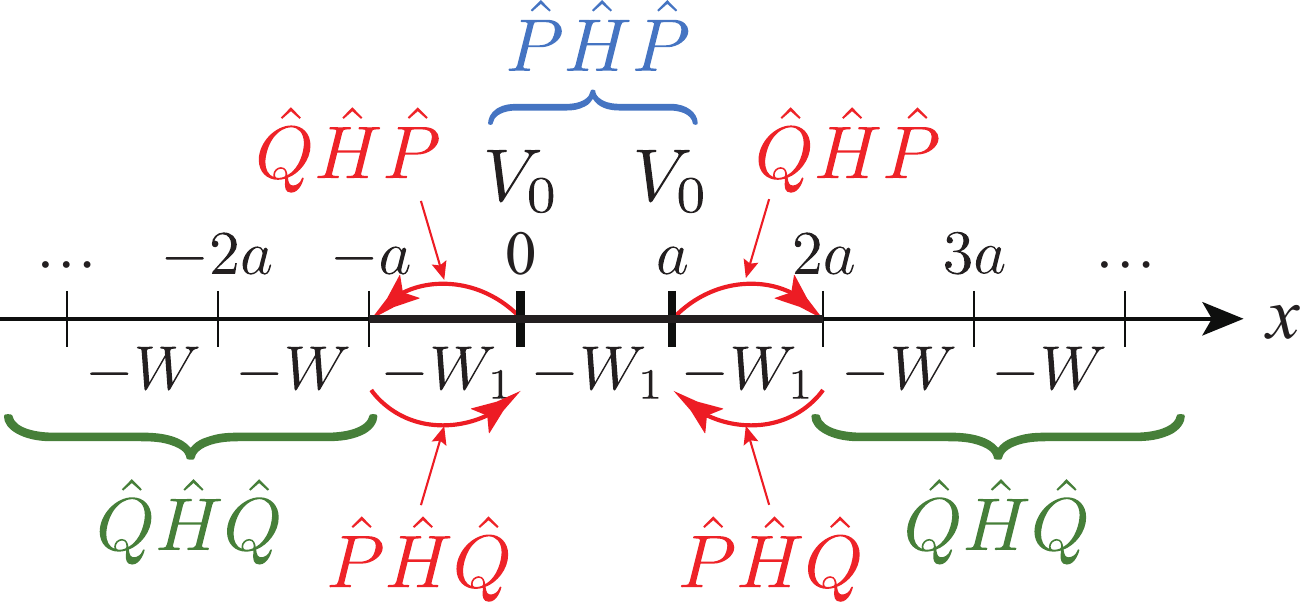}
\caption{The four parts of the Hamiltonian given in Eq.~\eqref{eq800} are indicated in the schematic view Fig.~\ref{fig10} of the tight-binding Hamiltonian~\eqref{eq420}.}
\label{fig13}
\end{figure}

In the specific cases of the tight-binding model~\eqref{eq420} with the projection operators~\eqref{eq710}, we find
\begin{subequations}\label{eq800}
\begin{align}\label{eq800a}
\PHP&=\begin{pmatrix}
V_0 & -W_1 \\
-W_1 & V_0
\end{pmatrix}
\intertext{for the basis set $\qty{\ket{0},\ket{a}}$, and}
\label{eq800b}
\PHQ&=-W_1\dyad{0}{-a}-W_1\dyad{a}{2a},\\
\label{eq800c}
\QHP&=-W_1\dyad{-a}{0}-W_1\dyad{2a}{a},\\
\label{eq800d}
\QHQ&=-W\sum_{n=-\infty}^{-2}\qty(\dyad{(n+1)a}{na}+\dyad{na}{(n+1)a})
\nonumber\\
&-W\sum_{n=+2}^{+\infty}\qty(\dyad{(n+1)a}{na}+\dyad{na}{(n+1)a}).
\end{align}
\end{subequations}
Figure~\ref{fig13} shows these parts of the Hamiltonian indicated originally in Fig.~\ref{fig10}.

The comparison of Eq.~\eqref{eq800a} with Eq.~\eqref{eq580} along with Eq.~\eqref{eq790} reveals that the self-energy~\eqref{eq570} corresponds to the second term on the right-hand side of Eq.~\eqref{eq790}.
We can indeed show it by directly calculating the Green's function $(E-\QHQ)^{-1}$, as described in App.~\ref{appD}:
\begin{align}\label{eq810}
\PHQ\frac{1}{E-\QHQ}\QHP=\begin{pmatrix}
\Sigma & 0\\
0 & \Sigma 
\end{pmatrix}
\end{align}
for the basis set $\qty{\ket{0},\ket{a}}$, where the self-energy $\Sigma$ is defined by Eq.~\eqref{eq570}, and therefore the effective Hamiltonian~\eqref{eq790} is equivalent to Eq.~\eqref{eq580}.
Since $\Sigma$ is generally a complex number, the matrix~\eqref{eq810} is generally non-Hermitian.
The calculation in App.~\ref{appD} also reveals that the reason why the seemingly Hermitian operator~\eqref{eq790} is non-Hermitian is the fact that the $\QHQ$ part of the Hamiltonian is semi-infinite.
As we stressed above, the non-Hermiticity of the open quantum system arises only when the environment continues to infinity.

To define a Green's function of a Hamiltonian that has a continuum spectrum, as is the case of $\QHQ$, we need to introduce an infinitesimal $\pm\ii\varepsilon$ to the denominator to avoid the continuous singularities.
This complex infinitesimal makes the effective Hamiltonian~\eqref{eq790} non-Hermitian.
Avoiding the continuous singularities to the lower side in the complex energy plane yields a retarded Green's function, while avoiding them to the upper side yields an advanced Green's function.
The retarded Green's function describes how the initial condition set to $\ket{2a}$ and $\ket{-a}$ in Fig.~\ref{fig13} evolves in time $t>0$ towards the positive and negative infinities, respectively.
This indeed corresponds to the Siegert boundary condition~\eqref{eq540} with outgoing waves only, that is, $\Re K>0$.
The advanced Green's function, on the other hand, describes how the incoming waves from the positive and negative infinities evolve in time $t<0$ and end up at $\ket{2a}$ and $\ket{-a}$ at $t=0$, respectively.
This corresponds to the Siegert boundary condition~\eqref{eq540} with incoming waves only, as in $\Re K<0$.
We can see these points explicitly in the straightforward calculation of the Green's function in App.~\ref{appD}.

\begin{figure}
\includegraphics[width=0.7\columnwidth]{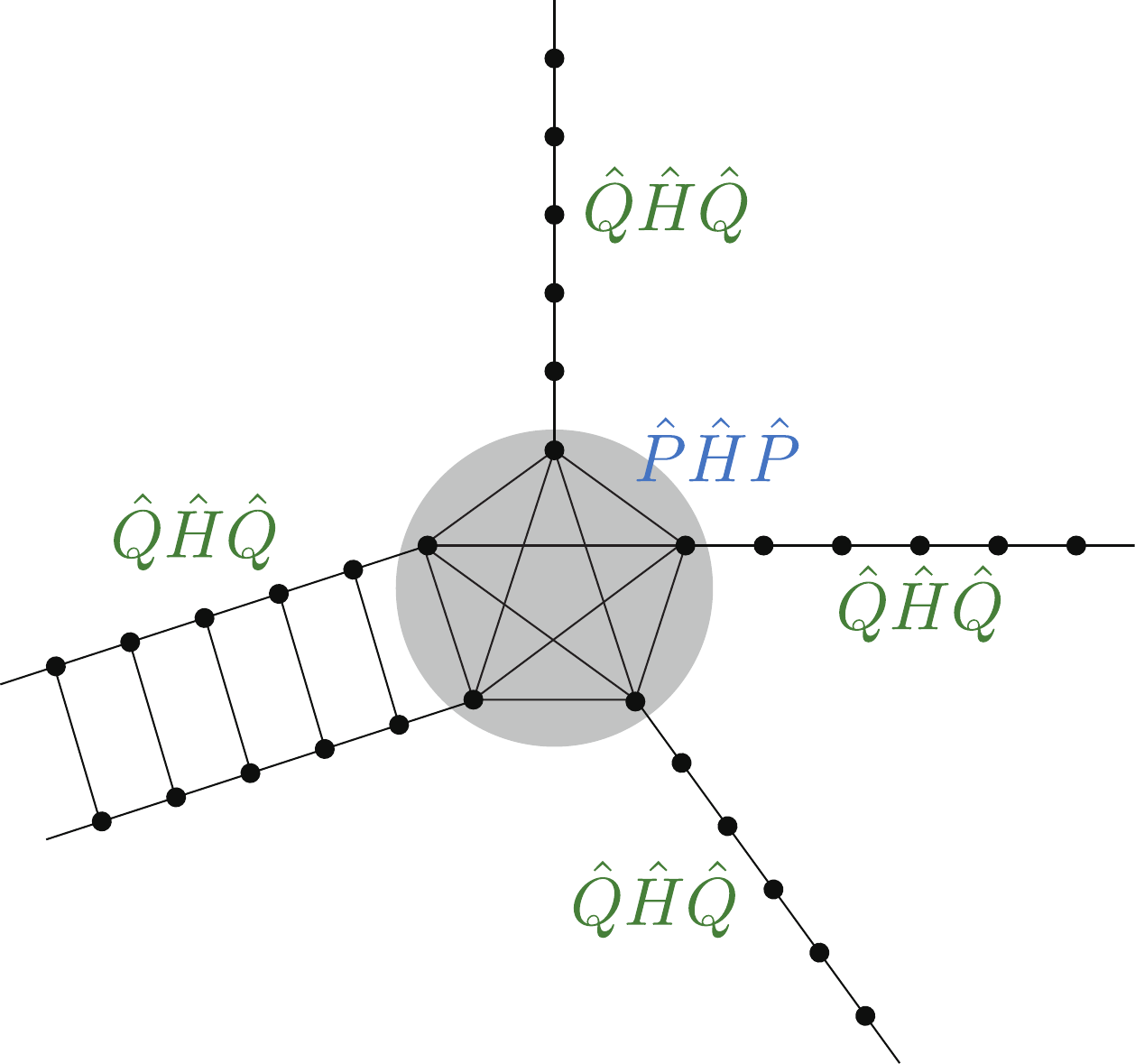}
\caption{An illustrative example of possible extension of the analysis in Sec.~\ref{sec2}.
The gray area is taken for the system with the Hamiltonian $\PHP$ and the rest is taken for the environment with the Hamiltonian $\QHQ$.}
\label{fig14}
\end{figure}

The example in Fig.~\ref{fig13} suggests the possibility of extension to more general setups.
The system described by $\PHP$ can take a more complicated form as illustrated in Fig.~\ref{fig14}.
The environment described by $\QHQ$ obviously can consist of more than two semi-infinite lines.
In fact, it can even be generalized to semi-one-dimensional systems, as also illustrated in Fig.~\ref{fig14}~\cite{Sasada08}.

\subsection{A new complete set involving the resonant states}
\label{subsec3-2}

We now delve into the most nontrivial result in the present article.
We reveal a new complete set of bases for potential scattering, using all discrete eigenvalues, including complex ones~\cite{Hatano14}.

Although the result is general, we again use the specific example in Fig.~\ref{fig13} for explanatory purposes.
We start from Eq.~\eqref{eq760}, which is the effective eigenvalue problem for the system part of the wave function.
It specifically reads
\begin{align}\label{eq820}
\begin{pmatrix}
V_0+\Sigma & -W_1 \\
-W_1 & V_0+\Sigma
\end{pmatrix}
\begin{pmatrix}
\psi_0 \\ \psi_1
\end{pmatrix}
=
E
\begin{pmatrix}
\psi_0 \\ \psi_1
\end{pmatrix},
\end{align}
as was given in Eq.~\eqref{eq560}.
It is an eigenvalue equation that is nonlinear in $E$, because the term
\begin{align}
\Sigma=-\frac{{W_1}^2}{W}\ee^{+\ii Ka}
\end{align}
in the left-hand side of Eq.~\eqref{eq820} depends on $E=-2W\cos(Ka)$ through $K$.

We can simplify the problem by using the energy-related variable $\lambda:=\ee^{\ii Ka}$.
Equation~\eqref{eq820} is then reduced to
\begin{align}
\begin{pmatrix}
v_0-{w_1}^2\lambda & -w_1 \\
-w_1 & v_0-{w_1}^2\lambda 
\end{pmatrix}
\begin{pmatrix}
\psi_0 \\ \psi_1
\end{pmatrix}
=
-\qty(\lambda+\frac{1}{\lambda})
\begin{pmatrix}
\psi_0 \\ \psi_1
\end{pmatrix},
\end{align}
and further to
\begin{align}\label{eq850}
\qty[
\begin{pmatrix}
\theta & 0 \\
0 & \theta 
\end{pmatrix}
\lambda^2
+
\begin{pmatrix}
v_0 & -w_1 \\
-w_1 & v_0 
\end{pmatrix}
\lambda
+\begin{pmatrix}
1 & 0\\
0 & 1
\end{pmatrix}
]\begin{pmatrix}
\psi_0 \\ \psi_1
\end{pmatrix}=0,
\end{align}
where $v_0=V_0/W$, $w_1=W_1/W$, and $\theta=1-{w_1}^2$, as defined previously in Eq.~\eqref{eq537}.
This is a quadratic eigenvalue equation in the sense that the left-hand side is a second-order polynomial in $\lambda$.
For more general setups, the equation reads
\begin{align}\label{eq860}
\qty[\hat{\Theta}\lambda^2+\Hsys\lambda+\iden_N]\hatP\ket{\psi}=0,
\end{align}
where $N$ denotes the number of sites in the $\hatP$ subspace ($N=5$ in the specific case of Fig.~\ref{fig14}, for example), $\iden_N$ denotes the identity matrix of dimension $N$, and
\begin{subequations}
\begin{align}
\hat{\Theta}&:=\iden_N-\frac{\PHQ \hat{H}\hat{P}}{W},\\
\Hsys&:=\frac{\PHP}{W}
\end{align}
\end{subequations}
are $N\times N$ matrices.
Note that it is still the second order in $\lambda$.
We also note the identity
\begin{align}\label{eq865}
\hat{\Theta}\lambda^2+\Hsys\lambda+\iden_N=-\frac{\lambda}{W}\qty(E\iden_N-\Heff),
\end{align}
which we will use below.

There is a standard way of solving the quadratic eigenvalue equation~\cite{Tisseur01}.
Going back to the example~\eqref{eq850}, we rewrite it in the following way:
\begin{align}\label{eq870}
\left(
\begin{array}{cc|cc}
-\lambda & 0 & 1 & 0 \\
0 & -\lambda & 0 & 1 \\
\hline
1 & 0 & v_0+\theta\lambda & -w_1 \\
0 & 1 & -w_1 & v_0+\theta\lambda
\end{array}
\right)
\begin{pmatrix}
\psi_0 \\ \psi_1 \\ 
\hline
\lambda\psi_0 \\ \lambda\psi_1
\end{pmatrix}
=
\begin{pmatrix}
0 \\ 0 \\
\hline
0 \\ 0
\end{pmatrix}.
\end{align}
The upper half of the $4\times 4$ matrix on the left-hand side of Eq.~\eqref{eq870} guarantees that the eigenvector takes the specific form of the lower half being equal to the upper half with an extra factor $\lambda$.
The lower half of the matrix is equivalent to Eq.~\eqref{eq850}.
The point is that we have linearized the eigenvalue equation as follows:
\begin{widetext}
\begin{align}\label{eq880}
\left(
\begin{array}{cc|cc}
0 & 0 & 1 & 0 \\
0 & 0 & 0 & 1 \\
\hline
1 & 0 & v_0 & -w_1 \\
0 & 1 & -w_1 & v_0
\end{array}
\right)
\begin{pmatrix}
\psi_0 \\ \psi_1 \\ 
\hline
\lambda\psi_0 \\ \lambda\psi_1
\end{pmatrix}
=\lambda
\left(
\begin{array}{cc|cc}
1 & 0 & 0 & 0 \\
0 & 1 & 0 & 0 \\
\hline
0 & 0 & -\theta & 0 \\
0 & 0 & 0 & -\theta\
\end{array}
\right)
\begin{pmatrix}
\psi_0 \\ \psi_1 \\ 
\hline
\lambda\psi_0 \\ \lambda\psi_1
\end{pmatrix}.
\end{align}
\end{widetext}
For the general expression~\eqref{eq860}, we have
\begin{align}\label{eq890}
\begin{pmatrix}
\zero_N & \iden_N \\
\iden_N & \Hsys
\end{pmatrix}
\begin{pmatrix}
\ket{\psi} \\ \lambda\ket{\psi}
\end{pmatrix}
=\lambda
\begin{pmatrix}
\iden_N & \zero_N \\
\zero_N & -\hat{\Theta}
\end{pmatrix}
\begin{pmatrix}
\ket{\psi} \\ \lambda\ket{\psi}
\end{pmatrix},
\end{align}
where $\zero_N$ denotes the $N\times N$ zero matrix.
Equations~\eqref{eq880} and~\eqref{eq890} are linear eigenvalue problems in the sense that they are linear in $\lambda$ and they are generalized eigenvalue problems in the sense that they have an extra matrix on the right-hand side.

Let us introduce a short-hand notation
\begin{align}\label{eq900}
\hatA\ket{\Psi}=\lambda \hatB\ket{\Psi},
\end{align}
where
\begin{subequations}
\begin{align}
\hatA&:=
\left(
\begin{array}{cc|cc}
0 & 0 & 1 & 0 \\
0 & 0 & 0 & 1 \\
\hline
1 & 0 & v_0 & -w_1 \\
0 & 1 & -w_1 & v_0
\end{array}
\right),
\\
\hatB&:=
\left(
\begin{array}{cc|cc}
1 & 0 & 0 & 0 \\
0 & 1 & 0 & 0 \\
\hline
0 & 0 & -\theta & 0 \\
0 & 0 & 0 & -\theta\
\end{array}
\right),
\\
\ket{\Psi}&:=
\begin{pmatrix}
\psi_0 \\ \psi_1 \\ 
\hline
\lambda\psi_0 \\ \lambda\psi_1
\end{pmatrix},
\end{align}
\end{subequations}
or for the more general expression~\eqref{eq890}, $(2N)\times(2N)$ matrices and the $(2N)$-dimensional vector
\begin{subequations}
\begin{align}
\hatA&=
\begin{pmatrix}
\zero_N & \iden_N \\
\iden_N & \Hsys
\end{pmatrix},
\\
\hatB&=
\begin{pmatrix}
\iden_N & \zero_N \\
\zero_N & -\hat{\Theta}
\end{pmatrix},
\\
\ket{\Psi}&=
\begin{pmatrix}
\ket{\psi} \\ \lambda\ket{\psi}
\end{pmatrix}.
\end{align}
\end{subequations}
The eigenvalues are then found by
\begin{align}
\det(\hatA-\lambda \hatB)=0,
\end{align}
which indeed produces the four solutions in Eq.~\eqref{eq530}. 
We thereby realize that we have generally $2N$ discrete eigenvalues, not just $N$ eigenvalues, for an open tight-binding model with $N$ sites,
because the effective eigenvalue problem~\eqref{eq760} is a nonlinear one.
The system exemplified in Fig.~\ref{fig14} should have ten discrete eigenvalues, not five.

Let the four eigenvalues in Eq.~\eqref{eq530} denoted by $\lambda_n$ and the corresponding right-eigenvector by $\ket{\Psi_n}$ as in
\begin{align}
\hatA\ket{\Psi_n}=\lambda_n\hatB\ket{\Psi_n}
\end{align}
for $n=1,2,3,4$.
Each eigenvector should take the form
\begin{align}
\ket{\Psi_n}=\begin{pmatrix}
\psi_0^{(n)} \\
\psi_1^{(n)} \\
\hline
\lambda_n\psi_0^{(n)} \\
\lambda_n\psi_1^{(n)} 
\end{pmatrix}.
\end{align}
The eigenvectors that we want are only the upper half of $\ket{\Psi_n}$.
We thereby introduce the operator 
\begin{align}\label{eq975}
\hatC:=
\begin{pmatrix}
1 & 0  \\
0 & 1  \\
\hline
0 & 0  \\
0 & 0  
\end{pmatrix}
\end{align}
to extract the relevant information as in
\begin{align}
\hatC^T\ket{\Psi_n}=
\begin{pmatrix}
\psi_0^{(n)} \\
\psi_1^{(n)} \\
\end{pmatrix}.
\end{align}

Since the matrices $A$ and $B$ are generally symmetric, even in the general case of Eq.~\eqref{eq890}, we find that 
each left-eigenvector is the transpose of the corresponding right-eigenvector $\ket{\Psi_n}$.
We let the left-eigenvectors denoted by $\bra{\tilde{\Psi}_n}:=\ket{\Psi_n}^T$ to emphasize that it is \textit{not} the Hermitian conjugate of $\ket{\Psi_n}$:
\begin{align}
\bra{\tilde{\Psi}_n}=
\left(
\begin{array}{cc|cc}
\psi_0^{(n)} &
\psi_1^{(n)} &
\lambda_n\psi_0^{(n)} &
\lambda_n\psi_1^{(n)} 
\end{array}
\right).
\end{align}
We thus have
\begin{align}
\bra{\tilde{\Psi}_n}\hatA=\lambda_n\bra{\tilde{\Psi}_n}\hatB
\end{align}
for all $n$, and therefore
\begin{align}
\braket{\tilde{\Psi}_m|\hatA|\Psi_n}&=\lambda_m\braket{\tilde{\Psi}_m|\hatB|\Psi_n}
\nonumber\\
&=\lambda_n\braket{\tilde{\Psi}_m|\hatB|\Psi_n}.
\end{align}
We thereby conclude that
\begin{subequations}\label{eq980}
\begin{align}\label{eq980a}
\braket{\tilde{\Psi}_m|\hatA|\Psi_n}&=\lambda_n\delta_{mn},
\\\label{eq980b}
\braket{\tilde{\Psi}_m|\hatB|\Psi_n}&=\delta_{mn}
\end{align}
\end{subequations}
if there is no degeneracy of the eigenvalues and if we normalize the eigenvectors according to $\braket{\tilde{\Psi}_n|\hatB|\Psi_n}=1$ for all $n$.

We finally consider the complete set.
By applying $\ket{\Psi_m}$ to Eq.~\eqref{eq980b} from the left and sum it up with all $m$, we have
\begin{align}
\sum_{m=1}^4\ket{\Psi_m}\!\!\braket{\tilde{\Psi}_m|\hatB|\Psi_n}=\ket{\Psi_n}.
\end{align}
for any $n$. Similarly, by applying $\bra{\tilde{\Psi}_n}$  to Eq.~\eqref{eq980b} from the right and sum it up with all $n$, we obtain
\begin{align}
\sum_{n=1}^4\braket{\tilde{\Psi}_m|\hatB|\Psi_n}\!\!\bra{\tilde{\Psi}_n}=\bra{\tilde{\Psi}_m}
\end{align}
for any $m$.
They indicate that
\begin{align}
\sum_{m=1}^4\ket{\Psi_m}\!\!\bra{\tilde{\Psi}_m}\hatB
=\hatB\sum_{n=1}^4\ket{\Psi_n}\!\!\bra{\tilde{\Psi}_n}=\iden_4.
\end{align}

Since the matrix $\hatB$ has the identity matrix on the upper-left block, while the upper-right and lower-left blocks are zero blocks, we obtain $\hatB\hatC=\hatC^T\hatB=\iden_2$, and hence
\begin{align}
\sum_{n=1}^4\hatC^T\ket{\Psi_n}\!\!\bra{\tilde{\Psi}_n}\hatC=\iden_2.
\end{align}
or more straightforwardly, we conclude
\begin{align}\label{eq1065}
\sum_{n=1}^4
\begin{pmatrix}
\psi_0^{(n)} &
\psi_1^{(n)} 
\end{pmatrix}
\begin{pmatrix}
\psi_0^{(n)} \\
\psi_1^{(n)} 
\end{pmatrix}
=\begin{pmatrix}
1 & 0 \\
0 & 1
\end{pmatrix}.
\end{align}

In more general case of Eqs.~\eqref{eq760} and~\eqref{eq860}, we obtain the ``completeness" relation
\begin{align}\label{eq1066}
\sum_{n=1}^{2N} \hatP\ket{\psi_n}\!\!\bra{\tilde{\psi}_n}\hatP=\hatP,
\end{align}
although the $N$-dimensional identity of $\hatP$ is produced out of the summation of the $2N$ pieces of eigenstates.
(Strictly speaking, Eq.~\eqref{eq1066} expresses the equality in the whole Hilbert space, which is different from Eq.~\eqref{eq1065}, but we loosely generalized Eq.~\eqref{eq1065} to Eq.~\eqref{eq1066}.
We will take the same jump in the expression in the following too.)  
Since the space $\hatQ=\comp-\hatP$ is spanned by the plane waves because it is a flat space without any potentials, we finally conclude that~\cite{Hatano14}
\begin{align}\label{eq1070}
\sum_{n=1}^{2N} \hatP\ket{\psi_n}\!\!\bra{\tilde{\psi}_n}\hatP
+\int_{-\pi}^\pi \dyad{\phi_k}\dd{k}=\comp,
\end{align}
where $\ket{\phi_k}$ denotes a plane wave of the wave number $k$.

Note that R.~Newton proves a well-known complete set~\cite{Newton60,Newton82}:
\begin{align}\label{eq1080}
\sum_{n: \textrm{b.s.}}\dyad{\psi_n}
+\int_{-\pi}^\pi \dyad{\psi_k}\dd{k}=\comp,
\end{align}
where the summation in the first term on the left-hand side is taken over the bound states both on the imaginary axis $\Re K=0$ and on the line $\Re K=\pi/a$, while the integral in the second term contains all scattering states, \textit{not} the plane waves.
(R.~Newton's original proof was given for the problem of potential scattering in the continuum space.
We here reformulated it into Eq.~\eqref{eq1080} for the tight-binding model.)

Comparing R.~Newton's conventional complete set~\eqref{eq1080} with our new complete set~\eqref{eq1070}, we notice a couple of differences: 
\begin{enumerate}
\renewcommand{\labelenumi}{\theenumi}
\renewcommand{\theenumi}{(\roman{enumi})}
\item All the scattering information is concentrated in the summation over all states with discrete spectra in our new complete set~\eqref{eq1070}. In contrast, all states on the left-hand side, including the scattering states with a continuum spectrum, contain the scattering information in the conventional complete set~\eqref{eq1080}.
\item Decaying resonant states, as in Fig.~\ref{fig8}(a), as well as growing anti-resonant states, as in Fig.~\ref{fig8}(b), explicitly appear in our new complete set~\eqref{eq1070} in a time-reversal symmetric way, whereas these states are hidden in the conventional complete set~\eqref{eq1080}.
\end{enumerate}
We will underscore the differences in the example given in the next Subsec.~\ref{subsec3-3}.

\subsection{Breaking down quantum dynamics into the new complete set}
\label{subsec3-3}

\begin{figure*}
\includegraphics[width=0.75\textwidth]{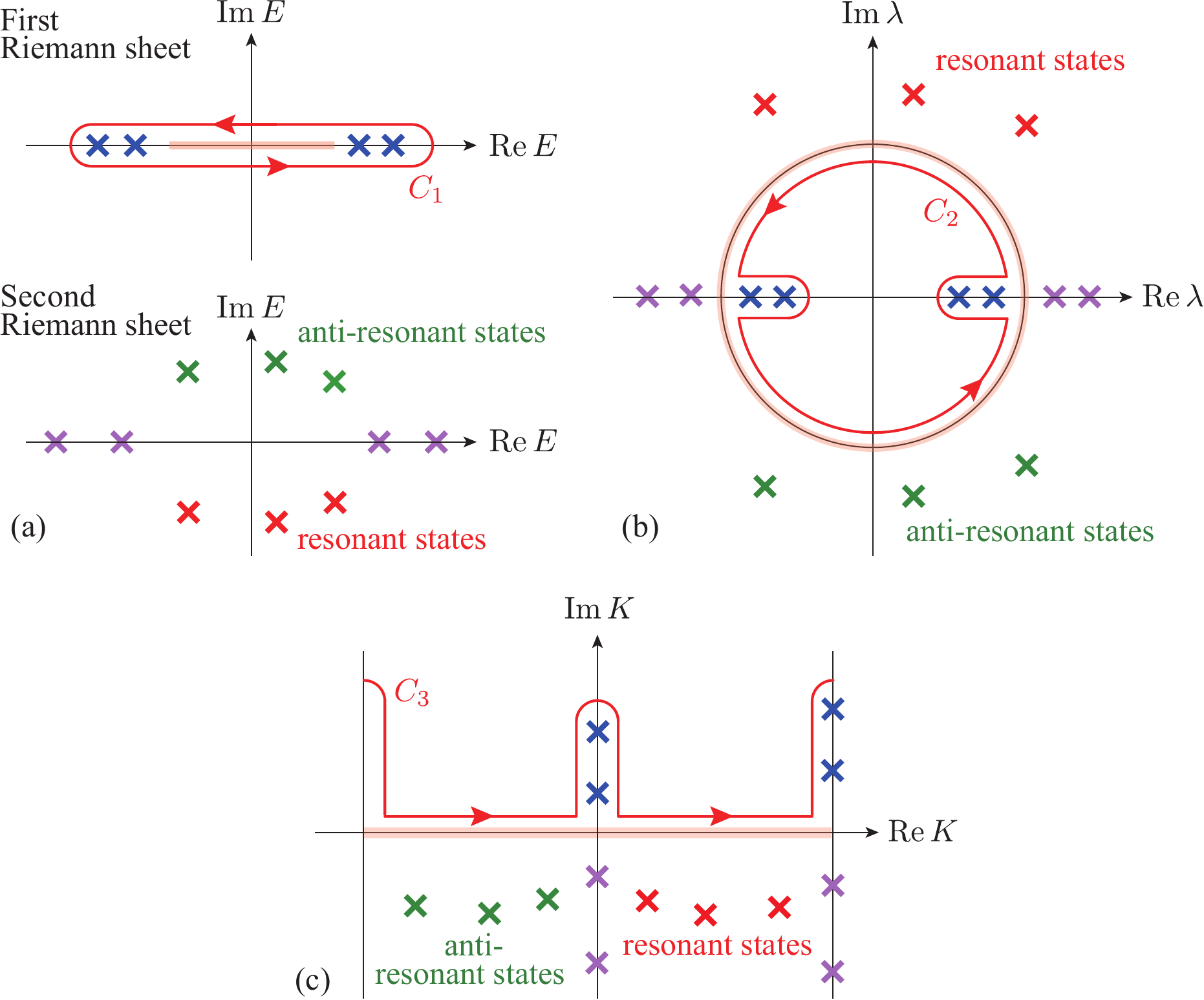}
\caption{The integral contours $C_1$, $C_2$, and $C_3$ used in Eqs.~\eqref{eq1090}, \eqref{eq1180}, and~\eqref{eq1190}, respectively. 
(a) The contour $C_1$ encircles the bound states (blue crosses) and the continuum states (orange thick line) on the real axis of the first Riemann sheet of the complex energy plane.
On the second Riemann sheet, the resonant states are in the lower half-plane, while the anti-resonant states are in the upper half-plane.
(b) The contour $C_2$ runs around just inside the unit circle (thin line), which coincides with the scattering states (orange thick circle), in the complex $\lambda$ plane, encircling the bound states on the real axis inside the unit circle.
Outside the unit circle, the resonant states are on the upper half plane, while the anti-resonant states are on the lower half plane.
(c) The contour $C_3$ runs just above the real axis, which coincides with the scattering states (orange thick line), in the complex $k$ plane, encircling the bound states on the positive part of the imaginary axis as well as on the positive part of the line $\Re k=\pi/a$.}
\label{fig15}
\end{figure*}

We here demonstrate how the new complete set~\eqref{eq1070} can break down the time evolution into discrete states, including resonant and anti-resonant states~\cite{Hatano14,Ordonez17a,Ordonez17b}.
We start our calculation from the Fourier transform of the time-evolution operator of the total Hamiltonian $H$:
\begin{align}\label{eq1090}
\ee^{-\ii \hatH t/\hbar}=\frac{1}{2\pi\ii}\int_{C_1}\ee^{-\ii Et/\hbar}\frac{1}{E-\hatH}\dd{E},
\end{align}
where the integration contour $C_1$ is specified in Fig.~\ref{fig15}(a).
The contour $C_1$ picks all poles due to the bound and scattering states, which constitute R.~Newton's complete set~\eqref{eq1080}, and puts them into the exponent of $\ee^{-\ii Et/\hbar}$ as residues, and hence produces the standard expansion of the time-evolution operator
\begin{align}
\ee^{-\ii \hatH t/\hbar}&=
\sum_{n: \textrm{b.s.}}\ket{\psi_n}\ee^{-\ii E_n t/\hbar}\bra{\psi_n}
\notag\\
&+\int_{-\pi}^\pi \ket{\psi_k}\ee^{-\ii E(k) t/\hbar}\bra{\psi_k}\dd{k}.
\end{align}
This justifies the Fourier transform~\eqref{eq1090}.

Since we are interested in the dynamics of the system, we sandwich Eq.~\eqref{eq1090} with the projection operator $\hatP$ from both sides:
\begin{align}\label{eq1115}
\hatP\ee^{-\ii \hatH t/\hbar}\hatP=\frac{1}{2\pi\ii}\int_{C_1}\ee^{-\ii Et/\hbar}\hatP\frac{1}{E-\hatH}\hatP\dd{E}.
\end{align}
We then employ a useful formula~\cite{Hatano14}
\begin{align}\label{eq1120}
\hatP\frac{1}{E-\hatH}\hatP=\hatP\frac{1}{E-\Heff(E)}\hatP,
\end{align}
where $\Heff(E)$ is the effective Hamiltonian given in Eq.~\eqref{eq790}; see App.~\ref{appE1} for an algebraic proof.
We further show the eigenstate expansion of the Green's function~\eqref{eq1120} of the form
\begin{align}\label{eq1160}
\hatP\frac{1}{E-\Heff(E)}\hatP
=\frac{1}{W}\sum_{n=1}^{2N}\hatP\ket{\psi_n}\frac{1}{{\lambda_n}^{-1}-\lambda^{-1}}\bra{\tilde{\psi}_n}\hatP.
\end{align}
We derive this equality in App.~\ref{appE2}.

Combining Eqs.~\eqref{eq1115}--\eqref{eq1160}, and 
further transferring the integration variable from $E$ to $\lambda$ with
\begin{subequations}
\begin{align}
E&=-W\qty(\lambda+\frac{1}{\lambda}),\\
\dd{E}&=-W\qty(1-\frac{1}{\lambda^2})\dd\lambda,
\end{align}
\end{subequations}
we finally arrive at
\begin{widetext}
\begin{align}\label{eq1180}
\hatP\ee^{-\ii \hatH t/\hbar}\hatP=\frac{1}{2\pi\ii }\sum_{n=1}^{2N}\int_{C_2}\exp\qty[\frac{\ii W}{\hbar} \qty(\lambda+\frac{1}{\lambda}) t]
\hatP\ket{\psi_n}\frac{1}{\lambda^{-1}-{\lambda_n}^{-1}}\bra{\tilde{\psi}_n}\hatP
\qty(-1+\frac{1}{\lambda^2})\dd\lambda,
\end{align}
\end{widetext}
where the integration contour $C_2$ is specified in Fig.~\ref{fig15}(b).
Note that we inverted the direction of the contour upon converting the variable from $E$ to $\lambda$.
We can also transform the integration variable from $\lambda=\ee^{\ii K a}$ to $K$, as in
\begin{align}\label{eq1190}
\hatP\ee^{-\ii \hatH t/\hbar}\hatP&=\frac{a}{\pi\ii }\sum_{n=1}^{2N}
\hatP\ket{\psi_n}\!\!\bra{\tilde{\psi}_n}\hatP
\notag\\
&\times\int_{C_3}
\frac{\ee^{-\ii E(K) t /\hbar}\sin(Ka)}{\ee^{-\ii Ka}-\ee^{-\ii K_n a}}
\ \dd K
\end{align}
where the integration contour $C_3$ is specified in Fig.~\ref{fig15}(c) and $E(K)=-2W\cos(Ka)$.

We stress here that these eigenstate expansions~\eqref{eq1180} and~\eqref{eq1190} still keep the time-reversal symmetry;
the time-reversal pairs of the resonant and anti-resonant states remain in the expansions as they are.
We will show below that, for positive time $t>0$, we naturally choose the resonant decaying poles, whereas for negative time $t<0$, we naturally choose the anti-resonant growing poles. 
We emphasize that these choices are \textit{not} arbitrary; the symmetry is spontaneously broken.
We will show that this is the essential difference from the previous approach~\cite{Berggren82}.

Before analyzing spontaneous symmetry breaking theoretically, we present a numerical example to illustrate our point in the simple case shown in Fig.~\ref{fig10}.
For this purpose, we use a specific parameter set $v_0=0$, $w_1=1/2$, and $\theta=3/4$.
We know from Fig.~\ref{fig11} that there are only two pairs of resonant and anti-resonant states; there are no bound and anti-bound states; see Table~\ref{tab1} for the eigenvalues for this specific parameter set.
\begin{table*}
\caption{The eigenvalues and eigenstates that exist for the parameter set  $v_0=0$, $w_1=1/2$ and $\theta=3/4$  with $\hbar=W=a=1$.
Note that the numbering of $n$ is arbitrary; we have chosen it only for convenience here.}
\label{tab1}
\begin{tabular}{lllccc}
\hline
$n$ $\quad$ & types & parity $\quad$ & $\lambda_n$ & $K_n$ & $E_n$ \\
\hline \hline
$1$ $\quad$ & resonant $\quad$ & even 		& $\frac{1}{3}\qty(+1+\ii\sqrt{11})$ $\quad$ & $+1.27795-\ii 0.143841$$\quad$ & $\frac{1}{12}\qty(-7-\ii\sqrt{11})$ \\
$2$ $\quad$ & anti-resonant $\quad$ & even 	& $\frac{1}{3}\qty(+1-\ii\sqrt{11})$ $\quad$ & $-1.27795-\ii 0.143841$$\quad$ & $\frac{1}{12}\qty(-7+\ii\sqrt{11})$ \\
$3$ $\quad$ & resonant $\quad$ & odd 		& $\frac{1}{3}\qty(-1+\ii\sqrt{11})$ $\quad$ & $+1.86364-\ii 0.143841$$\quad$ & $\frac{1}{12}\qty(+7-\ii\sqrt{11})$ \\
$4$ $\quad$ & anti-resonant $\quad$ & odd 	& $\frac{1}{3}\qty(-1-\ii\sqrt{11})$ $\quad$ & $-1.86364-\ii 0.143841$$\quad$ & $\frac{1}{12}\qty(+7+\ii\sqrt{11})$ \\
\hline
\end{tabular}
\end{table*}

We then evaluate the following probability of quantum dynamics. 
We set the initial state of the time evolution as
\begin{align}\label{eq1205}
\ket{\Psi(0)}=\begin{pmatrix}
\psi_0(0) \\
\psi_1(0)
\end{pmatrix}=
\frac{1}{\sqrt{2}}
\begin{pmatrix}
1\\
1
\end{pmatrix}.
\end{align}
We let it evolve in time for $t$, which can be positive or negative, according to $H$, and evaluate the probability that the state is still in the initial state~\eqref{eq1205}.
This survival probability is then expressed in the form
\begin{align}\label{eq1210}
\Ps(t)=\abs{\braket{\Psi(0)|\ee^{-\ii \hat{H} t/\hbar}|\Psi(0)}}^2.
\end{align}

\begin{figure}
\includegraphics[width=0.9\columnwidth]{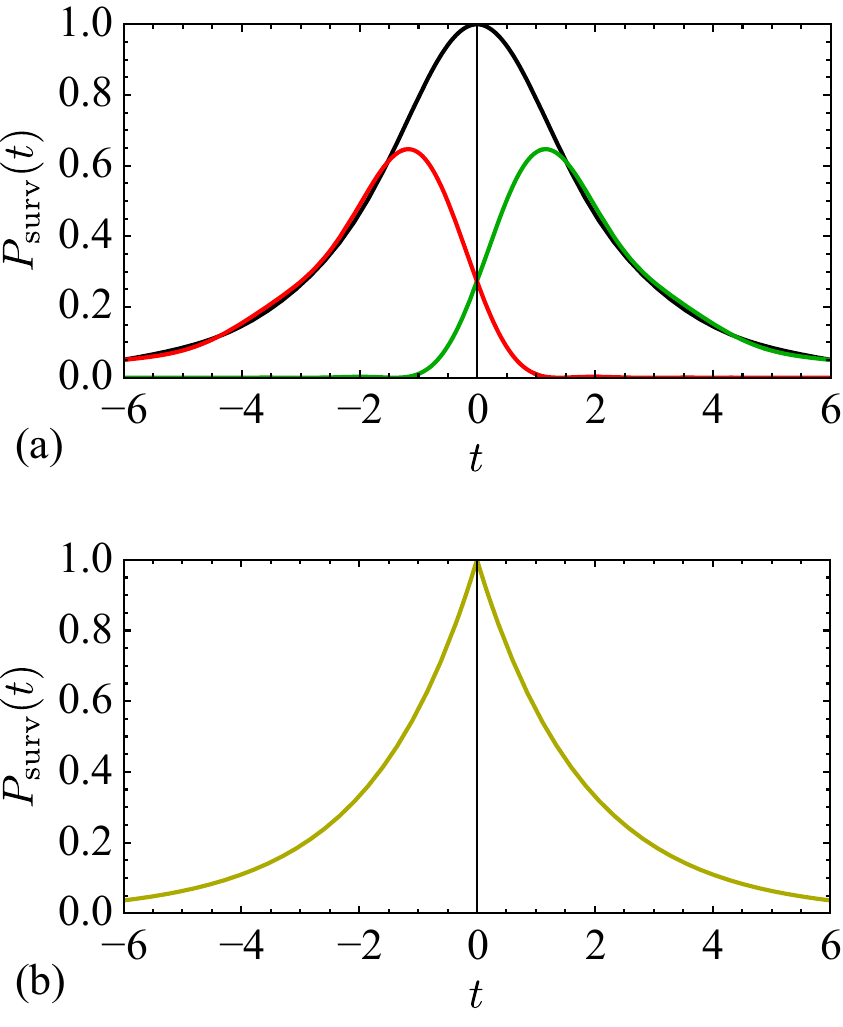}
\caption{The time dependence of the survival probability~\eqref{eq1210}. We here set $v_0=0$ and $w_1=1/2$ with $\hbar=W=a=1$. (a) The results of the numerical integration of Eq.~\eqref{eq1230}. The overall black curve indicates the survival probability $\Ps(t)=\abs{c_1(t)+c_2(t)}^2$, while the green curve on the right indicates the resonant-state component $\abs{c_1(t)}^2$ and the red curve on the left indicates the anti-resonant-state component $\abs{c_2(t)}^2$. 
(b) A possible evaluation according to other approaches with only pole components. We used the two energy eigenvalues $E_1$ and $E_2$ in Table~\eqref{tab1}. }
\label{fig16}
\end{figure}

Since the initial state~\eqref{eq1205} is in the $\hatP$ subspace, we can use either of the resonant-state expansions~\eqref{eq1180} and~\eqref{eq1190}.
For the numerical calculations in Fig.~\ref{fig16}(a), we used the latter.
The choice of the initial state~\eqref{eq1205} picks only the even parity solutions $n=1$ and $n=2$ in Table~\ref{tab1}.
We therefore have
\begin{align}
\braket{\Psi(0)|\ee^{-\ii \hat{H} t/\hbar}|\Psi(0)}&=c_1(t)+c_2(t),
\end{align}
where
\begin{align}\label{eq1230} 
c_n(t)&:=\frac{a}{\pi\ii}\qty( \braket{\Psi(0)|\psi_n})^2
\int_{-\pi/a}^{\pi/a}\frac{\ee^{-\ii E(k)t/\hbar}\sin(ka)}{\ee^{-\ii ka}-\ee^{-\ii K_na}}\dd k;
\end{align}
note that we used the fact $\bra{\tilde{\psi}_n}=\ket{\psi_n}^T$.
Figure~\ref{fig16}(a) shows the time evolution of $\Ps(t)=\abs{c_1(t)+c_2(t)}^2$ along with $\abs{c_1(t)}^2$ and $\abs{c_2(t)}^2$.

Let us note three crucial points in Fig.~\ref{fig16}(a). 
First, the survival probability~\eqref{eq1210} is an even function of $t$~\cite{Ordonez17a}.
We will show the symmetry by using the time-reversal operator $\hatT$, which is an anti-linear operator that commutes with the Hamiltonian $\hatH$, as we discussed in Eq.~\eqref{eq370}.
Applying it to the time-evolved state, we have
\begin{align}\label{eq1231}
\hatT\ket{\Psi(t)}=\hatT\ee^{-\ii H t}\ket{\Psi(0)}
=\ee^{\ii H t}\hatT\ket{\Psi(0)},
\end{align}
where we used the fact that the anti-linear operator flips the sign of $\ii$ to $-\ii$ but it commutes with $H$.
Note that it does not flip the sign of $t$ in the exponent because $t$ is only a parameter here.
Assuming the simple case of $\hatT^2=1$, we have
\begin{align}\label{eq1232}
\ket{\Psi(t)}=\hatT\ee^{\ii H t}\hatT\ket{\Psi(0)}.
\end{align}
This means that a state that evolves forward in time, which is on the left-hand side, is obtained, as on the right-hand side, by first flipping the direction of time of the initial state $\ket{\Psi(0)}$, letting it evolve backward in time for $-t$, and finally flipping the direction of the time again.
Particularly in the case of the initial state~\eqref{eq1205}, we have $\hatT\ket{\Psi(0)}=\ket{\Psi(0)}$.
Therefore, Eq.~\eqref{eq1232} now reads
\begin{align}
\ket{\Psi(t)}=\hatT\ee^{\ii H t}\ket{\Psi(0)}=\hatT\ket{\Psi(-t)}=\ket{\Psi(-t)}^\ast,
\end{align}
and hence we have
\begin{align}
\braket{\Psi(0)|\Psi(t)}=\braket{\Psi(0)|\Psi(-t)}^\ast,
\end{align}
which dictates that the survival probability~\eqref{eq1210} must be an even function of $t$, as in $\Ps(t)=\Ps(-t)$.

The second important point to note in Fig.~\ref{fig16}(a) is that the resonant-state component is dominant for $t>0$, describing decay from the initial state~\eqref{eq1205}, while the anti-resonant-state component is dominant for $t>0$, describing growth towards the ``terminal" state~\eqref{eq1205}~\cite{Ordonez17a,Ordonez17b,Hatano19}.
The symmetry between the resonant-state component and the anti-resonant-state component with respect to $t=0$ reflects the fact that our decomposition in Eqs.~\eqref{eq1180} and~\eqref{eq1190} keeps the decaying resonant state and the growing anti-resonant state as a pair.
This is in marked contrast to some of the previous approaches.

Since the well-known complete set~\eqref{eq1080} does not contain any decaying components directly, to describe the decay for $t>0$, one would pull down the positive-$k$ part of the integration contour to pick the resonance poles in the fourth quadrant in Fig.~\ref{fig12}(a) ``by hand"~\cite{Berggren82}.
To describe the dynamics for $t<0$, one would then have to switch "by hand" from picking the resonance poles to picking the anti-resonance poles in the third quadrant by pulling down this time the negative-$k$ part of the integration contour.
There have been other approaches that completely separate the decaying dynamics for $t>0$ from the growing dynamics for $t<0$~\cite{Prigogine73,Petrosky91}.
These approaches would give the dynamics of the survival probability~\eqref{eq1210} in the form schematically shown in Fig.~\ref{fig16}(b).

The final important point to note in Fig.~\ref{fig16}(a) is that the survival probability $\Ps(t)$ smoothly continues from the negative part to the positive part, that is, from the past to the future, not forming a cusp as in Fig.~\ref{fig16}(b), which would separate the past and the future with a singularity.
This difference is underscored by the fact that the anti-resonant component gradually transitions to the resonant component in Fig.~\ref{fig16}(a), rather than suddenly, as in Fig.~\ref{fig16}(b).
We will discuss the smoothness around $t=0$ in the next Sec.~\ref{sec4}.

We go back to the expansion~\eqref{eq1180} and show the first point, namely the reason why the resonant-state component is dominant for $t>0$, while the anti-resonant state one is for $t<0$~\cite{Hatano14}.
In evaluating Eq.~\eqref{eq1180} analytically, the most serious difficulty comes from the essential singularities at $\lambda=0$ and $\lambda=\infty$ in the exponential function in the integrand.
We suppress these essential singularities as follows, and along the way, we break the time-reversal symmetry unavoidably.

\begin{figure}
\includegraphics[width=0.75\columnwidth]{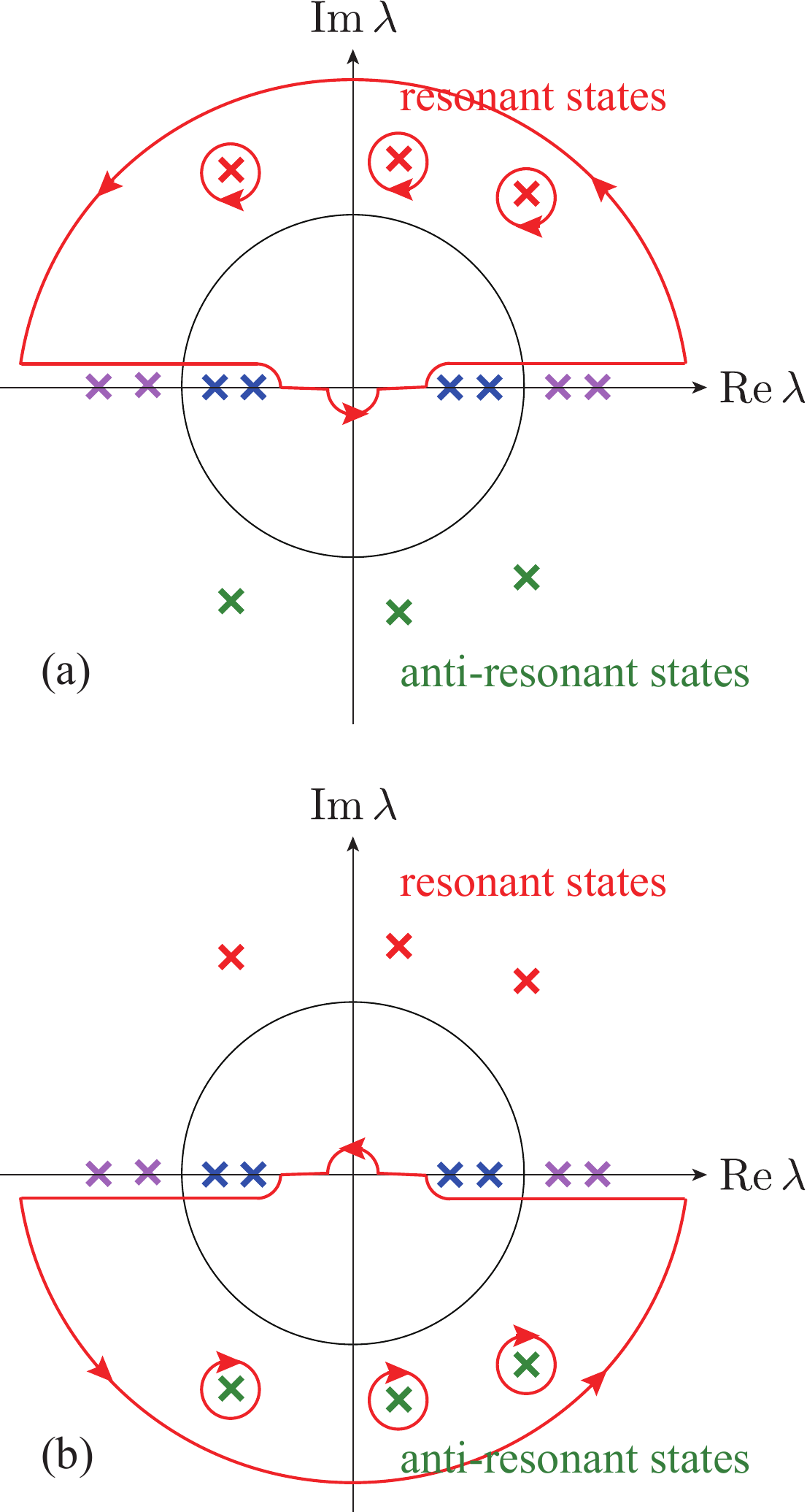}
\caption{The contour modification of $C_2$ in Fig.~\ref{fig15}(b) for Eq.~\eqref{eq1180}, to eliminate the essential singularities (a) for $t>0$ and (b) for $t<0$.}
\label{fig17}
\end{figure}

For $t>0$, we modify the contour as shown in Fig.~\ref{fig17}(a);
we circle the essential singularity at $\lambda=0$ on its lower side, while we expand the contour to the infinite circle on the upper side.
The first small half circle is given by $\lambda=\epsilon \exp(\ii \varphi)$ with $-\pi<\varphi<0$.
The contribution from this circle vanishes in the limit $\epsilon\to0$ because
\begin{align}
\abs{\exp\qty[\frac{\ii W}{\hbar}\qty(\lambda+\frac{1}{\lambda})t]}
\simeq \exp\qty[\frac{W}{\hbar}\qty(\epsilon -\frac{1}{\epsilon})t]
\end{align}
The second large half circle is given by $\lambda=R \exp(\ii \varphi)$ with $0<\varphi<\pi$.
The contribution from this circle vanishes in the limit $R\to\infty$ because
\begin{align}
\abs{\exp\qty[\frac{\ii W}{\hbar}\qty(\lambda+\frac{1}{\lambda})t]}
\simeq \exp\qty[\frac{W}{\hbar}\qty(-R +\frac{1}{R})t]
\end{align}
We are left with three contributions:
the residues from the resonance poles in the upper half plane outside the unit circle;
the half of the residues from the bound and anti-bound states on the real axis;
the principal value of the integral on the real axis.
The first contribution from resonant states naturally gives exponential decays for $t>0$.
Each component $n$ in the summation of Eq.~\eqref{eq1180} picks the corresponding pole $\lambda=\lambda_n$, but only in the upper-half plane for the contour in Fig.~\ref{fig17}(a), and its residue contains the exponential decay due to the imaginary part of the energy eigenvalue $E_n=-W(\lambda_n+1/\lambda_n)$ in $\exp(-\ii E_n t/\hbar)$.

Only this contribution would produce the time-dependence in Fig.~\ref{fig16}(b), without the smoothness around $t=0$ demonstrated in Fig.~\ref{fig16}(a).
In fact, the last contribution, namely the integral on the real axis of $\lambda$, gives the deviations from the pure exponential decay, both in the short- and long-time regions, due to its non-Markovian nature.
We will describe this in the next Sec.~\ref{sec4}.

For $t<0$, we modify the contour as shown in Fig.~\ref{fig17}(b);
we circle the essential singularity at $\lambda=0$ on its upper side, while we expand the contour to the infinite circle on the lower side.
We can similarly prove that the contributions from the small half circle around $\lambda=0$ and the large half circle in the lower half plane vanish.
We are then left with three contributions again, but for $t<0$, the residues from the anti-resonance poles in the lower half plane are dominant, instead of the resonance poles for $t>0$.
This naturally gives exponential growth for $t<0$ towards the state at $t=0$.

We again stress that switching from the anti-resonant to the resonant component occurs automatically, not manipulated ``by hand." 
This is because our decomposition to all discrete eigenstates based on the Feshbach formalism keeps all resonant and anti-resonant poles as pairs.
We believe that this is a new approach to the dynamics of open quantum systems.

\subsection{Calculation of the effective Hamiltonian in continuous models}
\label{subsec3-4}

So far, we have described the Siegert boundary condition for models in continuous space, whereas we have applied the Feshbach formalism to tight-binding models. 
The latter led to the effective Hamiltonian in Eq.~\eqref{eq820} and the quadratic eigenvalue equation~\eqref{eq860}. 
For continuous models, the calculation of the effective Hamiltonian using the Feshbach formalism seems nontrivial because of singularities that arise in operators such as $\PHQ$. 

In this subsection, we will show that the effective Hamiltonian in a simple continuous model can be obtained from its discrete counterpart, which is a tight-binding model, as is reviewed in the second half of App.~\ref{appC}.  
We will also show that this effective Hamiltonian leads to the same quadratic eigenvalue equation derived for the ``Siegert states for cutoff potentials'' in Ref.~\cite{Tolstikhin98}.

We will start by reviewing the latter formulation of Siegert states. We consider the one-dimensional continuous Hamiltonian
\begin{align}
\hatH =  \int_{-\infty}^\infty \dd x\,\ket{x}\qty(-\frac{\hbar^2}{2m} \dv[2]{x} + V(x))\bra{x}.
\end{align}
The original article~\cite{Tolstikhin98} considered the semi-infinite space, which corresponds to the radial coordinate, but here we will reformulate it in the infinite one-dimensional space.
Let us again assume that the potential is a real function that has a finite support $\abs{x}<\ell$ such that $V(x)= 0$ for $\abs{x}\ge \ell$.
We now try to find the Siegert states that satisfy the eigenvalue equation
\begin{align}\label{eqG1310}
\braket{x|\hatH|\psi} = E \braket{x|\psi}
\end{align}
under the boundary conditions
\begin{align}
\eval{\dv{x}\psi(x)}_{x=\pm\ell} &= \pm \ii K \psi(\pm\ell),
\label{eqG90}
\end{align}
where $\psi(x)=\braket{x|\psi}$, $E = (\hbar K)^2/(2m)$, and the signs are taken in the same order.
The boundary conditions at $x=\pm\ell$ in Eq.~\eqref{eqG90} are equivalent to outgoing free-wave propagation for $\abs{x}>\ell$, namely the Siegert boundary condition~\eqref{eq30}.
We show below that limiting the space to $\abs{x}\le\ell$ leads to a quadratic eigenvalue equation~\cite{Tolstikhin98}.

The Hamiltonian $\hatH$ is not Hermitian when it is in the restricted space $ \abs{x}\le \ell$.
Let us define
\begin{align}\label{eqG133}
\hatH_\ell =  \int_{-\ell}^\ell \dd x\,\ket{x}\qty(-\frac{\hbar^2}{2m} \dv[2]{x} + V(x))\bra{x}.
\end{align}
As was indeed demonstrated in Eq.~\eqref{eq210} in a different context,
we find
\begin{align}\label{eqG1330}
&\bra{\psi}\qty(-\frac{\hbar^2}{2m}\dv[2]{x})\ket{\psi}_\ell
\nonumber\\
&=\int_{-\ell}^\ell \dd x\,\psi(x)^\ast \qty(-\frac{\hbar^2}{2m}\dv[2]{x})\psi(x) 
\nonumber\\
&=-\frac{\hbar^2}{2m}\qty[\psi(x)^\ast\dv{x}\psi(x)]_{x=-\ell}^\ell+\frac{\hbar^2}{2m}\int_{-\ell}^\ell \dd x\,\abs{\dv{x}\psi(x)}^2.
\end{align}
The second term on the right-hand side is real for arbitrary functions $\psi(x)$ defined in $\abs{x}\le\ell$, whereas the first boundary term is generally complex;
particularly in the present case of the Siegert boundary condition~\eqref{eqG90}, the boundary term reads
\begin{align}
-\ii\frac{\hbar^2 K}{2m}\qty(\abs{\psi(\ell)}^2+\abs{\psi(-\ell)}^2)
\end{align}
This implies that the Hamiltonian $\hatH_\ell$ in Eq.~\eqref{eqG133} is not Hermitian.

To make it Hermitian, we can use the Bloch operator given by~\cite{Tolstikhin98}
\begin{align}
\hatL = \frac{\hbar^2}{2m} \qty(\eval{\ket{x}\dv{x}\bra{x}}_{x=\ell}-\eval{\ket{x}\dv{x}\bra{x}}_{x=-\ell}) .
 \label{eqG80}
\end{align}
The operator $\Hham: =  \hatH_\ell + \hatL$ is indeed Hermitian because the matrix element of the Bloch operator
\begin{align}\label{eqG1360}
&\braket{\psi| \hatL|\psi}_\ell
\nonumber\\
&=\frac{\hbar^2}{2m}\qty(\psi(x)^\ast\eval{\dv{x}\psi(x)}_{x=\ell}
-\psi(x)^\ast\eval{\dv{x}\psi(x)}_{x=-\ell})
\end{align}
exactly cancels out the boundary terms in Eq.~\eqref{eqG1330} and leaves only the second term on the right-hand side.
We thus have
\begin{align}
\braket{\psi|\Hham|\psi}_\ell=\int_{-\ell}^\ell \dd x\,\qty(\abs{\dv{x}\psi(x)}^2+V(x)\abs{\psi(x)}^2)\in\mathbb{R}
\end{align}
for an arbitrary function $\psi(x)$ defined in $\abs{x}\le\ell$, and therefore conclude that the operator $\Hham$ is Hermitian in this restricted space.
This is, in fact, a standard procedure to make a second-differential operator in a restricted space Hermitian and numerically tractable~\cite{Bloch57,Robson69,Milnikov01,Milnikov07}.


Instead of the eigenvalue equation~\eqref{eqG1310} in the full space, 
we now restrict the space to $\abs{x}\le \ell$ and replace $\hatH$ with $\Hham-\hatL$ as in
\begin{align}\label{eqG1380}
\bra{x}\qty(\Hham-\hatL-E)\ket{\psi}_\ell=0.
\end{align}
We thus separate the system space from the environmental space.
Using again the Siegert boundary condition~\eqref{eqG90} to convert Eq.~\eqref{eqG1360} to the form
\begin{align}
\braket{\psi|\hatL|\psi}_\ell = \ii \frac{\hbar^2}{2m} K\qty(\abs{\psi(\ell)}^2+\abs{\psi(-\ell)}^2)
\end{align}
and the dispersion relation $E=\hbar^2K^2/(2m)$, we can transform Eq.~\eqref{eqG1380} to the  eigenvalue equation~\cite{Tolstikhin98}
\begin{align}
\bra{x} \qty[ \Hham -\ii \frac{\hbar^2}{2m} K \qty( \dyad{\ell} +\dyad{-\ell})- \frac{\hbar^2}{2m} K^2  ]\ket{\psi}_\ell = 0,
\label{eqG100}
 \end{align}
which is quadratic in $K$.
We can use the same standard method~\cite{Tisseur01} as in Subsec.~\ref{subsec3-2} to solve this quadratic eigenvalue equation.


In the following, we will show that Eq.~\eqref{eqG100} can be obtained from the quadratic eigenvalue problem for a tight-binding model by introducing a discretized version of the continuous model. 
We show how the effective Hamiltonian in the continuous model can emerge from its discretized version. 
We will see that $(1+\ii Ka)$ will replace $\lambda:=\ee^{\ii Ka}$ in Eq.~\eqref{eq860}. 
This is plausible because the continuous model arises from the expansion of the tight-binding model with respect to $Ka$ up to second order, as we show in App.~\ref{appC}.

We start with the discretized second derivative, as introduced in Eq.~\eqref{eqC70},
\begin{align}\label{eqG142}
\dv[2]{x}\psi(x) \simeq \frac{\psi(x+a)+\psi(x - a)-2\psi(x)}{a^2},
\end{align}
and obtain the discrete version of the continuous Hamiltonian
\begin{align}\label{eqG1410}
\hatH_a &= -W a \sum_{n=-\infty}^\infty \qty( \dyad{n+1}{n} + \dyad{n}{n+1} - 2 \dyad{n}{n} ) \nonumber\\
&+  a \sum_{n=-\infty}^\infty \ket{n} V(na)\bra{n}
\end{align}
such that $\hatH=  \lim_{a \to 0} H_a$, where $\braket{x|n}=\psi(na)$ and $W = \hbar^2/(2ma^2)$.
This discrete Hamiltonian~\eqref{eqG1410} has a similar form to the tight-binding Hamiltonian of Sec.~\ref{subsec3-1} except for the addition of the onsite energy $2W$ and the extra factor $a$ in front of the sums.
We have the former because we did not shift the ground-state energy by $2W$ as in Eq.~\eqref{eqC90d}
and we put the latter so that we can take the continuum limit under proper normalization.
In fact, we normalize the states $\ket{n}$ as
\begin{align}
\braket{ m|n} = \frac{1}{a} \delta_{mn}
\end{align}
so that we can have the following relations in the limit  $a\to 0$:
\begin{align}
\braket{ m|n} = \frac{1}{a} \delta_{mn} &\longrightarrow \delta(x'- x),\\
a\sum_{n=-\infty}^{\infty} \dyad{n}{n} =1& \longrightarrow \int_{-\infty}^{\infty} \dyad{x}{ x} \dd x = 1,
\end{align}
where we made the correspondence $x:=na$ and $x':=ma$.

To apply the Feshbach formalism to the discrete model~\eqref{eqG1410}, we start by defining the operators $\hatP$ and $\hatQ$. We let the operator $\hatP$ project the states to the region $\abs{x}\le \ell$ where the potential $V(x)$ is present, while we let the operator $\hatQ$ project states to the region $\abs{x}>\ell$ where the potential vanishes. 
In other words, we choose
\begin{subequations}
\begin{align}
\hatP &=a \sum_{n=-n_\ell}^{n_\ell} \dyad{n},\\
\hatQ &= a \qty(\sum_{-\infty}^{-n_\ell-1}+\sum_{n=n_\ell+1}^{\infty} )\dyad{n},
\end{align}
\end{subequations}
where $n_\ell: = \ell/a$. Defining
\begin{align}
\hatH_n &:=  -W\qty( \dyad{n+1}{n} +\dyad{n}{n+1} - 2 \dyad{n}) \nonumber\\
&+ \ket{n} V(na)\bra{n},
\end{align}
we have
\begin{subequations}
\begin{align}
\label{eqG110a}
\PHP &= a \sum_{n=-n_\ell+1}^{n_\ell-1}  \hatH_n + 2W a\qty(\dyad{n_\ell}+\dyad{-n_\ell}),\\
\QHQ &= a \qty(\sum_{-\infty}^{-n_\ell-1}+\sum_{n=n_\ell+1}^{\infty} ) \hatH_n ,\\
\QHP &= -W a \dyad{n_\ell+1}{n_\ell} ,\\
\PHQ &= -W  a \dyad{n_\ell}{n_{\ell+1} } .
\end{align}
\end{subequations}
Note that an extra term appears on the right-hand side of Eq.~\eqref{eqG110a}.

A calculation similar to the one in App.~\ref{appD} shows that 
\begin{align}
 &\PHQ\frac{1}{E-\QHQ}\QHP 
 \nonumber\\
& = -W a\, \ee^{\ii Ka} \qty(\dyad{n_\ell}+\dyad{-n_\ell})
\end{align}
with the dispersion relation
\begin{align}
E = 2W\qty[1-\cos(K a)].
\end{align}
The complete effective discrete Hamiltonian~\eqref{eq790} is then given by
\begin{subequations}
\begin{align}\label{eqG152a}
&\Heff(E) =\PHP + \PHQ\frac{1}{E-\QHQ}\QHP \\
\label{eqG152b}
&=  a \sum_{n=-n_\ell+1}^{n_\ell-1}  \hatH_n 
+ W a (2-\ee^{\ii Ka}) \qty(\dyad{n_\ell}+\dyad{-n_\ell}),
\end{align}
\end{subequations}
and hence the  eigenvalue equation 
\begin{widetext}
\begin{subequations}
\begin{align}\label{eqG1520a}
0&=\left(\Heff(E)-E\right)\ket{\psi}
\\\label{eqG1520b}
&=\left[ a \sum_{n=-n_\ell+1}^{n_\ell-1}  \hatH_n + W a (2-\ee^{\ii Ka})\qty(\dyad{n_\ell}+\dyad{-n_\ell})\right.
 -W\qty(2-\ee^{\ii Ka}-\ee^{-\ii Ka})\Bigg]\ket{\psi}
\end{align}
\end{subequations}
is quadratic in $\lambda=\ee^{\ii Ka}$.

In taking the limit $a\to0$, we will make the following identifications between Eqs.~\eqref{eqG100} and~\eqref{eqG1520b}:
\begin{subequations}
\begin{align}
\label{eqG1530a}
\Hham&\longleftrightarrow \Hham':=a\sum_{n=-n_\ell+1}^{n_\ell-1}  \hatH_n +Wa\qty(\dyad{n_\ell}+\dyad{-n_\ell}),\\
\label{eqG1530b}
-\ii \frac{\hbar^2}{2m}K\qty(\dyad{\ell}+\dyad{-\ell})
&\longleftrightarrow Wa\qty(1-\ee^{\ii Ka})\qty(\dyad{n_\ell}+\dyad{-n_\ell}),\\
\label{eqG1530c}
\frac{\hbar^2}{2m}K^2
&\longleftrightarrow W\qty(2-\ee^{\ii Ka}-\ee^{-\ii Ka})
\end{align}
\end{subequations}
\end{widetext}
We understand the identification in the second and third equations straightforwardly from the correspondence $W\longleftrightarrow \hbar^2/(2ma^2)$ in Eq.~\eqref{eqC90a}

Let us show the identification in the first equation~\eqref{eqG1530a}.
Remember $\Hham=\hatH_\ell-\hatL$ on its left-hand side.
On its right-hand side, we have
\begin{widetext}
\begin{align}\label{eq1550}
\braket{n|\Hham'|\psi}=
\begin{cases}
-Wa\qty(\braket{n-1|\psi}+\braket{n+1|\psi}-2\braket{n|\psi})+aV(na)\braket{n|\psi}\quad& \mbox{for $\abs{n}<n_\ell$},\\
0 & \mbox{for $\abs{n}>n_\ell$}.
\end{cases}
\end{align}
\end{widetext}
The first line in Eq.~\eqref{eq1550} indeed converges to $\braket{x|\hatH_\ell|\psi}$ in the limit $a\to 0$ because of Eq.~\eqref{eqG142}, and we have
\begin{align}
\sum_{n=-n_\ell+1}^{n_\ell-1}\braket{n|\Hham'|\psi}\longrightarrow \int_{-\ell}^\ell \braket{x|\hatH_\ell|\psi} \dd x.
\end{align}
The second line in Eq.~\eqref{eq1550} is consistent with the fact that $\Hham$ is restricted to the space $\abs{x}<\ell$.
We should be careful only when $\abs{n}=n_\ell$:
\begin{subequations}\label{eqG157}
\begin{align}
\braket{+n_\ell|\Hham'|\psi}=-Wa\qty(\braket{+n_\ell-1|\psi}-\braket{+n_\ell|\psi}),\\
\braket{-n_\ell|\Hham'|\psi}=-Wa\qty(\braket{-n_\ell+1|\psi}-\braket{-n_\ell|\psi}).
\end{align}
\end{subequations}
They converge to 
\begin{subequations}
\begin{align}
+\frac{\hbar^2}{2m}\eval{\dv{x}\psi(x)}_{x=\ell},\\
-\frac{\hbar^2}{2m}\eval{\dv{x}\psi(x)}_{x=\ell},
\end{align}
\end{subequations}
respectively, because of the discretization of the first-order derivative.
Therefore, the elements in Eq.~\eqref{eqG157} lead to the matrix element $\braket{x|\hatL|\psi}$ of the Bloch operator.
Therefore, the effective Hamiltonian~\eqref{eqG152b} takes the following form in the continuous limit:
\begin{align}
\Heff(E)=\Hham-\ii\frac{\hbar^2}{2m}K\qty(\dyad{\ell}+\dyad{-\ell}).
\end{align}
This completes showing the correspondence between Eqs.~\eqref{eqG100} and~\eqref{eqG1520a}.

In summary, we have shown that the effective Hamiltonian in the continuous model can be obtained from the Feshbach formulation of the corresponding discrete model and verified that the quadratic eigenvalue equation of the effective Hamiltonian agrees with the one given in Ref.~\cite{Tolstikhin98}.  
It is an interesting problem to prove the completeness of the solutions of the quadratic eigenvalue equation~\eqref{eqG100}, in analogy to our proof of the new complete set in Subsec.~\ref{subsec3-2}.

It is important to stress that in the present argument we assume that the potential vanishes in the region $x\ge\ell$. 
For potentials with tails outside the region, we would need to cut off the potential at $x=\ell$ and neglect the tail outside that region. 
The validity of the approximation depends on the specific form of the tail of the potential in this region, particularly when it is a long-range potential, such as the Coulomb potential.


\section{Non-Markovian dynamics of open quantum systems}
\label{sec4}

In the present section, we discuss more about the dynamics of open quantum systems
based on what we have discussed in the preceding sections, particularly in Sec.~\ref{sec3}.
We first show how the non-Markovianity arises in the dynamics, using the Feshbach formalism for the time-dependent \Sch equation.
We then evaluate deviations from exponential decay using the Feshbach formalism, particularly in the short- and long-time regimes.

\subsection{Feshbach formalism for the time-dependent \Sch equation}
\label{subsec4-1}

We show here how the non-Markovianity emerges in open quantum systems, even in the one-body problem.
We apply the Feshbach formalism which we used in Subsec.~\ref{subsec3-1} for the time-independent \Sch equation~\eqref{eq730},  this time to the time-dependent \Sch equation:
\begin{align}
\ii\hbar\dv{t}\ket{\Psi(t)}=\hatH\ket{\Psi(t)}.
\end{align}

We apply the projection operators $\hatP$ and $\hatQ$ to it, obtaining
\begin{subequations}
\begin{align}
\ii\hbar\dv{t}\hatP\ket{\Psi(t)}&=\hatP\hatH\ket{\Psi(t)},\\
\ii\hbar\dv{t}\hatQ\ket{\Psi(t)}&=\hatQ\hatH\ket{\Psi(t)}.
\end{align}
\end{subequations}
Using Eq.~\eqref{eq720} again, we have 
\begin{subequations}
\begin{align}\label{eq1280a}
\ii\hbar\dv{t}\qty(\hatP\ket{\Psi(t)})&=\PHP\qty(\hatP\ket{\Psi(t)})+\PHQ\qty(\hatQ\ket{\Psi(t)}),\\
\label{eq1280b}
\ii\hbar\dv{t}\qty(\hatQ\ket{\Psi(t)})&=\QHP\qty(\hatP\ket{\Psi(t)})+\QHQ\qty(\hatQ\ket{\Psi(t)}).
\end{align}
\end{subequations}
We solve the second equation with respect to $\hatQ\ket{\Psi(t)}$ and insert it into the first equation, making the first one an equation for $\hatP\ket{\Psi(t)}$. 

The second equation~\eqref{eq1280b} is an inhomogeneous differential equation.
The homogeneous term for  $\hatQ\ket{\Psi(t)}$ yields
\begin{align}
\hatQ\ket{\Psi(t)} =\ee^{-\ii \QHQ t/\hbar}  \hatQ\ket{\Psi(0)},
\end{align}
while the inhomogeneous term adds a term, resulting in
\begin{align}\label{eq1300}
\hatQ\ket{\Psi(t)} &=\ee^{-\ii \QHQ t/\hbar}  \hatQ\ket{\Psi(0)}
\notag\\
&+\frac{1}{\ii\hbar}\int_0^t\dd\tau\ \ee^{-\ii \QHQ (t-\tau)/\hbar}\QHP\qty(  \hatP\ket{\Psi(\tau)}).
\end{align}
Note that this solution itself holds for all real $t$.

\begin{figure}
\includegraphics[width=0.4\columnwidth]{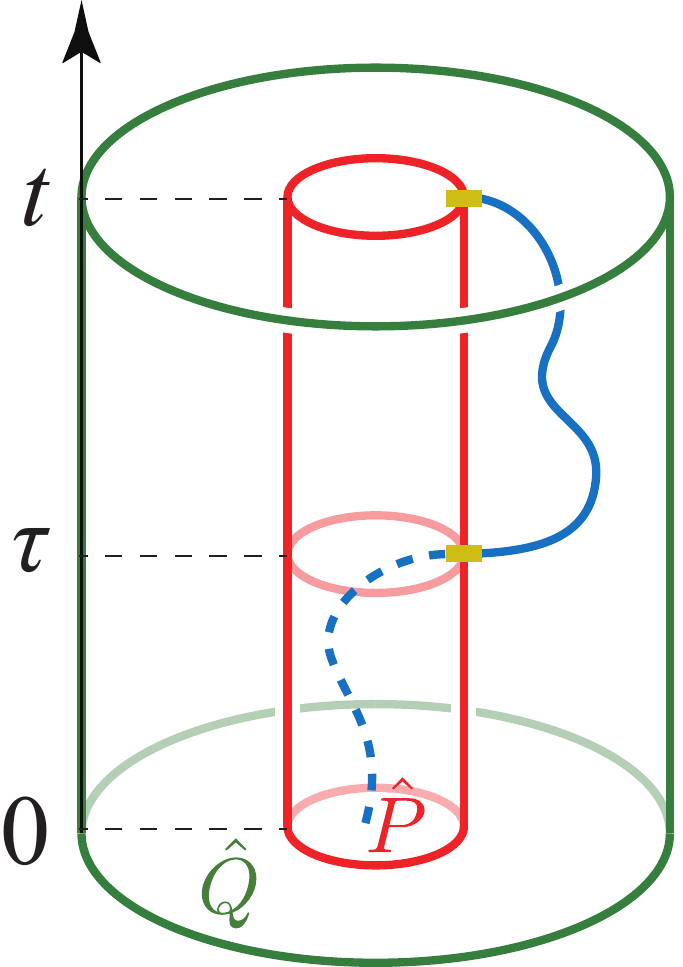}
\caption{A schematic view of time evolution described by the second term on the right-hand side of Eq.~\eqref{eq1310}.}
\label{fig18}
\end{figure}

Let us assume for explanatory purposes that the state is concentrated in the $\hatP$ subspace at $t=0$, and hence $\hatQ\ket{\Psi(0)}=0$.
This eliminates the first term on the right-hand side of Eq.~\eqref{eq1300}.
Inserting its second term into the second term on the right-hand side of Eq.~\eqref{eq1280a}, we arrive at
\begin{align}\label{eq1310}
&\ii\hbar\dv{t}\qty(\hatP\ket{\Psi(t)})=\PHP\qty(\hatP\ket{\Psi(t)})
\notag\\
&+\frac{1}{\ii\hbar}\int_0^t\dd\tau\ \PHQ\ee^{-\ii \QHQ (t-\tau)/\hbar}\QHP\qty(  \hatP\ket{\Psi(\tau)}).
\end{align}
We can interpret the second term on the right-hand side of this integro-differential equation as follows; see Fig.~\ref{fig18}.
The particle in the state $\ket{\Psi(\tau)}$, which has evolved in time up to $\tau$, goes out of the system (the $\hatP$ subspace) to the environment (the $\hatQ$ subspace) due to the element $\QHP$, evolves in the environment due to the Hamiltonian $\QHQ$ during the time $(t-\tau)$, and comes back to the system due to the element $\PHQ$ at time $t$.
From the point of view of the $\hatP$ subspace, this means that the time evolution of the system at time $t$ depends on the states of the system in the past.
In other words, the dynamics is non-Markovian because the environment serves as a memory.

This non-Markovianity emerges in the dynamics of the survival probability~\eqref{eq1210} as the integral on the real axis in Fig.~\ref{fig17}.
For example, $\QHQ$ in Eq.~\eqref{eq800d} is Fourier-transformed to the form
\begin{align}
\int_{-\pi/a}^{\pi/a} \sin^2 (ka) \dd k\  E(k)\dyad{k},
\end{align}
and hence the memory kernel in Eq.~\eqref{eq1300} reads
\begin{align}
\int_{-\pi/a}^{\pi/a} \sin^2 (ka)\dd k \  \ee^{-\ii E(k) (t-\tau))/\hbar},
\end{align}
giving a representation in terms of the Bessel function, which we will describe in the next Subsec.~\ref{subsec4-2}.

On the other hand, if the environment consisted of a one-dimensional massless Dirac particle with the dispersion relation $E(k)=ck$, where $c$ is the speed of light, instead of the tight-binding particle, the memory kernel would read
\begin{align}
\int_{-\infty}^\infty \dd k \ \ee^{-\ii ck (t-\tau)/\hbar}=\frac{\hbar}{c} \delta(t-\tau),
\end{align}
and hence the non-Markovian memory effect would exactly vanish.

\subsection{Short-time and long-time deviation from the Markovian decay}
\label{subsec4-2}


Let us now examine the dynamics exemplified in Fig.~\ref{fig16}(a) further.
The simplest explanation of the smooth behavior around $t=0$ for the tight-binding model is given as follows~\cite{Zeno}.
In the expression of the survival probability~\eqref{eq1210}, we expand the time-evolution operator with respect to $t$ up to the second order.
Assuming that $\braket{\Psi(0)|\Psi(0)}=1$, while both $\braket{\Psi(0)|\hatH|\Psi(0)}$ and $\braket{\Psi(0)|{\hatH}^2|\Psi(0)}$ are real and finite, we have the expansion
\begin{align}
\braket{\Psi(0)|\ee^{-\ii \hatH t/\hbar}|\Psi(0)}&\simeq 1-\ii\frac{t}{\hbar}\braket{\Psi(0)|\hatH|\Psi(0)}
\notag\\
&-\frac{t^2}{2\hbar^2}\braket{\Psi(0)|{\hatH}^2|\Psi(0)}+\cdots.
\end{align}
Since the first and third terms on the right-hand side are real, while the second term is imaginary, we obtain
\begin{align}\label{eq1360}
&\Ps=\abs{\braket{\Psi(0)|\ee^{-\ii \hatH t/\hbar}|\Psi(0)}}^2
\notag\\
&\simeq\qty(1-\frac{t^2}{2\hbar^2}\braket{\Psi(0)|{\hatH}^2|\Psi(0)})^2+\qty(\frac{t}{\hbar}\braket{\Psi(0)|\hatH|\Psi(0)})^2
\notag\\
&\simeq 1-\frac{t^2}{\hbar^2}\qty(\braket{\Psi(0)|{\hatH}^2|\Psi(0)}-\braket{\Psi(0)|\hatH|\Psi(0)}^2)
\end{align}
We thereby conclude that the curve of $\Ps$ in Fig.~\ref{fig16}(a) is quadratic in $t$~\cite{Zeno};
note that the quantity in the parentheses on the final line is positive.

\begin{figure}
\includegraphics[width=0.9\columnwidth]{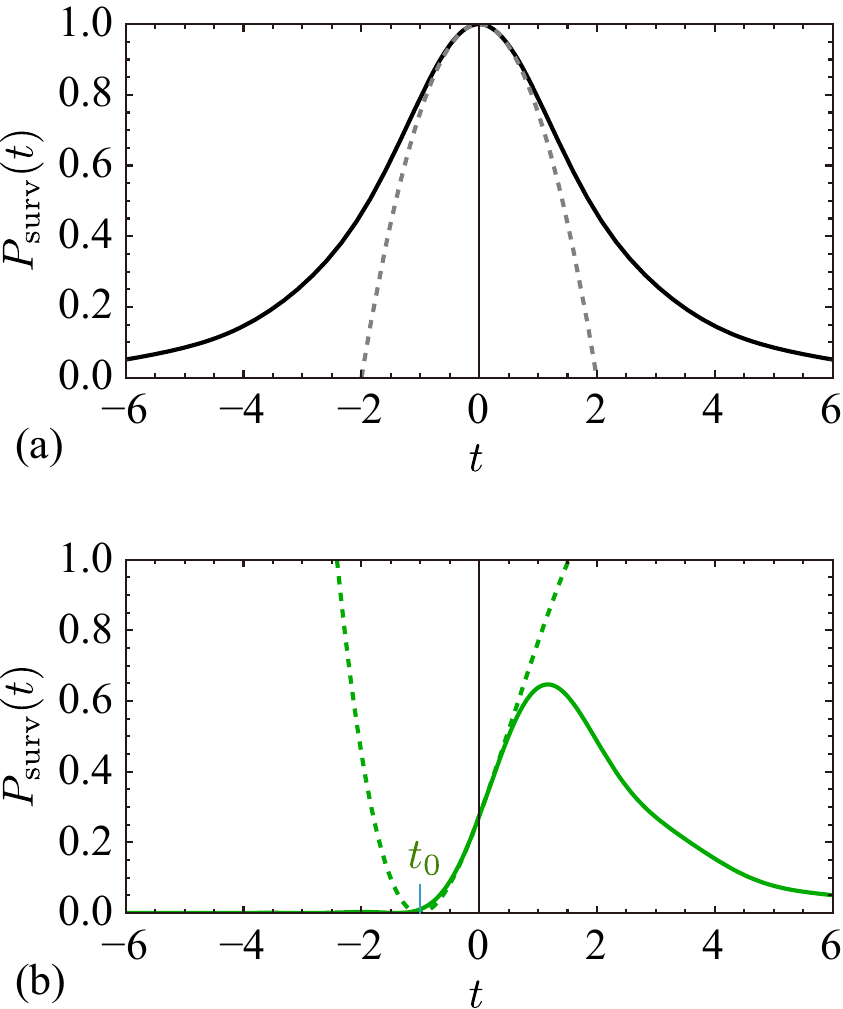}
\caption{(a) The total survival probability $\Ps$, as plotted in Fig.~\ref{fig16}(a) as a black solid curve, and the quadratic curve $1-{w_1}^2t^2$ based on Eq.~\eqref{eq1360}.
(b) The resonant component $\abs{c_1(t)}^2$, as plotted in  Fig.~\ref{fig16}(a) as a green solid curve on the right, and its short-time approximation based on Eq.~\eqref{eq1380}.}
\label{fig19}
\end{figure}

In the case of our specific model of Eq.~\eqref{eq800}, the approximate expression~\eqref{eq1360} of the survival probability with the initial state~\eqref{eq1205} reads $1-{w_1}^2t^2$, which we plot in Fig.~\ref{fig19}(a) with a broken curve.
We can see that the top part is indeed quadratic, as given in Eq.~\eqref{eq1360}.
This quadratic behavior is essential for the occurrence of the quantum Zeno effect~\cite{Zeno}; see App.~\ref{appF} for a brief introduction to the quantum Zeno effect.

We next examine the smooth exchange between the anti-resonant and resonant components in Fig.~\ref{fig16}(a).
In App.~\ref{appE3}, we present yet another expression of the time-evolution operator in Eqs.~\eqref{eq1180} and~\eqref{eq1190}.
We here give an expression only for the resonant component $n$~\cite{Ordonez17a}:
\begin{align}\label{eq1200}
&\frac{a}{\pi\ii }
\hatP\dyad{\psi_n}\hatP\ \ee^{-\ii E_n t/\hbar}
\notag\\
&\times 
\qty[1-\ii\ee^{+\ii K_na}\int_0^t\dd t'\ee^{\ii E_n t' /\hbar}\frac{J_1(2Wt'/\hbar)}{t'}],
\end{align}
where $J_1$ is the Bessel function of the first kind.
In the short-time region, we may approximate the Bessel function with $J_1(2Wt'/\hbar)\simeq Wt'/\hbar$, obtaining
\begin{align}\label{eq1380}
&\frac{a}{\pi\ii }
\hatP\dyad{\psi_n}\hatP\ 
 \ee^{-\ii E_n t/\hbar}
\qty[1-\frac{W\ee^{+\ii K_na}}{E_n}\qty(\ee^{\ii E_n t /\hbar}-1)].
\end{align}
The component $n=1$ of this expression gives an approximate estimate of the resonant component of the survival probability, $\abs{c_1(t)}^2$, plotted in Fig.~\ref{fig16}(a).

The result of the approximation~\eqref{eq1380} is plotted in Fig.~\ref{fig19}(b) with a broken curve.
It approximates the short-time behavior quite well.
Indeed, its minimum allows us to estimate the time at which the resonant component first appears in the negative time domain.
(Note that the numerical evaluation reveals that the broken curve in Fig.~\ref{fig19}(b) barely touches the real axis.)
Let us define this time scale, which is denoted by $-t_0$ here, in the form
\begin{align}
1-\frac{W\ee^{+\ii K_1a}}{E_n}\qty(\ee^{\ii E_1 t_0 /\hbar}-1)=0,
\end{align}
which yields
\begin{align}
-t_0&=\frac{\ii\hbar}{ E_1}\ln\qty(1+\frac{E_1}{W\ee^{+\ii K_1a}})
\notag\\
&=\frac{\ii \hbar}{E_1}\ln\qty(-\ee^{-2\ii K_1a}),
\end{align}
because $E_1=-W(\ee^{+\ii K_1a}+\ee^{-\ii K_1a})$.
For the phase of $\ln(-1)$, we choose $+\pi\ii$ to match the result in Fig.~\ref{fig19}(b), having~\cite{Ordonez17a}
\begin{align}\label{eq1410}
-t_0=\frac{\hbar(2K_1a-\pi)}{E_1}.
\end{align}
Using the values $K_1$ and $E_1$ in Table~\ref{tab1} for the specific parameter set $v_0=0$ and $w_1=1/2$ with $\hbar=W=a=1$, we evaluate Eq.~\eqref{eq1410} as $1.01079 + \ii 0.0142551$. Ignoring the small imaginary part, we have a good estimate of the minimum of the broken green curve in Fig.~\ref{fig19}(b), as indicated by a tick on the real axis.

We note that the estimate~\eqref{eq1410} is of the same order of magnitude as the ``Zeno time" defined by $t_\textrm{Z}=1/E$ in Ref.~\cite{Petrosky02}, where $E$ is the unperturbed energy of the excited state.
The Zeno time $t_\textrm{Z}$ marks the time scale of the initial non-Markovian dynamics, which makes the quantum Zeno effect possible, as we emphasize in App.~\ref{appF}.

Let us finally analyze non-Markovian dynamics in the long-time regime.
One way to see it is to modify the expression~\eqref{eq1200} to
\begin{align}\label{eq1420}
&\frac{a}{\pi\ii }
\hatP\dyad{\psi_n}\hatP\ \ee^{-\ii E_n t/\hbar}
\notag\\
&\times 
\qty[1+\frac{\ii\ee^{+\ii K_na}}{1-\ee^{+2\ii K_na}}\int_t^\infty\dd t'\ee^{\ii E_n t' /\hbar}\frac{J_1(2Wt'/\hbar)}{t'}].
\end{align}
Because the Bessel function has a power $J_1(z)\sim1/\sqrt{z}$ for large $z$, we can guess that the long-time behavior of the survival probability is~\cite{Garmon13,Hatano14}
\begin{align}\label{eq1425}
\Ps(t)\sim t^{-3}.
\end{align}
We can indeed find this power from the saddle-point approximation of the integral over the real line of $\lambda$ of Eq.~\eqref{eq1180};
see App.~\ref{appG}. 

\begin{figure}
\includegraphics[width=0.75\columnwidth]{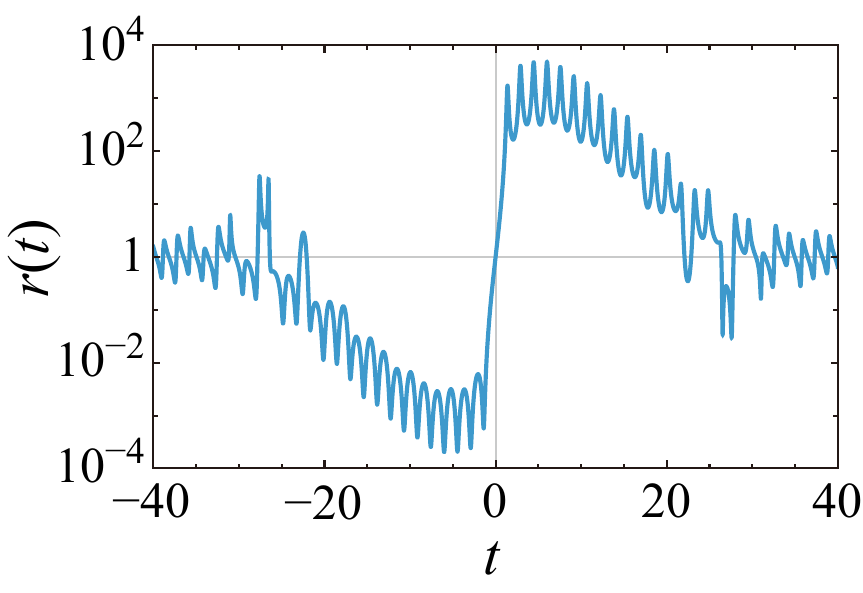}
\caption{A semi-logarithmic plot of the time dependence of the ratio between the resonant and anti-resonant components. 
We used the same parameter values as in Fig.~\ref{fig16}(a); $v_0=0$ and $w_1=1/2$ with $\hbar=W=a=1$.}
\label{fig20}
\end{figure}

The power $t^{-3}$ on the long-time regime is usually very difficult to observe experimentally, because it usually emerges only after the exponential decay.
We show in Fig.~\ref{fig20} the time dependence of the ratio between the resonant and anti-resonant components plotted in Fig.~\ref{fig16}(a):
\begin{align}
r(t):=\frac{\abs{c_1(t)}^2}{\abs{c_2(t)}^2}.
\end{align}
This shows the following two points: it confirms the gradual switching from the anti-resonant component to the resonant component over the time scale $2\abs{t_0}$, as demonstrated in Fig.~\ref{fig19}(b); 
it also shows that the balance between the resonant and anti-resonant components is restored in the long-time regime, around $\abs{t}>30$ in Fig.~\ref{fig20}.
This is the region where the long-time power law emerges.

\section{Conclusions}
\label{sec5}

In this article, we have overviewed the physics of open quantum systems within the one-body problem.
We first defined the resonant and anti-resonant states using the Siegert boundary condition, and found that they have complex eigenvalues, although the Hamiltonian is seemingly Hermitian. 
We stress again that the Hermiticity of the Hamiltonian depends on the functional space.
Indeed, the resonant and anti-resonant states are outside the space of normalizable functions, and therefore legitimately have complex eigenvalues.
We also showed a physical view of the resonant and anti-resonant states, based on which we proved the conservation of probability in spite of the fact that the eigenfunctions diverge in space.

We next introduced the Feshbach formalism, which enabled us to find the resonant and anti-resonant eigenvalues as complex eigenvalues of the effective Hamiltonian of the central system of a finite size.
Transforming the quadratic eigenvalue equation for the effective Hamiltonian, we proved completeness of the set that consists of all discrete eigenvalues, including the resonant and anti-resonant states.
This allows us to break down non-Markovian dynamics in a time-reversal-symmetric way.

We hope that this article convinced readers that interesting issues remain unexplored even in the domain of the one-body problem.
We also hope that this review stimulates new developments in formulating theories of open quantum systems in strong-coupling regimes with interactions.
In the Introduction, we have classified problems in open quantum systems with arbitrarily strong couplings between the system and the environment into the following three categories: 
(i) the class of no interactions within the system or the environment, which we have explored in the present article;
(ii) the class of strong interactions within the system, but no interactions within the environment;
(iii) the class of strong interactions, both within the system and the environment.

In fact, solutions to problems in the second class are emerging.
A.~Nishino and one of the present authors (N.H.)~\cite{Nishino24,Nishino26} formulated the Siegert boundary condition for a double quantum dot with Columb interactions, and obtained an effective Hamiltonian in a similar way to obtain Eq.~\eqref{eq580}.
This implies that the Feshbach formalism applies to the second class of the problem, yielding a new complete set that includes the resonance poles.

For the third class of problems, we may obtain a density matrix $\rho$ rather than a pure state after eliminating the environmental degrees of freedom.
We then have to extend the present analysis to the Liouville-von Neumann equation:
\begin{align}
\ii\dv{t}\rho(t)=\comm{\hatH}{\rho(t)}=:\mathcal{L}\rho,
\end{align}
where $\mathcal{L}$ is called the Liouvillian.
Since the state is a matrix rather than a vector, the extension is not straightforward, but the present approach may provide a clue;
the Liouvillian is also a linear operator, anyway.

The Nakajima-Zwanzig formalism~\cite{Nakajima58,Zwanzig60} is one way of extension using the  projection operator $\hatP$ in the form $\hatP\rho=(\Tr_\textrm{env}\rho)\otimes\rho_\textrm{eq}$,
where $\Tr_\textrm{env}$ represents the trace operation over the environmental degrees of freedom and $\rho_\textrm{eq}$ denotes the equilibrium density matrix of the environment.
This formalism, however, cannot be continued exactly as we did in the one-body problem, and one must resort to some form of Markovian approximation in the end.
A breakthrough choice of the projection operators $\hatP$ and $\hatQ$ would be crucial in a successful extension of the present formalism to this class of problems; see \textit{e.g.} Ref.~\cite{Kinkawa24}.

\section*{Acknowledgments}
The present authors gratefully appreciate intimate discussions with Dr.~Tomio~Petrosky.
N.H.~acknowledges financial support by JSPS KAKENHI Grant Numbers JP24K00545, P23K22411, JP21H01005, and JP19H00658.

\appendix
\renewcommand{\thefigure}{\thesection\arabic{figure}}

\section{Finding scattering solutions of the potential~\eqref{eq50}}
\label{appA-1}
\setcounter{figure}{0}

Here we solve the scattering problem of the \Sch equation~\eqref{eq10} with the potential~\eqref{eq50}.
We find solutions of the form~\eqref{eq40}.
We further assume a wave function in the potential range in the form
\begin{align}\label{eqA10}
\psi(x)=\begin{cases}
J\ee^{\ii kx}+M\ee^{-\ii k x} &\quad\mbox{for $-\ell<x<0$},\\
F\ee^{\ii kx}+G\ee^{-\ii k x} &\quad\mbox{for $0<x<+\ell$},
\end{cases}
\end{align}
and set the connection conditions at $x=0$ and $x=\pm \ell$ as follows:
\begin{subequations}\label{eqA20}
\begin{align}
\psi(-\ell-\epsilon)&=\psi(-\ell+\epsilon),\\
\psi'(-\ell-\epsilon)&=\psi'(-\ell+\epsilon)-\frac{2mV_1}{\hbar^2}\psi(-\ell),\\
\psi(-\epsilon)&=\psi(+\epsilon),\\
\psi'(-\epsilon)&=\psi'(+\epsilon)+\frac{2mV_0}{\hbar^2}\psi(0),\\
\psi(+\ell-\epsilon)&=\psi(+\ell+\epsilon),\\
\psi'(+\ell-\epsilon)&=\psi'(+\ell+\epsilon)-\frac{2mV_1}{\hbar^2}\psi(+\ell),
\end{align}
\end{subequations}
where $\epsilon$ is an infinitesimal positive number.
The equations for the first derivatives are given by integrating the \Sch equation over an infinitesimal range around each delta potential.

Using the connection conditions~\eqref{eqA20}, we have
\begin{widetext}
\begin{subequations}\label{eqA30}
\begin{align}
A\ee^{-\ii k \ell}+B\ee^{+\ii k \ell}&=J\ee^{-\ii k \ell}+M\ee^{+\ii k \ell},\\
\ii k A\ee^{-\ii k \ell}-\ii k B\ee^{+\ii k \ell}&=\ii k J\ee^{-\ii k \ell}-\ii k M\ee^{+\ii k \ell}
-v_1\qty(A\ee^{-\ii k \ell}+B\ee^{+\ii k \ell}+J\ee^{-\ii k \ell}+M\ee^{+\ii k \ell}),\\
J+M&=F+G,\\
\ii k J-\ii k M&=\ii k F-\ii k G+v_0\qty(J+M+F+G),\\
F\ee^{\ii k \ell}+G\ee^{-\ii k \ell}&=C\ee^{+\ii k \ell},\\
\ii k F\ee^{+\ii k \ell}-\ii k G\ee^{-\ii k \ell}&=\ii k C\ee^{+\ii k \ell}
-v_1\qty(F\ee^{\ii k \ell}+G\ee^{-\ii k \ell}+C\ee^{+\ii k \ell}),
\end{align}
\end{subequations}
\end{widetext}
where
\begin{align}
v_i:=\frac{mV_i}{\hbar^2}\quad\mbox{for}\quad i=1,2.
\end{align}
We can cast the connection conditions~\eqref{eqA30} into the following matrix equations:
\begin{widetext}
\begin{subequations}\label{eqA50}
\begin{align}
\begin{pmatrix}
\ee^{-\ii k \ell} & \ee^{+\ii k \ell} \\
(\ii k+v_1) \ee^{-\ii k \ell} & (-\ii k+v_1) \ee^{+\ii k \ell}
\end{pmatrix}
\begin{pmatrix}
A \\ B
\end{pmatrix}
&=
\begin{pmatrix}
\ee^{-\ii k \ell} & \ee^{+\ii k \ell} \\
(\ii k-v_1) \ee^{-\ii k \ell} & (-\ii k-v_1) \ee^{+\ii k \ell}
\end{pmatrix}
\begin{pmatrix}
J \\ M
\end{pmatrix},
\\
\begin{pmatrix}
1 & 1 \\
\ii k-v_0  & -\ii k-v_0 
\end{pmatrix}
\begin{pmatrix}
J \\ M
\end{pmatrix}
&=
\begin{pmatrix}
1 & 1 \\
\ii k+v_0 & -\ii k+v_0
\end{pmatrix}
\begin{pmatrix}
F \\ G
\end{pmatrix},
\\\label{eqA50c}
\begin{pmatrix}
\ee^{+\ii k \ell} & \ee^{-\ii k \ell} \\
(\ii k+v_1) \ee^{+\ii k \ell} & (-\ii k+v_1) \ee^{-\ii k \ell}
\end{pmatrix}
\begin{pmatrix}
F \\ G
\end{pmatrix}
&=
\begin{pmatrix}
\ee^{+\ii k \ell} & \ee^{-\ii k \ell} \\
(\ii k-v_1) \ee^{+\ii k \ell} & (-\ii k-v_1) \ee^{-\ii k \ell}
\end{pmatrix}
\begin{pmatrix}
C \\ 0
\end{pmatrix},
\end{align}
\end{subequations}
where we have taken the right-hand side of the third equation~\eqref{eqA50c} in its present form to obtain the transfer matrix in a full form.
We thereby find
\begin{align}
\begin{pmatrix}
A \\ B
\end{pmatrix}
&=
\frac{1}{\qty(2\ii k)^3}
\begin{pmatrix}
(\ii k-v_1) \ee^{+\ii k \ell} & \ee^{+\ii k \ell} \\
(\ii k+v_1) \ee^{-\ii k \ell} & -\ee^{-\ii k \ell} 
\end{pmatrix}
\begin{pmatrix}
\ee^{-\ii k \ell} & \ee^{+\ii k \ell} \\
(\ii k-v_1) \ee^{-\ii k \ell} & -(\ii k+v_1) \ee^{+\ii k \ell}
\end{pmatrix}
\nonumber\\
&\times
\begin{pmatrix}
\ii k+v_0 & 1 \\
\ii k-v_0  & -1  
\end{pmatrix}
\begin{pmatrix}
1 & 1 \\
\ii k+v_0 & -(\ii k-v_0)
\end{pmatrix}
\nonumber\\
&\times
\begin{pmatrix}
(\ii k-v_1) \ee^{-\ii k \ell} & \ee^{-\ii k \ell} \\
(\ii k+v_1) \ee^{+\ii k \ell} & -\ee^{+\ii k \ell} 
\end{pmatrix}
\begin{pmatrix}
\ee^{+\ii k \ell} & \ee^{-\ii k \ell} \\
(\ii k-v_1) \ee^{+\ii k \ell} & -(\ii k+v_1) \ee^{-\ii k \ell}
\end{pmatrix}
\begin{pmatrix}
C \\ 0
\end{pmatrix}.
\end{align}

After a straightforward algebra, we find the transfer matrix $T$ in the form
\begin{align}\label{eqA70}
\begin{pmatrix}
A \\ B
\end{pmatrix}
&=T
\begin{pmatrix}
C \\ 0
\end{pmatrix}
\end{align}
with
\begin{align}\label{eqA80}
T_{11}=\frac{\ii}{k^3}\qty[\qty(\ii k-v_1)^2\qty(\ii k+v_0)+2v_0v_1\qty(\ii k-v_1)\ee^{2\ii k \ell}-{v_1}^2(\ii k-v_0)\ee^{4\ii k \ell}].
\end{align}
\end{widetext}
%
We thereby obtain the transmission amplitude in the form
\begin{align}
t_\textrm{amp}&:=\frac{C}{A}=\frac{1}{T_{11}}
\end{align}
and the transmission coefficient in the form
\begin{align}
T&:=\abs{\frac{C}{A}}^2=\frac{1}{\abs{T_{11}}^2}.
\end{align}

\section{Transision from a resonant-anti-resonant pair to a bound-anti-bound pair}
\label{appB}
\setcounter{figure}{0}

\begin{figure}
\includegraphics[width=0.95\columnwidth]{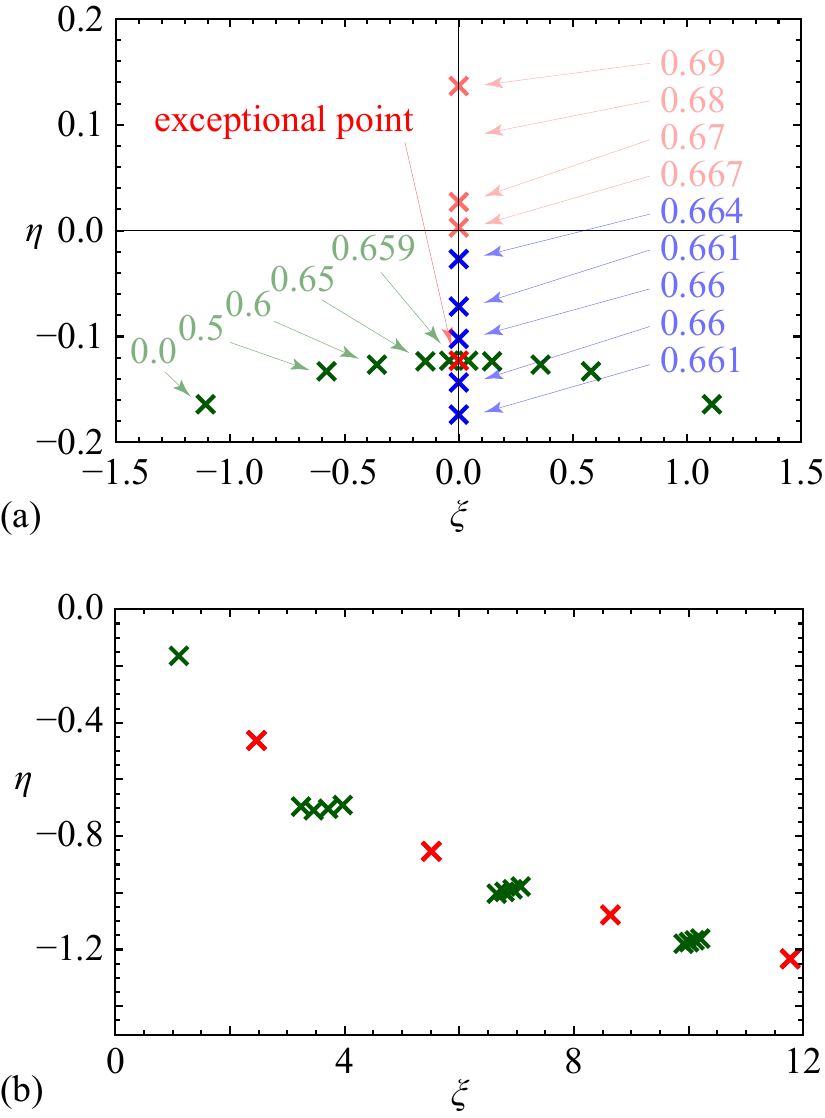}
\caption{The movement of the discrete eigenvalues. (a) The eigenvalues closest to the imaginary axis of even parity. The numbers indicate the value of $\alpha_0$, while $\alpha_1=1$ is fixed.
For $0\le\alpha_0\le0.659$, a pair of resonant and anti-resonant eigenvalues of even parity (green crosses) approach the imaginary part.
They collide on the imaginary axis at $\eta=-0.1228574213\cdots$ for $\alpha_0=0.6598057357\cdots$ (a red cross), which is a second-order exceptional point.
For $0.66\le\alpha_0< 3/2$, they stay at two anti-bound states (blue crosses). For $\alpha=2/3$, one of the anti-bound state climbing on the imaginary axis passes the zero, and turns to a bound state (orange crosses) for $\alpha_0>2/3$.
(b) Other eigenvalues. The eigenvalues that move to the left are of even parity (green crosses), for $\alpha_0=0$, $1$, $2$, and $3$, while $\alpha_1=1$ is fixed. The eigenvalues of odd parity (red crosses) do not depend on $\alpha_0$.}
\label{figB1}
\end{figure}

We describe here in more detail what happens between the situation in Fig.~\ref{fig4} with $\alpha_0=0$ and $\alpha_1$ and the one in Fig.~\ref{fig6} with $\alpha_0=3$ and $\alpha_1=1$. 
We first show in Fig.~\ref{figB1}(a) that, as we increase $\alpha_0$ from zero to a positive value, a pair of resonant and anti-resonant states of even parity collides on the negative imaginary axis, which is a second-order exceptional point, and then turns into a pair of two anti-bound states.
One of the anti-bound states climbs along the imaginary axis and eventually becomes a bound state.

If we reverse the order of the description, a bound state for a deep attractive potential, as we make the potential shallower, turns first into an anti-bound state, collides with another anti-bound state on the negative imaginary axis, and splits into a pair of resonant and anti-resonant states. 
When a bound state vanishes upon making the potential shallower, one might imagine it would turn into a scattering state, but it is not correct.
The truth is that the number of discrete eigenvalues is conserved, except at the exceptional point, where the two eigenvalues coalesce, and we have one less rank of the functional space.

\begin{figure}
\includegraphics[width=0.95\columnwidth]{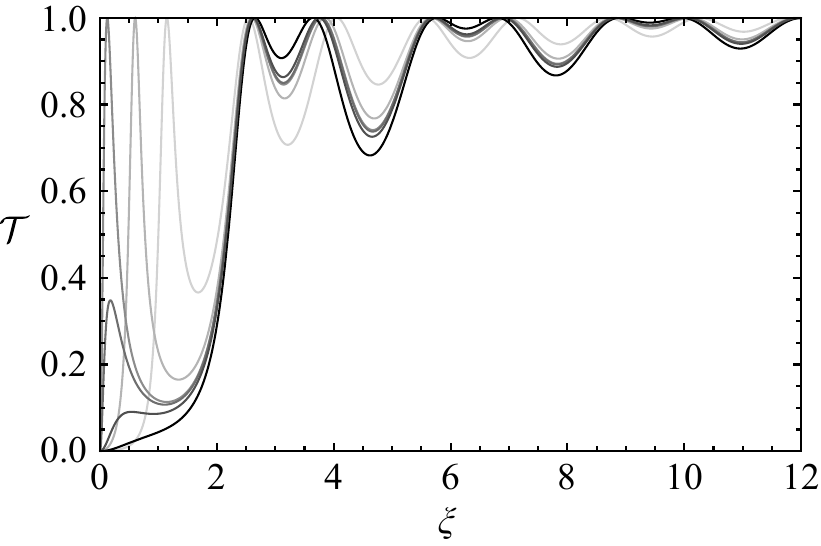}
\caption{The variation of the transmission coefficient $\mathcal{T}(\xi)$ due to the change of $\alpha_0$ from $0$ to $0.5$, the exceptional point $0.6598057357\cdots$, $0.68$, $0.75$, and $1.0$ (gradation from thin gray to black), with $\alpha_1=1$ fixed.}
\label{figB2}
\end{figure}

Figure~\ref{figB1}(b) shows that other eigenvalues of even parity also move towards the imaginary axis as we increase the attractive potential $\alpha_0$, approaching the neighboring eigenvalues of odd parity, which do not move.
This results in the resonance peaks of even parity merging those of odd parity, as demonstrated in Fig.~\ref{figB2}.
We note that the change is quite rapid in parallel with the process of starting from the collision of the resonant and anti-resonant states and with the appearance of the bound state.
We also note that for a very large $\alpha_0$, Eq.~\eqref{eq120} reduces to Eq.~\eqref{eqA120} with the $+$ sign, and hence the poles of even parity converge to the poles of odd parity.

\section{Introduction to the tight-binding model}
\label{appC}
\setcounter{figure}{0}

We briefly overview the basics of the tight-binding model, which we intensively use in Subsec.~\ref{subsec2-4} and Sec.~\ref{sec3}.
We focus on the one-dimensional model.
In the main text, we omit for brevity the subscripts TB and Sch used in the present Appendix.

\begin{figure}
\includegraphics[width=0.95\columnwidth]{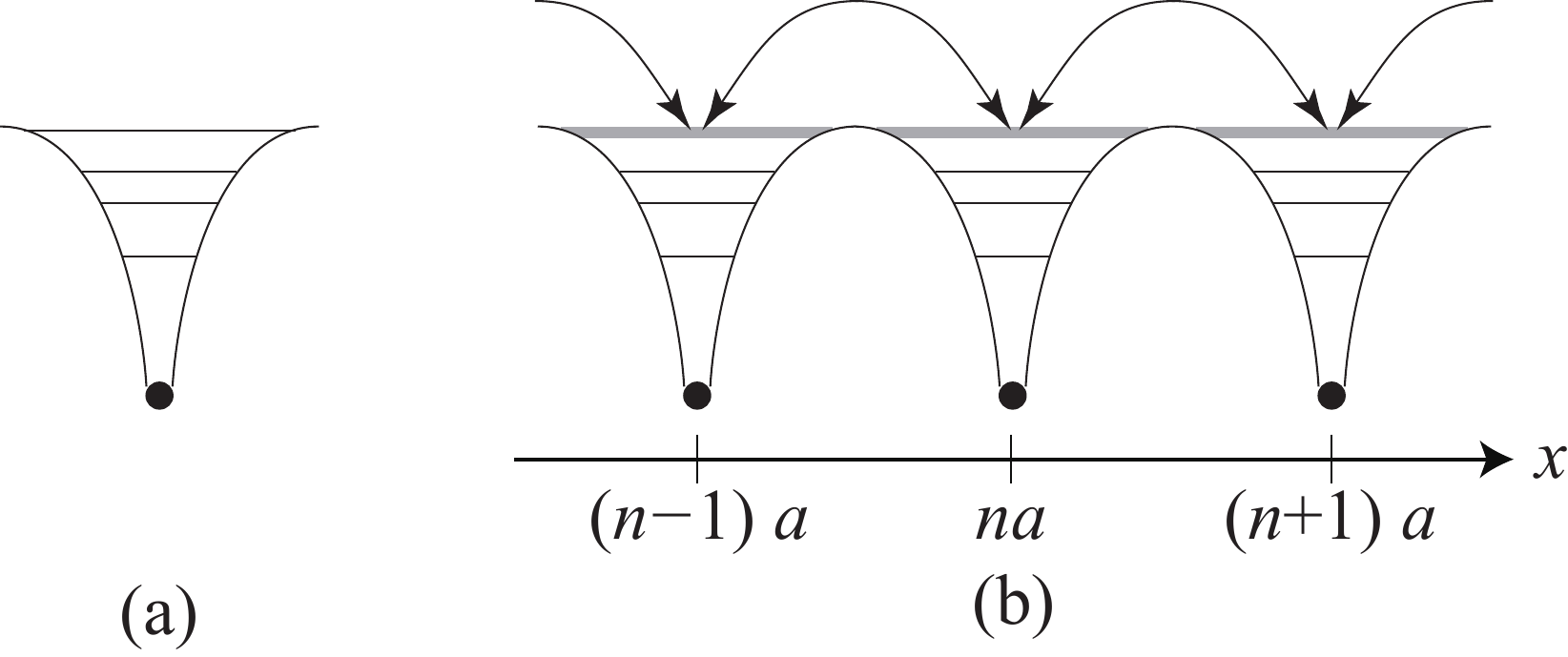}
\caption{(a) One nucleus with several bound states. (b) A series of nuclei in a solid state. The uppermost bound state of a nuclide (indicated by grey thick lines) may now tunnel to the state of the next nucleus.}
\label{figC1}
\end{figure}

The tight-binding model was originally introduced to describe a solid-state system.
Consider first one nucleus in vacuum, as schematically shown in one dimension in Fig.~\ref{figC1}(a).
Suppose that the Coulomb interaction is strong and the nucleus can have states that are tightly bound to it.

For a series of nuclei in a solid state, the uppermost bound state of each nucleus may tunnel to the corresponding state of the nearest neighboring nuclei, as is shown in Fig.~\ref{figC1}(b).
This tunneling can make the series of uppermost bound states conductive.
The effective Hamiltonian of this situation may be given by
\begin{align}\label{eqC10}
\hatH_\textrm{TB}:=-W\sum_{n=-\infty}^\infty\qty(\ketbra{n+1}{n}+\ketbra{n}{n+1}),
\end{align}
where $\ket{n}$ represents an electronic state localized at the $n$th nucleus at $x=na$. 
This is called the (one-dimensional) tight-binding model, because the electronic state is tightly bound to each nucleus and can hop only to the nearest neighbors.

We can solve the eigenvalue equation $\hatH_\textrm{TB}\ket{\psi}=E_\textrm{TB}\ket{\psi}$ in the following way.
Because of the discrete translational symmetry, the eigenstates are given by
\begin{align}\label{eqC20}
\ket{k}:=\frac{1}{\sqrt{2\pi}}\sum_{n=-\infty}^\infty \ee^{\ii k na}\ket{n}.
\end{align}
The straightforward calculation reveals that its energy eigenvalue is 
\begin{align}\label{eqC30}
E_\textrm{TB}=-2W \cos(ka),
\end{align}
which is often called the cosine band, because the energy eigenvalues only exist in the energy band $-2W\leq E_\textrm{TB}\leq 2W$.
Note that the wave number $k$ only in the first Brillouin zone $-\pi/a<k\leq\pi/a$ is relevant, because the state with $k+2\pi/a$ in Eq.~\eqref{eqC20} reduces to the state with $k$.

There is another view of the tight-binding model.
For clarity, we write down the equation for the wave function.
In the equation
\begin{align}\label{eqC40}
\braket{n|\hatH_\textrm{TB}|\psi}=E_\textrm{TB}\braket{n|\psi},
\end{align}
we introduce the notation $\psi_n:=\braket{n|\psi}$.
Putting the Hamiltonian~\eqref{eqC10} into Eq.~\eqref{eqC40} leads to
\begin{align}\label{eqC50}
-W\qty(\psi_{n-1}+\psi_{n+1})=E_\textrm{TB}\psi_n.
\end{align}
This equation is regarded as a discretization of the \Sch equation
\begin{align}\label{eqC60}
-\frac{\hbar^2}{2m}\dv[2]{x}\psi(x)=E_\textrm{Sch}\psi(x),
\end{align}
where
\begin{align}\label{eqC65}
E_\textrm{Sch}:=\frac{\hbar^2}{2m}k^2.
\end{align}
The second-order derivative with respect to $x$ is discretized as in
\begin{align}\label{eqC70}
-\frac{\hbar^2}{2m}
\frac{\displaystyle \frac{\psi(x+a)-\psi(x)}{a}
-\frac{\displaystyle \psi(x)-\psi(x-a)}{a}}{a}
=E_\textrm{Sch}\psi(x),
\end{align}
which is reduced to the form
\begin{align}\label{eqC80}
-\frac{\hbar^2}{2ma^2}\qty(\psi(x-a)+\psi(x+a))=\qty(E_\textrm{Sch}-\frac{\hbar^2}{ma^2})\psi(x).
\end{align}
Under the identification
\begin{subequations}
\begin{align}\label{eqC90a}
W&\longleftrightarrow \frac{\hbar^2}{2ma^2},\\
\label{eqC90b}
\psi_n&\longleftrightarrow \psi(na),\\
\label{eqC90c}
E_\textrm{TB}&\longleftrightarrow E_\textrm{Sch}-\frac{\hbar^2}{ma^2},\\
\intertext{or equivalently}
\label{eqC90d}
E_\textrm{TB}+2W&\longleftrightarrow E_\textrm{Sch},
\end{align}
\end{subequations}
Eqs.~\eqref{eqC50} and~\eqref{eqC80} are equivalent to each other.
The correspondence~\eqref{eqC90a} is the reason why we put the negative sign in front of $W$ in Eq.~\eqref{eqC10}.

Note that the lattice constant $a$ in the tight-binding model is a small discretization parameter in Eq.~\eqref{eqC70}.
Indeed, expanding the energy eigenvalue~\eqref{eqC30} with respect to $a$ up to the second order, we have the identification
\begin{align}
E_\textrm{TB}\simeq -2W +W a^2 k^2 \longleftrightarrow  -\frac{\hbar^2}{ma^2}+E_\textrm{Sch}.
\end{align}
We thereby realize that the bottom of the cosine band~\eqref{eqC30} of the tight-binding model approximates the quadratic dispersion~\eqref{eqC65} of the free particle in the continuum space.

Introducing the potential to the tight-binding Hamiltonian~\eqref{eqC10} is therefore straightforward.
Discretizing the \Sch equation with the potential
\begin{align}\label{eqC120}
\qty(-\frac{\hbar^2}{2m}\dv[2]{x}+V(x))\psi(x)=E_\textrm{Sch}\psi(x)
\end{align}
instead of Eq.~\eqref{eqC60}, we have the wave equation
\begin{align}
-W(\psi_{n-1}+\psi_{n+1})+V_n\psi_n=E_\textrm{TB}\psi_n
\end{align}
instead of Eq.~\eqref{eqC50}, and hence the tight-binding Hamiltonian
\begin{align}
\hatH_\textrm{TB}&=-W\sum_{n=-\infty}^\infty\qty(\ketbra{n+1}{n}+\ketbra{n}{n+1})
\nonumber\\
&+\sum_{n=-\infty}^\infty V_n\ketbra{n}
\end{align}
instead of Eq.~\eqref{eqC10}.

Finally, we mention the sublattice symmetry of the one-dimensional tight-binding model.
Suppose that we found an eigenfunction $\{\psi_n\}$ that satisfies Eq.~\eqref{eqC50}. 
We then find that the following function is also an eigenfunction, but with the energy eigenvalue $-E_\textrm{TB}$.
\begin{align}\label{eqC150}
\psi'_n:=(-1)^n\psi_n.
\end{align}
In fact, the transformation~\eqref{eqC150} is equivalent to shift the wave number $k$ by $\pi/a$ in Eq.~\eqref{eqC20},
which indeed flips the sign of the energy eigenvalue in Eq.~\eqref{eqC30}.

Consider next the case in which we have an impurity potential only at the origin:
\begin{align}\label{eqC160}
\hatH_\textrm{TB}&=-W\sum_{n=-\infty}^\infty\qty(\ketbra{n+1}{n}+\ketbra{n}{n+1})
+V_0\ketbra{0}.
\end{align}
If $V_0$ is negatively large enough, we can imagine from the analogue to the \Sch equation~\eqref{eqC120} that the model~\eqref{eqC160} has a bound state around the impurity at the origin,
and its energy eigenvalue is less than $-2W$, which corresponds to the zero energy in the \Sch equation.
In the complex wave-number plane, it exists on the positive side of the imaginary axis: $\Re K=0$ and $\Im K>0$.

Flipping the sign of $V_0$ along with the transformation~\eqref{eqC150} then yields a bound state around the impurity, but with an energy eigenvalue greater than $+2W$.
In the complex wave-number plane, it exists on the axis $\Re K= \pi/a$ with $\Im K>0$, because the transformation~\eqref{eqC150} shifts the wave number by $\pi/a$.
The appearance of a bound state with the eigen-wave-number $\Re K=\pi/a$ never happens in the \Sch equation, for which all bound states exist only on the positive side of the imaginary axis.

\section{Calculation of the Green's function in Eq.~\eqref{eq790}}
\label{appD}
\setcounter{figure}{0}

We calculate the second term on the right-hand side of Eq.~\eqref{eq790} straightforwardly here.
Because the environmental Hamiltonian $\QHQ$ is separated into the right and left sides of the system, as seen in Fig.~\ref{fig13}, let us focus on the right side. 
The matrix $E-\QHQ$ is a semi-infinite-dimensional matrix of the form
\begin{align}
E-QHQ:=
\begin{pmatrix}
E&W&&&&& \\
W&E&W&&&\text{\huge{0}}& \\
&W&E&W&&& \\
&&W&E&W&& \\
&&&W&E&\ddots& \\
&\text{\huge{0}}&&&\ddots&\ddots&
\end{pmatrix}.
\end{align}
Since the Green's function $(E-\QHQ)^{-1}$ is sandwiched by the operators $\PHQ$ in Eq.~\eqref{eq800b} from the left and $\QHP$ in Eq.~\eqref{eq800c} from the right, we only need the element $\braket{2a|(E-\QHQ)^{-1}|2a}$.

\begin{figure}
\includegraphics[width=0.8\columnwidth]{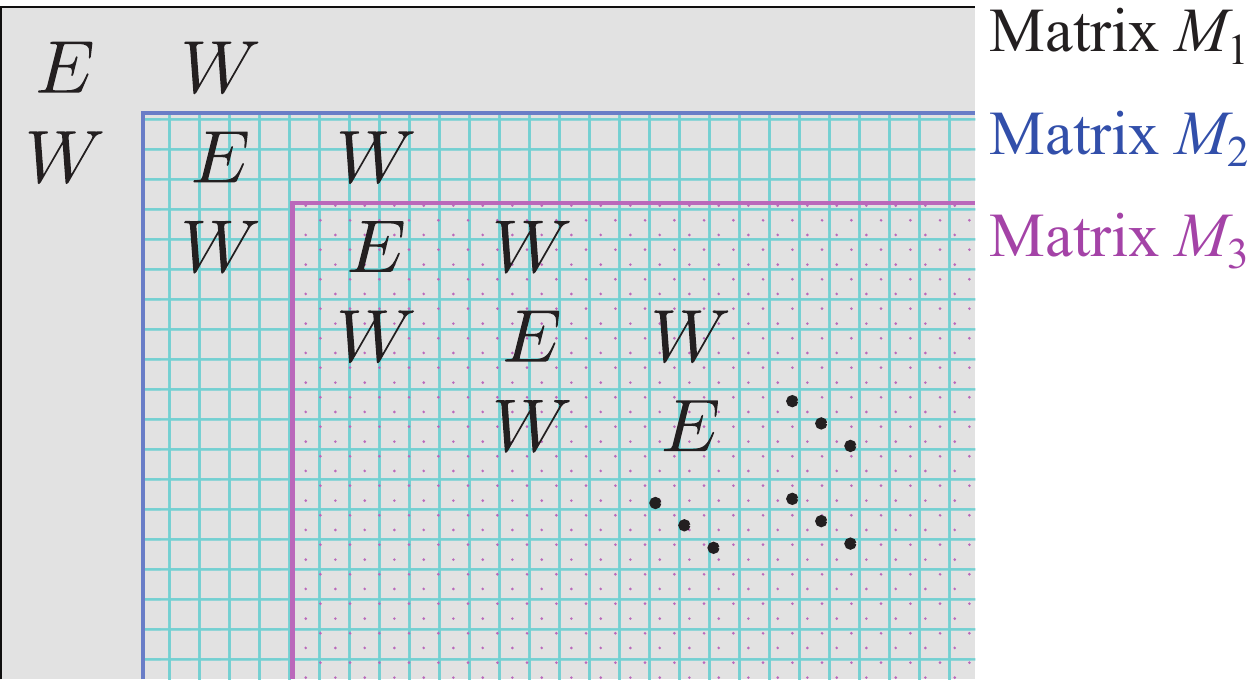}
\caption{Gray area, cross-hatched area, and dotted area of the matrix $E-\QHQ$ are denoted by $M_1$, $M_2$, and $M_3$, respectively.}
\label{figD1}
\end{figure}

For explanatory purposes, we let the matrices in Fig.~\ref{figD1} be denoted by $M_1$, $M_2$, and $M_3$.
We thereby calculate the element $({M_1}^{-1})_{11}$, which is given by
\begin{align}\label{eqD20}
G:=({M_1}^{-1})_{11}=\frac{\det M_2}{\det M_1}.
\end{align}
We next rewrite the denominator $\det M_1$ using the cofactor expansion with respect to the first row of the matrix $M_1$.
We find
\begin{align}
\det M_1=E\det M_2 -W^2\det M_3,
\end{align}
which is followed by
\begin{align}\label{eqD40}
G=\frac{1}{\displaystyle E-W^2\frac{\displaystyle \det M_3}{\displaystyle \det M_2}}.
\end{align}
We now assume that
\begin{align}
\frac{\displaystyle \det M_3}{\displaystyle \det M_2}
\end{align}
is equal to $G$ defined by Eq.~\eqref{eqD20}, because all matrices $M_1$, $M_2$, and $M_3$ are semi-infinite-dimensional.

Therefore, we cast Eq.~\eqref{eqD40} into the form
\begin{align}
G=\frac{1}{E-W^2 G},
\end{align}
which produces the second-order equation of $G$:
\begin{align}
W^2 G^2 -EG+1=0.
\end{align}
The solution
\begin{align}
G=\frac{E\pm\sqrt{E^2-4W^2}}{2W^2}
\end{align}
is transformed to
\begin{align}\label{eqD90}
G=-\frac{1}{W}\qty[\cos (Ka)\mp \ii \sin(Ka)]=-\frac{1}{W}\ee^{\mp\ii Ka}
\end{align}
when we used the dispersion relation $E=-2W\cos(Ka)$.
The upper minus sign on the right-hand side of Eq.~\eqref{eqD90} gives the incoming wave, or the advanced Green's function, while the lower positive sign gives the outgoing wave, or the retarded Green's function.

Adopting the retarded one to obtain the outgoing-wave boundary condition, we find
\begin{align}
\braket{2a|\frac{1}{E-\QHQ}|2a}=-\frac{1}{W}\ee^{+\ii Ka}
\end{align}
for the right side of the system, and similarly
\begin{align}
\braket{-a|\frac{1}{E-\QHQ}|-a}=-\frac{1}{W}\ee^{+\ii Ka}
\end{align}
for the left side of the system. Summarizing them, we have
\begin{widetext}
\begin{align}
\PHQ\frac{1}{E-\QHQ}\QHP&=
{W_1}^2\dyad{0}{-a}\frac{1}{E-\QHQ}\dyad{-a}{0}
+{W_1}^2\dyad{a}{2a}\frac{1}{E-\QHQ}\dyad{2a}{a}
\nonumber\\
&=-\frac{{W_1}^2}{W}\ee^{+\ii Ka}\qty(\dyad{0}+\dyad{a}),
\end{align}
\end{widetext}
which gives the terms of $\Sigma$ in Eq.~\eqref{eq810}.

\clearpage

\section{Analytic calculations in SubSecs.~\ref{subsec3-3} and~\ref{subsec4-2}}
\label{appE}
\setcounter{figure}{0}

\subsection{An algebraic proof of the formula~\eqref{eq1120}}
\label{appE1}

We here prove Eq.~\eqref{eq1120} using the resolvent expansion:
\begin{align}\label{eqE10}
\frac{1}{\hatA-\hatB}=\frac{1}{\hatA}+\frac{1}{\hatA}\hatB\frac{1}{\hatA}+\frac{1}{\hatA}\hatB\frac{1}{\hatA}\hatB\frac{1}{\hatA}+\cdots.
\end{align}
This expansion is a consequence of a simple version of the Dyson equation:
\begin{align}\label{eqE20}
\frac{1}{\hatA-\hatB}=\frac{1}{\hatA}+\frac{1}{\hatA}\hatB\frac{1}{\hatA-\hatB}.
\end{align}
We can prove Eq.~\eqref{eqE20} by multiplying its both sides by $\hatA$ from the left and $\hatA-\hatB$ from the right,
which reduces Eq.~\eqref{eqE20} to a trivial equation $\hatA=(\hatA-\hatB)+\hatB$.
Then, by repeatedly inserting the left-hand side of the Dyson equation~\eqref{eqE20} into the last term on the right-hand side, we obtain the resolvent expansion~\eqref{eqE10}.

Let us split the total Hamiltonian $\hatH$ into the two parts
\begin{subequations}
\begin{align}\label{eqE30a}
\hatH_0&:=\PHP+\QHQ,
\\
\label{eqE30b}
\hatH_1&:=\PHQ+\QHP.
\end{align}
\end{subequations}
We then analyze the resolvent expansion
\begin{widetext}
\begin{align}\label{eqE40}
\hatP\frac{1}{E-\hatH}\hatP
=\hatP\frac{1}{E-\hatH_0}\hatP
+\hatP\frac{1}{E-\hatH_0}\hatH_1\frac{1}{E-\hatH_0}\hatP
+\hatP\frac{1}{E-\hatH_0}\hatH_1\frac{1}{E-\hatH_0}\hatH_1\frac{1}{E-\hatH_0}\hatP
+\cdots.
\end{align}
\end{widetext}
In the first term on the right-hand side, only $\PHP$ survives in $\hatH_0$ in Eq.~\eqref{eqE30a} in the denominator because it is sandwiched by $\hatP$ from the both sides:
\begin{align}
\hatP\frac{1}{E-\PHP}\hatP.
\end{align}
In the second term on the right-hand side of Eq.~\eqref{eqE40}, again only $\PHP$ would survive in the first Green's function, and hence $\PHQ$ would survive in $\hatH_1$ in Eq.~\eqref{eqE30b}, which then results in the second Green's function sandwiched by $\hatQ$ from the left and $\hatP$ from the right.
Consequently, this term vanishes. 
In the third term on the right-hand side of Eq.~\eqref{eqE40}, again only $\PHP$ survives in the first Green's function, and again $\PHQ$ survives in the next $\hatH_1$. This time, in the denominator of the second Green's function, $\QHQ$ survives out of $\hatH_0$, and therefore $\QHP$ survives in the next $\hatH_1$, and finally in the denominator of the third Green's function, $\PHP$ survives out of $\hatH_0$.
As a consequence, this term is reduced to
\begin{align}
\hatP\frac{1}{E-\PHP}\PHQ\frac{1}{E-\QHQ}\QHP\frac{1}{E-\PHP}\hatP.
\end{align}
We notice that the self-energy term~\eqref{eq810} emerges in the middle, as in
\begin{align}
\hatP\frac{1}{E-\PHP}\hat{\Sigma}\frac{1}{E-\PHP}\hatP,
\end{align}
where we redefine it as the operator
\begin{align}
\hat{\Sigma}:=\PHQ\frac{1}{E-\QHQ}\QHP,
\end{align}
simplifying a generalized expression of Eq.~\eqref{eq810}

Repeating the above argument term by term, we notice that only the terms of even orders of $\hatH_1$ survive in the expansion~\eqref{eqE40}, and we have
\begin{widetext}
\begin{align}\label{eqE90}
\hatP\frac{1}{E-\hatH}\hatP
=\hatP\frac{1}{E-\hatH_0}\hatP
+\hatP\frac{1}{E-\hatH_0}\hat{\Sigma}\frac{1}{E-\hatH_0}\hatP
+\hatP\frac{1}{E-\hatH_0}\hat{\Sigma}\frac{1}{E-\hatH_0}\hat{\Sigma}\frac{1}{E-\hatH_0}\hatP
+\cdots,
\end{align}
\end{widetext}
which we can sum up to
\begin{align}\label{eqE100}
\hatP\frac{1}{E-\hatH}\hatP
=\hatP\frac{1}{E-\PHP-\hat{\Sigma}}\hatP,
\end{align}
using the resolvent expansion~\eqref{eqE10} in the reverse way.
Equation~\eqref{eqE100} is equivalent to the formula~\eqref{eq1120} because $\Heff=\PHP+\hat{\Sigma}$.

\subsection{Derivation of Eq.~\eqref{eq1160}}
\label{appE2}

We here derive Eq.~\eqref{eq1160} by relating its left-hand side to the inverse of the matrix
\begin{align}
\hatA-\lambda\hatB=\begin{pmatrix}
-\lambda \iden_N & \iden_N \\
\iden_N & \Hsys+\lambda\hat{\Theta}
\end{pmatrix}.
\end{align}
This matrix is block-diagonalized as in
\begin{align}\label{eqE120}
\hatX(\lambda)\qty(\hatA-\lambda\hatB)\hatY(\lambda)
=\hatZ(\lambda) ,
\end{align}
where
\begin{subequations}
\begin{align}
\hatX(\lambda)&:=
\begin{pmatrix}
-\Hsys-\lambda\hat{\Theta} & \iden_N \\
\iden_N & \zero_N
\end{pmatrix},
\\
\hatY(\lambda)&:=
\begin{pmatrix}
\iden_N & \zero_N \\
\lambda \iden_N & \iden_N
\end{pmatrix},
\\
\hatZ(\lambda)&:=
\begin{pmatrix}
\hat{\Theta}\lambda^2+\Hsys\lambda+\iden_N & \zero_N \\
\zero_N & \iden_N
\end{pmatrix}.
\end{align}
\end{subequations}
Note here that the $(1,1)$ block of the $(2N)\times (2N)$ matrix $\hatZ(\lambda)$ is related to $E-\Heff(E)$ as in Eq.~\eqref{eq865}, and hence
\begin{align}\label{eqE140}
\hatP\frac{1}{E-\Heff(E)}\hatP&=
-\frac{\lambda}{W}\hatC^T\hatZ^{-1}\hatC,
\end{align}
where $\hatC$ is the $(2N)\times N$ matrix that extracts the $(1,1)$ block, 
as defined specifically in Eq.~\eqref{eq975}, and generally by
\begin{align}
\hatC:=\begin{pmatrix}
\iden_N  \\
\zero_N
\end{pmatrix}.
\end{align}
(The left-hand side of Eq.~\eqref{eqE140} is an expression in the whole Hilbert space, while its right-hand side is an expression of an $N\times N$ matrix, and hence we should not equate them, strictly speaking, but we here loosely use the equality because there is no possibility of misunderstanding.)

In order to find Eq.~\eqref{eq1120}, we therefore invert both sides of Eq.~\eqref{eqE120}. and obtain
\begin{align}\label{eqE141}
\hatP\frac{1}{E-\Heff(E)}\hatP
&=-\frac{\lambda}{W}\hatC^T\hatY^{-1}\frac{1}{\hatA-\lambda\hatB}\hatX^{-1}\hatC,
\end{align}
Since we have
\begin{subequations}
\begin{align}
\hatX^{-1}&=\begin{pmatrix}
\zero_N &\iden_N \\
\iden_N & \Hsys+\lambda\hat{\Theta} 
\end{pmatrix},
\\
\hatY^{-1}&=\begin{pmatrix}
\iden_N & \zero_N \\
-\lambda \iden_N & \iden_N
\end{pmatrix},
\end{align}
\end{subequations}
we find
\begin{subequations}
\begin{align}\label{eqE170a}
\hatX^{-1}\hatC&=\hatD:=\begin{pmatrix}
\zero_N \\
\iden_N
\end{pmatrix},\\
\hatC^T\hatY^{-1}&=\hatC^T,
\end{align}
\end{subequations}
and therefore we obtain
\begin{align}\label{eqE180}
\hatP\frac{1}{E-\Heff(E)}\hatP&=
-\frac{\lambda}{W}\hatC^T
\frac{1}{\hatA-\lambda\hatB}\hatD.
\end{align}

We finally expand the inverse of the $(2N)\times(2N)$ matrix $\hatA-\lambda\hatB$ with respect to the $2N$ eigenstates $\{\ket{\Psi_n}\}$.
For this purpose, we introduce the diagonalizing matrix 
\begin{subequations}
\begin{align}\label{eqE190a}
\hatU&:=\begin{pmatrix}
\ket{\Psi_1} & \ket{\Psi_2} & \cdots & \ket{\Psi_{2N}}
\end{pmatrix}
\notag\\
&=\begin{pmatrix}
\ket{\psi_1} & \ket{\psi_2} & \cdots & \ket{\psi_{2N}}\\
\lambda_1\ket{\psi_1} & \lambda_2\ket{\psi_2} & \cdots & \lambda_{2N}\ket{\psi_{2N}}
\end{pmatrix},
\\\label{eqE190b}
\hattilU&:=\begin{pmatrix}
\bra{\tilde{\Psi}_1} \\
\bra{\tilde{\Psi}_2} \\
\vdots\\
\bra{\tilde{\Psi}_{2N}} 
\end{pmatrix}
=\begin{pmatrix}
\bra{\tilde{\psi}_1} &\lambda_1\bra{\tilde{\psi}_1} \\
\bra{\tilde{\psi}_2} &\lambda_2\bra{\tilde{\psi}_2} \\
\vdots & \vdots\\
\bra{\tilde{\psi}_{2N}} & \lambda_{2N}\bra{\tilde{\psi}_{2N}} 
\end{pmatrix},
\end{align}
\end{subequations}
where we align the $2N$ column vectors in the former and the $2N$ row vectors in the latter.
Because of Eq.~\eqref{eq980}, we find
\begin{align}
\hattilU\qty(\hatA-\lambda\hatB)\hatU=\hat{\Lambda}-\lambda \iden_{2N},
\end{align}
where
\begin{align}
\hat{\Lambda}:=\begin{pmatrix}
\lambda_1 & && \text{\huge{0}} \\
& \lambda_2 & & \\
& & \ddots & \\
\text{\huge{0}}& & & \lambda_{2N}
\end{pmatrix}.
\end{align}
Therefore, the inverse reads
\begin{subequations}
\begin{align}
\frac{1}{\hatA-\lambda\hatB}
&=\hatU\frac{1}{\hat{\Lambda}-\lambda \iden_{2N}}\hattilU
\\\label{eqE220b}
&=\sum_{n=1}^{2N}\ket{\Psi_n}\frac{1}{\lambda_n-\lambda}\bra{\tilde{\Psi}_n}.
\end{align}
\end{subequations}
Applying $\hatC^T$ from the left of Eq.~\eqref{eqE220b} picks the first row of Eq.~\eqref{eqE190a}, while applying $\hatD$ in Eq.~\eqref{eqE170a} from the right picks the second column of Eq.~\eqref{eqE190b}, and hence Eq.~\eqref{eqE180} now reads
\begin{align}
\hatP\frac{1}{E-\Heff(E)}\hatP&=
-\frac{\lambda}{W}
\sum_{n=1}^{2N}\ket{\psi_n}\frac{1}{\lambda_n-\lambda}\lambda_n\bra{\tilde{\psi}_n}
\notag\\
&=\frac{1}{W}\sum_{n=1}^{2N}\ket{\psi_n}\frac{\lambda\lambda_n}{\lambda-\lambda_n}\bra{\tilde{\psi}_n}
\end{align}
This gives Eq.~\eqref{eq1160}.

\subsection{Derivation of Eq.~\eqref{eq1200}}
\label{appE3}

We here derive Eq.~\eqref{eq1200} from Eq.~\eqref{eq1190}.
The integration contour $C_3$ for Eq.~\eqref{eq1190} consists of the integration over the range $[-\pi,\pi]$ and the contributions of bound states, as shown in Fig.~\ref{fig15}(c).

We take the amplitude of one resonant component out of Eq.~\eqref{eq1190} and let it be denoted by
\begin{align}
\chi_n:=\frac{a}{\pi\ii}
\int_{-\pi/a}^{\pi/a}\dd k 
\frac{\ee^{-\ii E(k) t/\hbar}\sin(ka)}{\ee^{-\ii ka}-\ee^{-\ii K_n a}}.
\end{align}
Multiplying and dividing the integrand by the same factor $(\ee^{+\ii ka}-\ee^{-\ii K_n a})$, we obtain
\begin{align}
\chi_n=\frac{Wa}{\pi\ii}\int_{-\pi/a}^{\pi/a} \dd k
\frac{\ee^{-\ii E(k) t/\hbar}(\ee^{+\ii ka}\ee^{+\ii K_n a}-1)\sin(ka)}{E(k)-E_n}  ,
\end{align}
where $E_n=-2W\cos(K_na)=-W(\lambda_n-1/\lambda_n)$ is the energy eigenvalue of the resonant state.
Assuming $\Im E_n<0$ for a resonant state, we introduce an integral representation of the denominator, as in
\begin{align}\label{eqE260}
\chi_n&=\frac{Wa}{\pi\ii}\int_{-\pi/a}^{\pi/a}\dd k\ \ee^{-\ii E(k) t/\hbar}(\ee^{+\ii ka}\ee^{+\ii K_n a}-1)\sin(ka)
\notag\\
&\times\frac{-\ii}{\hbar}\int_0^\infty \dd \tau \ \ee^{\ii\tau (E(k)-E_n)/\hbar}.
\end{align}
If $E_n$ is the real eigenvalue of a bound state, we add an infinitesimal imaginary part $-\ii\varepsilon$ to $E_n$ to make the $\tau$ integral in Eq.~\eqref{eqE260} convergent, and then take the limit $\varepsilon\to0$.
If $\Im E_n>0$ for an anti-resonant state, the range of the $\tau$ integral in Eq.~\eqref{eqE260} should be from $-\infty$ to $0$.
We here proceed with the transformation, assuming the resonant state.

The odd terms with respect to $k$ in the integrand in Eq.~\eqref{eqE260} vanish, and thereby we have
\begin{align}\label{eqE270}
\chi_n&=\frac{Wa\ \ee^{+\ii K_n a}}{\pi\hbar\ii}\int_0^\infty \dd \tau \ \ee^{-\ii\tau E_n/\hbar}
\notag\\
&\times \int_{-\pi/a}^{\pi/a}\dd k\ \sin^2(ka)
\ \ee^{\ii(\tau-t) E(k)/\hbar}.
\end{align}
Using an integral representation of the Bessel function of the first kind (8.411.7 of Ref.~\cite{TableofIntegralsChap8}),
\begin{align}
J_\nu(z)=\frac{(z/2)^\nu}{\Gamma(\nu+1/2)\Gamma(1/2)}\int_0^\pi \ee^{\pm\ii z\cos\varphi}\sin^{2\nu}\varphi\ \dd\varphi
\end{align}
we find the $k$ integral in Eq.~\eqref{eqE270} in the form
\begin{align}
\int_{-\pi}^{\pi}\dd k\ \sin^2(ka)
&\ \ee^{2\ii(t-\tau) W\cos(ka)/\hbar}
\notag\\
&=\frac{\pi\hbar}{Wa} \frac{J_1(2W(t-\tau)/\hbar)}{t-\tau}.
\end{align}
We thus arrive at
\begin{align}
\chi_n=-\ii\ee^{+\ii K_na}I(E_n,t),
\end{align}
where
\begin{align}
I(E_n,t):=\int_0^\infty\dd\tau\ \ee^{-\ii \tau E_n/\hbar}\frac{J_1(2W(t-\tau)/\hbar)}{t-\tau}.
\end{align}

Let us further transform $I(E_n,t)$ by changing the integration variable from $\tau$ to $t'=t-\tau$:
\begin{align}
I(E_n,t)&=\int_{-\infty}^t \dd t'\ee^{-\ii (t-t') E_n/\hbar}\frac{J_1(2Wt'/\hbar)}{t'}
\notag\\
&=\ee^{-\ii E_nt/\hbar}\qty(\int_{-\infty}^0+\int_0^{t'}) \dd t' \ee^{\ii E_nt'/\hbar}\frac{J_1(2Wt'/\hbar)}{t'}.
\end{align}
The integral from $-\infty$ to $0$ on the right-hand side is simplified via an expression in terms of the hypergeometric function:
\begin{align}
\int_{-\infty}^0 \dd t'&\ee^{\ii E_nt'/\hbar}\frac{J_1(2Wt'/\hbar)}{t'}
\notag\\
&=\int_0^\infty \dd t'\ee^{\ii E_nt'/\hbar}\frac{J_1(2Wt'/\hbar)}{t'}
\notag\\
&=\frac{W}{\ii E_n}\text{}_2F_1\qty(\frac{1}{2},1;2;\frac{4W^2}{{E_n}^2})
\notag\\
&=\frac{W}{\ii E_n}\qty(\frac{1}{2}+\frac{1}{2}\sqrt{1-\frac{4W^2}{{E_n}^2}})^{-1}
\notag\\
&=\ii \ee^{-\ii Ka};
\end{align}
see Eqs.~(10.22.49) and (15.4.17) of \cite{NIST:DLMF}.
This leads to Eq.~\eqref{eq1200}.

\section{Introduction to the quantum Zeno effect}
\label{appF}
\setcounter{figure}{0}

The quantum Zeno effect~\cite{Zeno} occurs when the survival probability of an excited state decays quadratically, as in Eq.~\eqref{eq1360}.
For explanatory purposes, let us consider a two-level system immersed in an environment with the total Hamiltonian denoted by $\hatH$.
Let the ground and excited states of the two-level system be denoted by $\ket{\mathrm{g}}$ and $\ket{\mathrm{e}}$, respectively.
We first set the system to the excited state $\ket{\Psi(0)}=\ket{\mathrm{e}}$.
The survival probability is then given by
\begin{align}
\Ps =\abs{\braket{\mathrm{e}|\ee^{-\ii \hatH t/\hbar}|\mathrm{e}}}^2.
\end{align}
We assume that the expansion~\eqref{eq1360} is convergent for a short time, and let the quadratic decay be denoted by 
\begin{align}\label{eqF20}
\Ps\simeq 1-\gamma^2 t^2+\cdots.
\end{align}

We repeatedly perform the projection measurement on the two-level system and compute the probability that the system is found in the excited state at every measurement.
In the first projection measurement at $t=\Delta t$, the probability that we find the system to be in the excited state is $\Ps(\Delta t)\simeq 1-\gamma^2\Delta t^2$. 
With this probability, the system is projected back to the excited state at $t=\Delta t$.
In the second projection measurement at $t=2\Delta t$, the probability that we again find the system to be in the excited state is $\Ps(\Delta t)\simeq 1-\gamma^2\Delta t^2$. 
With this probability, the system is again projected back to the excited state at $t=2\Delta t$.
We repeat this $N$ times up to the time $T:=N\Delta t$.

Then the probability that we find the system to be in the excited state in every measurement is
\begin{align}\label{eqF30}
P_N(T):=\qty(1-\gamma^2\frac{T^2}{N^2})^N.
\end{align}
With this probability, the system is projected back into the excited state at every measurement.

In the limit $N\to\infty$, the probability $P_N(T)$ converges to unity, because the parentheses of Eq.~\eqref{eqF30} has $1/N^2$, not $1/N$.
Therefore, frequent measurements keep the two-level system in the excited state throughout the time.
This is called the quantum Zeno effect.
If $P_N(T)$ were $(1-\gamma T/N)^N$, the function would converge to the exponential decay $\exp(-\gamma T)$.
This shows that the quadratic decay~\eqref{eqF20} is essential for the occurrence of the quantum Zeno effect.
Since the effect may help control qubits in quantum computers, it is critical to recognize that the dynamics of open quantum systems are non-Markovian, deviating from purely exponential decay.

\section{Finding the power law $t^{-3}$ in the long-tome regime}
\label{appG}
\setcounter{figure}{0}

We here derive the long-time behavior~\eqref{eq1425} of the survival probability $\Ps(t)$ from the saddle-point approximation of Eq.~\eqref{eq1180}~\cite{Hatano14}.
After the contour modifications in Fig.~\ref{fig17}, in addition to pole contributions, we have the principal integral over the real axis of $\lambda$:
\begin{widetext}
\begin{align}\label{eqG10}
\omega_n:=\frac{1}{2\pi\ii}\mathrm{P}\int_{-\infty}^\infty\dd\lambda\ \exp\qty[\frac{\ii W}{\hbar}\qty(\lambda+\frac{1}{\lambda})]
\hatP\ket{\psi_n}\frac{1}{\lambda^{-1}-{\lambda_n}^{-1}}\bra{\tilde{\psi}_n}\hatP\qty(-1+\frac{1}{\lambda^2}),
\end{align}
\end{widetext}
where we chose the sign for the case of $t>0$ in Fig.~\ref{fig17}(a).

The saddle points of the exponent $\ii W(\lambda+1/\lambda)/\hbar$ are at $\lambda=\pm1$, which correspond to the energy minimum and maximum $E=-W(\lambda+1/\lambda)=\mp 2W$.
Indeed, the energy minimum and maximum are the branch points of the branch cut that connects the first and second Riemann sheets for the tight-binding model, as shown in Fig.~\ref{fig12}(b).
The integrals around the branch points are known to produce non-Markovian dynamics because they lack a characteristic timescale, leading to power-law decays in the long-time regime.

\begin{figure}
\includegraphics[width=0.55\columnwidth]{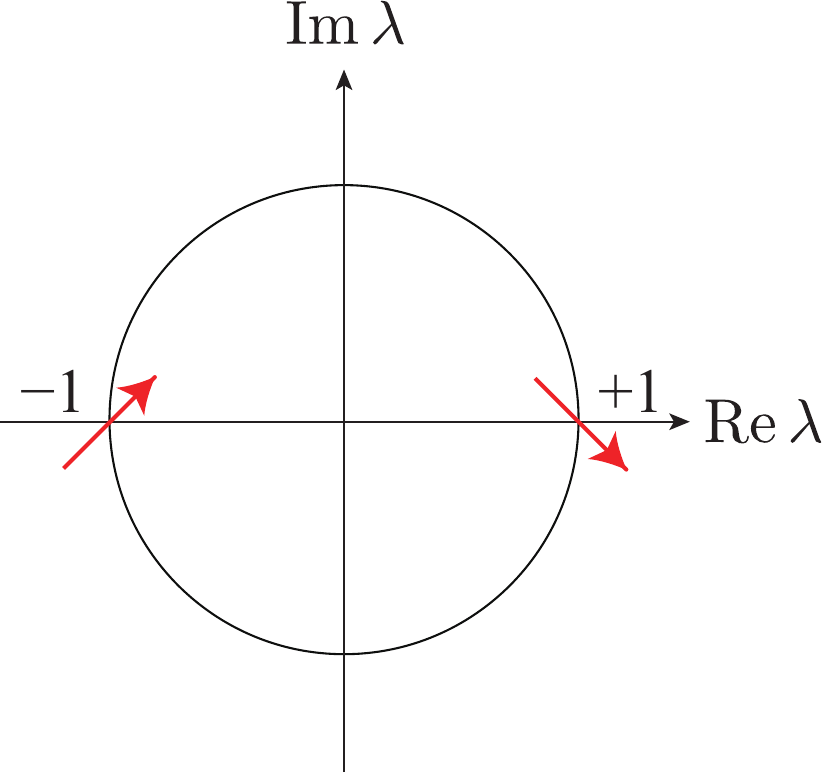}
\caption{The integration contour around the saddle points $\lambda=\pm1$ with respect to the new integration variable $s$. The circle indicates the unit circle.}
\label{figG1}
\end{figure}

We expand the exponent around each saddle point in the form
\begin{align}
\frac{\ii Wt}{\hbar}\qty(\lambda+\frac{1}{\lambda})=\frac{\ii Wt }{\hbar}\qty[\pm 2\pm \qty(\lambda\mp 1)^2+\cdots].
\end{align}
Leaving the first term on the right-hand side as a constant and taking the second term, we convert the exponential function in Eq.~\eqref{eqG10} to the Gaussian form $\exp(-s^2)$ by introducing an integration variable
\begin{align}
s:=\sqrt{\mp \frac{\ii Wt}{\hbar}}(\lambda\mp1)=\ee^{\mp\ii\pi/4}\sqrt{\frac{Wt}{\hbar}}(\lambda\mp1).
\end{align}
This rotates the integration contour with respect to $s$ as shown in Fig.~\ref{figG1}.
The resulting function is well approximated by the Gaussian $\exp(-s^2)$ only when its range, which is of the order of $1/\sqrt{t}$, is much narrower than the distance to the nearest pole, particularly the bound and anti-bound states.
In other words, the present saddle-point approximation is legitimate in a long-time regime.
In fact, Garmon \textit{et al.}~\cite{Garmon13} revealed that the proximity of bound and anti-bound states to the band edges $E=\pm 2W$ results in a different power $\Ps(t)\simeq t^{-1}$.

We thereby approximate $\omega_n$ in Eq.~\eqref{eqG10} as in
\begin{align}
&\exp\qty[\frac{\ii W}{\hbar}\qty(\lambda+\frac{1}{\lambda})]
\frac{\lambda_n}{\lambda_n-\lambda}\qty(-\lambda+\frac{1}{\lambda})
\notag\\
&\simeq
\frac{\lambda_n\ee^{-s^2}}{(\lambda_n\mp1)-\sqrt{\frac{\hbar}{Wt}}\ee^{\pm\ii\pi/4}s}\qty(-2\sqrt{\frac{\hbar}{Wt}}\ee^{\pm\ii\pi/4}s)
\notag\\
&\simeq-\frac{2\ee^{\pm\ii\pi/4}\lambda_n}{\lambda_n\mp1}\qty(1+\frac{\ee^{\pm\ii\pi/4}}{\lambda_n\mp1}\sqrt{\frac{\hbar}{Wt}}s)
\sqrt{\frac{\hbar}{Wt}}s\ee^{-s^2}
\end{align}
in the long-time limit $t\to\infty$.
Because the integral of $s\ee^{-s^2}$ vanishes, the remaining contribution is
\begin{align}
\omega_n&\simeq\ee^{\pm2\ii Wt/\hbar}\frac{\ee^{\pm\ii\pi/2}\lambda_n}{\pi\ii(\lambda_n\mp1)^2}\frac{\hbar}{Wt}\int_{-\infty}^\infty s^2\ee^{-s^2}\frac{\dd s}{\ee^{\mp\ii\pi/4}\sqrt{Wt/\hbar}}
\notag\\
&=\ee^{\pm2\ii Wt/\hbar}\frac{\sqrt{\pi}\ee^{\pm3\ii\pi/4}}{2\pi\ii(E_n\pm2)}\qty(\frac{\hbar}{Wt})^{3/2}.
\end{align}
This produces $\abs{\omega_n}^2\simeq t^{-3}$.
After all resonant contributions die out exponentially, this power law remains.

\bibliography{hatano}

\end{document}